\documentclass[useAMS,usenatbib]{mn2e}
\usepackage{graphicx,amsmath,epsfig,rotating,fleqn}

\newcommand{\figdir}
  {./}
\newlength{\figwidth}
\newlength{\thinfigwidth}
\newlength{\midfigwidth}
\newcommand{\poppars}
  {\vect{\phi}}
\newcommand{\detected}
  {{\rmn{det}}}

\newcommand{\ninitial}
  {893}
\newcommand{\nreal}
  {88}
\newcommand{\nqso}
  {seven}
\newcommand{\nqsonew}
  {four}

\newcommand{\nqsoprev}
  {three}
\newcommand{\nspec}
  {seven}

\newcommand{\nexp}
  {$\simm 10$}
\newcommand{\nsdssmain}
  {19}

\newcommand{\sdssukidss}
  {UKIDSS--SDSS}

\newcommand{\hzq}
  {HZQ}
\newcommand{\hzqs}
  {HZQs}

\newcommand{\unit}[1]
  {{\mbox{\rm\,\,#1}}}
\newcommand{\percent}
  {\,{\rmn{per\ cent}}}
\newcommand{\lya}
  {Ly\,$\alpha$}
\newcommand{\hi}
  {H\,{\sc{i}}}
\newcommand{\simm}
  {\sim \!}
\newcommand{\stonlong}
  {signal--to--noise ratio}
\newcommand{\ston}
  {S/N}
\newcommand{\stonmin}
  {S/N_{\rmn{min}}}
\newcommand{\limiting}
  {{\rmn{lim}}}
\newcommand{\qlf}
  {QLF}
\newcommand{\hatiprime}
  {\hat{i}}

\newcommand{\nbin}
  {N_{\rmn{bin}}}
\newcommand{\bin}
  {i}
\newcommand{\lf}
  {\Phi}
\newcommand{\dl}
  {D_{\rmn L}}

\newcommand{\nir}
  {NIR}

\newcommand{\uprime}
  {\mbox{$u$}}
\newcommand{\gprime}
  {\mbox{$g$}}
\newcommand{\rprime}
  {\mbox{$r$}}
\newcommand{\iprime}
  {\mbox{$i$}}
\newcommand{\zprime}
  {\mbox{$z$}}
\newcommand{\ab}
  { }
\newcommand{\vega}
  { }

\newcommand{\gab}
  {\mbox{$g\ab$}}

\newcommand{\iab}
  {\mbox{$i\ab$}}
\newcommand{\zab}
  {\mbox{$z\ab$}}
\newcommand{\yv}
  {\mbox{$Y\vega$}}
\newcommand{\jv}
  {\mbox{$J\vega$}}

\newcommand{\zmyval}
  {\mbox{$\zprime\ab$$-$$Y\vega$}}

\newcommand{\imyval}
  {\mbox{$\iprime\ab$$-$$Y\vega$}}

\newcommand{\imzobs}
  {\mbox{$\hatiprime\ab$$-$$\hat{\zprime}\ab$}}

\newcommand{\ymjobs}
  {\mbox{$\hat{Y}\vega$$\!\,-$$\hat{J}\vega$}}

\newcommand{\imyobs}
  {\mbox{$\hatiprime\ab$$-$$\hat{Y}\vega$}}

\newcommand{\imz}
  {\mbox{\iprime$-$\zprime}}
\newcommand{\zmy}
  {\mbox{\zprime$-$$Y$}}
\newcommand{\ymj}
  {\mbox{$Y$$\!\,-$$J$}}

\newcommand{\imy}
  {\mbox{\iprime$-$$Y$}}
\newcommand{\zmj}
  {\mbox{\zprime$-$$J$}}

\newcommand{\redshift}
  {z}
\newcommand{\mabsrest}
  {M}

\newcommand{\firstqso}
  {ULAS~J0203+0012}

\newcommand{\etc}
  {etc.}
\newcommand{\etal}
  {et al.}
\newcommand{\eg}
  {e.g.}
\newcommand{\cf}
  {cf.}
\newcommand{\ie}
  {i.e.}
\newcommand{\diff}
  {{\rmn{d}}}
\newcommand{\vect}[1]
  {\mbox{\boldmath ${#1}$}}

\newcommand{\eq}[1]
  {Eq.~(\ref{equation:#1})}
\newcommand{\eqs}[1]
  {Eqs~(\ref{equation:#1})}
\newcommand{\Eq}[1]
  {Equation~(\ref{equation:#1})}

\newcommand{\sect}[1]
  {Section~\ref{section:#1}}
\newcommand{\sects}[1]
  {Sections~\ref{section:#1}}

\newcommand{\apdx}[1]
  {{\mbox Appendix~\ref{section:#1}}}

\newcommand{\fig}[1]
  {Fig.~\ref{figure:#1}}
\newcommand{\figs}[1]
  {Figs.~\ref{figure:#1}}
\newcommand{\Fig}[1]
  {Figure~\ref{figure:#1}}

\newcommand{\qso}
  {{\rmn{q}}}
\newcommand{\str}
  {{\rmn{s}}}
\newcommand{\refval}
  {*}
\newcommand{\pop}
  {{t}}
\newcommand{\npop}
  {{N_{\rmn{\pop}}}}
\newcommand{\popprime}
  {{t^\prime}}

\newcommand{\prob}
  {{\rm{Pr}}}

\newcommand{\weight}
  {W}
\newcommand{\model}
  {\pop}
\newcommand{\modelprime}
  {\pop^\prime}
\renewcommand{\models}
  {\vect{\pop}}

\newcommand{\erf}
  {{\rmn{erf}}}
\newcommand{\pq}
  {P_\qso}

\newcommand{\noise}
  {\sigma}
\newcommand{\ndat}
  {N_{\rmn{d}}}
\newcommand{\nband}
  {N_{\rmn{b}}}

\newcommand{\band}
  {b}
\newcommand{\nparam}
  {N_t}

\newcommand{\parameters}
  {\vect{\theta}}
\newcommand{\parameter}  
  {\theta}
\newcommand{\datum}
  {d}
\newcommand{\data}
  {\vect{d}}
\newcommand{\flux}
  {F}
\newcommand{\fluxes}
  {\vect{\flux}}

\newcommand{\density}
  {\rho}

\newcommand{\dvol}
  {\diff \parameter_{\pop,1} \, 
   \diff \parameter_{\pop,2} \, 
   \ldots \, 
   \diff \parameter_{\pop,\nparam}}

\newcommand{\asinh}
  {{\rmn{asinh}}}
\newcommand{\sqdeg}
  {\mbox{${\rmn{deg}}^2$}}

\title
  [Probabilistic quasar selection]
  {Probabilistic selection of high-redshift quasars}
\author
  [D.\ J. Mortlock \etal]
  {Daniel J.\ Mortlock$^{1}$\thanks{E-mail: mortlock@ic.ac.uk},
  Mitesh Patel$^1$,
  Stephen J.\ Warren$^1$, 
  Paul C.\ Hewett$^2$,
  \newauthor
  Bram P.\ Venemans$^3$,
  Richard G.\ McMahon$^2$
  and 
  Chris J.\ Simpson$^4$
\vspace{3mm}\\
$^1$Astrophysics Group, Imperial College London, Blackett Laboratory,
  Prince Consort Road, London SW7 2AZ, U.K. \\
$^2$Institute of Astronomy, Madingley Road, Cambridge CB3 0HA, U.K. \\
$^3$European Southern Observatory, Karl-Schwarzschild Strasse 2, 85748 
  Garching bei M\"{u}nchen, Germany \\
$^4$Astrophysics Research Institute, Liverpool John Moores University, \\
  \,\, Twelve Quays House, Egerton Wharf, Birkenhead CH41 1LD, U.K. \\
  }
\begin{document}

\date{Received 2010 December 28} 

\pagerange{\pageref{firstpage}--\pageref{lastpage}} \pubyear{2011}

\maketitle

\label{firstpage}

\begin{abstract}
High redshift quasars (\hzqs) with redshifts of $\redshift \ga 6$ are so rare
that any photometrically-selected
sample of sources with HZQ-like colours 
is likely to be dominated by 
Galactic stars and brown dwarfs
scattered from the stellar locus.
It is impractical to reobserve all such candidates,
so an alternative approach was developed in which
Bayesian model comparison techniques are used to
calculate the probability that a candidate is a \hzq, $\pq$, by
combining models of the
quasar and star populations with the photometric measurements of the object.
This method was motivated specifically by the large number of 
\hzq\ candidates identified by cross-matching the 
UKIRT Infrared Deep Sky Survey (UKIDSS) Large Area Survey (LAS) 
to the Sloan Digital Sky Survey (SDSS):
in the $\simm 1900 \unit{\sqdeg}$ covered by the LAS in the 
UKIDSS Seventh Data Release (DR7)
there are 
$\simm 10^3$ real astronomical point--sources with the measured
colours of the target quasars,
of which only $\simm 10$ are expected to be \hzqs.
Applying Bayesian model comparison to the sample reveals that
most sources with \hzq-like colours have $\pq \la 0.1$
and can be confidently rejected without the need for any further observations.
In the case of the UKIDSS DR7 LAS,
there were just \nreal\ candidates with $\pq \geq 0.1$;
these object were prioritized for reobservation 
by ranking according to $\pq$ 
(and their likely redshift, which was also inferred from the photometric data).
Most candidates were rejected after one or two 
(moderate depth) photometric measurements 
by recalculating $\pq$ using the new data.
That left \nqso\ confirmed \hzqs, 
\nqsoprev\ of which were previously identified in the 
SDSS and \nqsonew\ of which were 
new UKIDSS discoveries.
The high efficiency of this Bayesian selection method 
suggests that it could usefully 
be extended to other \hzq\ surveys
(\eg\ searches by the
Panoramic Survey Telescope And Rapid Response System, Pan--STARRS,
or the Visible and Infrared Survey Telescope for Astronomy, VISTA)
as well as to other searches for rare objects.
\end{abstract}


%%%%%%%%%%%%%%%%%%%%%%%%%%%%%%%%%%%%%%%%%%%%%%%%%%%%%%%%%%%%%%%%%%%%%%%%%%%%%%

\section{Introduction}
\label{section:intro}

Quasars are the most luminous non-transient astronomical sources 
to redshifts of at least 
$\redshift \simeq 6.5$ and have,
ever since their discovery \citep{Schmidt:1963,Hazard_etal:1963},
been key cosmological probes (\eg\ \citealt{Schneider:1999}).
Most recently, 
observations of high-redshift quasars (\hzqs) with 
$z \simeq 6$ 
have revealed a marked
increase in the optical depth to neutral hydrogen (\hi) at redshifts
of $\redshift \ga 5.7$ \citep{Becker_etal:2001,Fan_etal:2002,Fan_etal:2006b},
which appears to mark the end of cosmological reionization
(see, \eg\ \citealt{Barkana_Loeb:2001}). 
Measurements of the 
quasar luminosity function (\qlf) at $\redshift \simeq 6$
also constrain the growth of structure and the
early formation of super-massive black holes 
in the first billion years of the Universe
(\eg\ \citealt{Jiang_etal:2008}).
There is thus a strong motivation to discover any new \hzqs,
and there is a particular premium on finding 
the most luminous 
quasars because most \hzq\ science requires
high signal--to--noise ratio spectroscopic data.
The possibility of making extensive spectroscopic observations 
of the brightest \hzqs\ also differentiates them 
from other high-redshift sources:
the $\redshift \simeq 7$ field galaxies 
found in deep surveys are too faint to obtain
high signal--to--noise ratio spectra (\eg\ \citealt{Stark_etal:2010});
and 
gamma ray bursts remain sufficiently bright for spectroscopy 
only for a few days (\eg\ \citealt{Gehrels_etal:2009}).

Despite the strong motivations for identifying bright \hzqs,  
only $\simm 50$ 
are known at present
(\eg\ \citealt{Fan_etal:2006b,Jiang_etal:2008,Willott_etal:2010}).
This is primarily because \hzqs\ are so rare:
the results of \cite{Jiang_etal:2008}
imply there are only
$\simm 450$ redshift $\redshift \geq 6.0$ quasars
brighter than $\zab = 21.0$ over the whole sky.
A direct implication is that the
bright quasars at $\redshift \ga 6$ are only likely to be found in
fairly shallow wide-area surveys.
The first such \hzq\ search was 
based on the Sloan Digital Sky Survey (SDSS; \citealt{York_etal:2000}),
which has a typical single-scan magnitude limit
of $\zab_\limiting \simeq 20.8$,
and has discovered \nsdssmain\ redshift $\ga 5.8$ quasars 
in $6600 \unit{deg}^2$
\citep{Fan_etal:2006b}.
Similar numbers of lower luminosity \hzqs\ have been found in the deeper 
SDSS Stripe 82 region 
\citep{Jiang_etal:2009}
and the 
Canada France High-$z$ Quasar Survey (CFHQS; \citealt{Willott_etal:2007}).
Continuing and future optical surveys will be able to increase the 
number of known $\redshift \simeq 6$ quasars
but will not be able to probe past
a redshift limit of $\redshift \simeq 6.4$.
Sources beyond this redshift are undetectable in optical surveys
as their
\lya\ emission is redshifted out of the optical bands
and all shorter wavelength photons 
are absorbed by intervening \hi\ (\eg\ \citealt{Gunn_Peterson:1965}).

Quasars with $\redshift \ga 6.4$ will eventually be 
identified by future radio surveys (\eg\ \citealt{Wyithe:2008}),
but the most immediate progress will be made 
by observing in the near-infrared (\nir).
The largest completed \nir\ survey,
the Two Micron All-Sky Survey (2MASS; \citealt{Skrutskie_etal:2006}),
has a detection limit
of $J_\limiting \simeq 16.6$ 
and so is too shallow to detect any \hzqs.
The partially complete 
UKIRT Infrared Deep Sky Survey (UKIDSS; \citealt{Lawrence_etal:2007})
includes a Large Area Survey (LAS) which
reaches typical depths of $\yv = 20.2$ and $\jv = 19.6$ 
  \citep{Warren_etal:2007} and has already 
yielded four new $z \simeq 6$ quasars 
(\citealt{Venemans_etal:2007,Mortlock_etal:2009a,Patel_etal:2011,
  Venemans_etal:2011}).
In the future, the 
Panoramic Survey Telescope And Rapid Response System
(Pan--STARRS; \citealt{Kaiser:2002})
and the 
Visible and Infrared Survey Telescope for Astronomy 
(VISTA; \citealt{Emerson_etal:2004}) should extend the UKIDSS results 
in both redshift and numbers,
and various planned satellite missions could increase the size 
of the \hzq\ samples by an order of magnitude
(\eg\ \citealt{Willott_etal:2010}).

The existence of surveys with the appropriate combinations of area, depth and
wavelength coverage is not, however, a sufficient condition for 
discovering \hzqs.
It is also necessary to be able to separate these rare objects of interest
from the far more numerous galaxies and Galactic stars that inevitably 
dominate the resultant source catalogues.
A survey to, \eg, $\zab \simeq 21$
would contain $\simm 10^6$ times as many Galactic stars (and galaxies)
as target \hzqs,
and it is hence
almost inevitable that the majority of sources 
which are consistent with being \hzqs\ are 
more common sources scattered by photometric noise.
This is true of 
candidate samples generated 
from low-resolution grism or objective prism spectra
(\eg\ \citealt{Schmidt_etal:1995,Hewett_etal:1995})
or from the
colour-based selection techniques 
(\eg\ \citealt{Warren_etal:1994,Fan_etal:2001,Willott_etal:2007})
considered here.

Given a sample of sources with measured colours,
how should the most promising objects be identified as quasar candidates? 
How should the candidates be prioritized for follow-up observations?
What sort of follow-up observations should be obtained 
 -- spectroscopy or photometry?
If photometric follow-up is chosen, in which band(s) and to what depth(s) 
should measurements be made?
The answers to these questions obviously depend on
the details of the survey -- 
in particular on how well the target \hzqs\ are
separated from the various contaminants in the survey's data space --
but
the basic aim of extracting as much as possible from the 
available data is generic.
In the context of rare object searches
such as \hzq\ surveys,
the primary goal is simply to maximize the number of discoveries
given finite observational resources,
although it is also desireable to make the 
search quantifiable 
to enable subsequent statistical studies of the underlying population.
In the case of \hzq\ surveys there is an additional premium 
on finding the highest-redshift objects 
(\eg\ it might be considered acceptable to miss
several $\redshift \simeq 6$ objects
if it resulted in the discovery of 
even one quasar with $\redshift \ga 6.5$). 

The most common approach to \hzq\ selection is to apply heuristic 
colour and magnitude cuts, chosen to ensure a manageable candidate
list that also (hopefully) includes most of the quasars in the survey 
(\eg\ \citealt{Warren_etal:1994,Fan_etal:2001,Willott_etal:2007}).
Both \cite{Fan_etal:2001} and \cite{Willott_etal:2007} 
selected their \hzq\ candidates in this way, 
applying carefully chosen cuts in the 
$i$, $z$ and $J$ bands to reject the vast majority of 
sources as Main Sequence stars and brown dwarfs.  
Aside from being a good pragmatic selection option,
the simplicity of hard colour cuts makes it very easy to 
quantify the completeness of the resultant quasar samples 
(\eg\ \citealt{Fan_etal:2003,Willott_etal:2010}).
Cut-based approaches do, however, have several short-comings:
the conversion from photometric measurements to a binary selection 
represents a significant loss of information;
and the most promising high signal--to--noise ratio candidates
are inevitably grouped with more numerous 
marginal candidates near the edges of the selection region.
Even if there are sufficient observational
resources to follow-up all the objects identified in this way,
it is inevitable that some worthy candidates will have been 
rejected\footnote{The limitations of
cut-based candidate selection were illustrated in the case of 
the $\redshift = 6.13$ quasar
ULAS~J131911.29$+$095051.4 \citep{Mortlock_etal:2009a}.
This source 
was detected with 
$\hatiprime = 22.83 \pm 0.32$ 
and 
$\hat{\zab} = 20.13 \pm 0.12$ in SDSS, and so satisfied
the \cite{Fan_etal:2001} requirements that 
$\hatiprime - \hat{\zab} > 2.2$
and $\hat{\zab} < 20.2$;
however it was observed in slightly worse than average conditions and
hence did not meet the additional requirement that the 
$\zprime$-band noise be $\sigma_z < 0.1$.}.

An alternative to applying hard data cuts is to adopt a probabilistic
approach to quasar selection,
replacing the construction of a definite candidate list
with a calculation of the probability,
$\pq$, that each source is a quasar.
While this idea has not been applied to $\redshift \ga 6$ quasar searches,
it has been used 
to generate large samples of lower-redshift quasars
(\citealt{Richards_etal:2004,BailerJones_etal:2008,Richards_etal:2009,
  Bovy_etal:2010}).
Most relevantly,
\cite{Richards_etal:2004}
applied kernel density estimation (KDE)
to training sets of spectroscopically-confirmed stars and quasars,
giving estimates of the observed distribution of both populations 
in SDSS colour space.
\cite{Bovy_etal:2010} used extreme deconvolution 
in place of KDE to estimate the instrinsic distribution.
In both cases, the second step was to apply 
Bayes's theorem to calculate
$\pq$ for each source in turn.
The power of these methods is adequately illustrated by 
the first result obtained by \cite{Richards_etal:2004}:
a photometric sample of $\simm 10^5$ quasars 
that is $95~\percent$ complete to $\gab \leq 21.0$
with only $5~\percent$ contamination.
The use of a prior to account for the fact that quasars 
are outnumbered by Galactic stars is important to all
the above probabilistic selection methods,
although most sources would have been classified 
decisively (and correctly) simply by comparing the 
normalised KDE quasar and star
density estimates at the sources' locations in colour space.
More critical was 
the availability of significant numbers of 
confirmed stars and quasars from
which their distributions in
the four-dimensional SDSS colour space could be inferred.
Unfortunately, 
the need for large training samples makes it problematic to 
use this selection method to search for rare objects as,
by definition, very few are known.
In the case of \hzqs, 
which have reasonably distinctive and predictable colours,
one option would be to generate a synthetic training set 
of simulated quasars.
However 
that would still not overcome the more fundamental problem 
that most sources with the observed colours of \hzqs\ are
extreme outliers from the stellar locus rather than distant quasars.
The fact that most such objects will be close to the survey's 
detection limit could be used to down-weight faint sources
with large photometric errors, although the algorithm described 
by \cite{Richards_etal:2004} would require non-trivial modifications
to account for this.
From an inferential point of view the problem is that,
by ignoring the photometric errors,
KDE of the observed colour distribution
does not utilise all the information contained in the data.
For brighter sources this should not be too important,
as there is still sufficient information to make a confident classification
in most cases;
but the inclusion of the photometric errors in the analysis
of fainter sources 
would prevent the overly optimistic identification of stellar outliers 
as strong \hzq\ candidates.

In principle, the ideal method of \hzq\ candidate selection
is to adopt a fully self-consistent Bayesian method which 
combines all the information available for each source in an optimal way.
This idea is explored in this paper,
starting from first principles by
adapting standard Bayesian model selection techniques 
to astronomical classification (\sect{prob}).
The resultant formalism is then
applied to the \sdssukidss\ \hzq\ search in \sect{example}.
These results and some future extensions
to this techinque are summarised in \sect{conc}.
Some technical issues relating to the evaluation 
of the likelihood for photometric data are explored in 
\apdx{likelihoodfaint},
and the method used to model the
stellar population is detailed in \apdx{starfit}.

All photometry is given in the native system of the 
relevant survey and explicitly subscripted 
wherever numerical values are quoted.
Thus SDSS \iprime\ and \zprime\ photometry is on the AB
system, whereas UKIDSS $Y$ and $J$ photometry is Vega-based.  
Under the assumption that Vega has zero magnitude in all passbands, 
the Vega to AB conversions
for these two UKIDSS filters are 
$Y_{\rm{AB}} = Y_{\rm{Vega}} + 0.634$ 
and 
$J_{\rm{AB}} = J_{\rm{Vega}} + 0.938$
\citep{Hewett_etal:2006}.  
All SDSS and follow-up photometry in the \iprime\ and \zprime\ bands
is reported using asinh magnitudes \citep{Lupton_etal:1999};
as a result model colours in these bands depend on the overall
flux level assumed.
Photometric measurements are denoted with a $\hat{}$ to 
emphasize that these are purely data-derived statistics.
All detections limits are given as the magnitude of a point--source
which would, on average, be measured with a signal--to--noise ratio
of $\ston = 5$ in such observations.
The rest-frame absolute magnitudes of quasars are given as
$M = M_{{\rm AB},1450}$ (\ie\ on the AB system at a wavelength of 
$\lambda = 0.1450 \unit{\micron}$).
Conversions between absolute and apparent magnitudes 
are performed assuming a fiducial 
flat cosmological model with
normalised matter density $\Omega_{\rm{m}} = 0.27$,
normalised vacuum density $\Omega_\Lambda = 0.73$,
and Hubble constant $H_0 = 71 \unit{km} \unit{s}^{-1} \unit{Mpc}^{-1}$
(\cf\ \citealt{Dunkley_etal:2009}).

%%%%%%%%%%%%%%%%%%%%%%%%%%%%%%%%%%%%%%%%%%%%%%%%%%%%%%%%%%%%%%%%%%%%%%%%%%%%%%

\section{Probabilistic classification of astronomical sources}
\label{section:prob}

Having made some measurements of an astronomical source,
what can be inferred about the type of object it is?
Assuming there are 
$\npop$ distinct populations of astronomical\footnote{It 
would also be possible to include 
various non-astronomical noise processes
(\eg\ bad pixels, cross--talk, noise peaks, \etc)
amongst the models that might explain the data,
an possibility which is especially relevant 
when searching for rare objects.
The difficulty in implementing this idea is that,
whereas most astrophysical populations are at least 
resonably well constrained, the huge variety of 
poorly understood noise processes make it far more difficult
to quantify these processes.
Nonetheless, it is a useful reminder that all probabilities
are conditional, and the model selection approach followed 
here is always predicated on the source being drawn from 
one of the astronomical populations that have explicitly been included 
in the calculation.}
objects,
$\models = \{ \model_1, \model_2, \ldots, \model_{\npop} \}$, 
under consideration,
the fullest answer to this question is to 
use the $\ndat$ available measurements,
$\data = \{ \datum_1, \datum_2, \ldots, \datum_{\nband} \}$, 
along with the fact that the object was detected in the first place,
to calculate the posterior probability\footnote{The notation
  $\prob(A|B)$ is used to 
  indicate the degree to which
  (the truth of) proposition $B$ implies (the truth of) proposition $A$.
  As such, 
  the probability $\prob(A|B)$ is not a mathematical function in the usual
  sense, although if
  $A$ and $B$ are mathematical in nature then
  formal expressions such as $\prob(x = x_0 | y = y_0)$
  are replaced by the less cumbersome,
  if occasionally ambiguous, shorthand $\prob(x|y)$.},
$\prob (\pop | \data, \detected, \models)$,
of each hypothesis $\pop$.
Applying Bayes's theorem 
yields the standard model comparison result 
(\eg\ \citealt{Jaynes:2003,Sivia_Skilling:2006}) that
\begin{equation}
\label{equation:pqsoev}
\prob(\pop | \data, \detected, \models)
  = \frac{\prob(\pop| \detected, \models) \, \prob(\data, \detected | \pop)}
  {\sum_{\popprime = 1}^\npop
    \prob(\popprime| \detected, \models) 
    \, \prob(\data, \detected | \modelprime)},
\end{equation}
where
$\prob(\pop | \detected, \models)$
is the prior probability that a detected source is of type $\pop$
and 
$\prob(\data, \detected | \pop)$
is the probability of detecting the source
and obtaining the observed data
under the $\pop$'th hypothesis.
Known as the evidence or the model likelihood, 
this is given by 
\[
\prob(\data, \detected | \pop)
\]
\begin{equation}
\mbox{}
  = \int
    \prob(\parameters_\pop | \pop) \,
    \prob(\data, \detected | \parameters_\pop, \pop)
    \, \dvol,
\end{equation}
where 
$\prob(\parameters_\pop | \pop)$
is the unit-normalised prior distribution of the 
$\nparam$ model parameters,
$\parameters_\pop$, that describe objects of type $\pop$,
and the likelihood,
$\prob(\data, \detected | \parameters_\pop, \pop)$,
is the probability of 
detecting the source and obtaining the observed data
given a particular value of those parameters.

\Eq{pqsoev} is a standard application of Bayes's theorem
but for the explicit statement that the source under consideration
has been detected.
The reason for its inclusion here is to 
ensure that the 
the prior distribution of each population's parameters
can be normalised unambiguously,
as well as to avoid the meaningless notion of an 
unconditional prior probability of the nature of a source.
Asked out of context, the question 
`What is the probability that a source is a quasar?'
is ill-posed and 
has no sensible answer.
This immediately implies that it is impossible to determine 
the prior probability of a source being of a certain type without 
at least some constraining information,
such as a range of fluxes or colours.
Thus the similar question `What is the probability that a source
with $\zab \leq 21.0$ is a quasar?'
does have a well-defined answer, 
the numerical value of which is given approximately 
by the observed numbers of quasars and stars down to the specified limit.
This would then be a reasonable emprical value for the quasar prior,
although even here the answer depends on 
various other factors, such as Galactic latitude.
The implication of the above arguments is that 
the prior
would have to be calculated independently for 
surveys with, \eg, different footprints on the sky or different depths,
a far from satisfactory situation.

These potentially troublesome ambiguities can be resolved
by combining the model and parameter priors with the likelihood
into a weighted evidence term, defined as
\[
\weight_\pop(\data, \detected)
\]
\begin{equation}
\label{equation:weight}
\mbox{}  = \int
    \density_\pop(\parameters_\pop)
    \prob(\detected | \parameters_\pop, \pop)
    \prob(\data | \parameters_\pop, \pop)
    \, \dvol,
\end{equation}
where $\density_\pop(\parameters_\pop)$ 
is the surface density of type $\pop$ sources on the sky
and $\prob(\detected | \parameters_\pop, \pop)$ is 
the probability that a source of type $\pop$ with parameters
$\parameters_\pop$ would have been detected in the survey.
The main benefit of using $\density_\pop(\parameters_\pop)$
instead of the necessarily-normalised
$\prob(\parameters_\pop | \detected, \pop)$ is that the source density 
has an empirical normalisation given by the number of observed 
number of sources per unit solid angle.
Not being dependent on generally arbitrary parameter space
boundaries it is independent of the details of the current experiment,
and need only be calculated once.
The trade-off is the need to introduce the detection 
probability, $\prob(\detected | \parameters_\pop, \pop)$,
although most sources are sufficiently brighter than the survey's
detection limit that 
$\prob(\detected | \parameters_\pop, \pop)$
is close to unity and can be ignored.
Using the weighted evidence, 
\eq{pqsoev} simplifies to
\begin{equation}
\label{equation:pqsoweight}
\prob(\pop | \data, \detected)
  = \frac{\weight_\pop(\data, \detected)}
    {\sum_{\popprime = 1}^{\npop} \weight_\popprime(\data, \detected)}.
\end{equation}

Equations~\ref{equation:weight} and (\ref{equation:pqsoweight})
describe a general method for probabilistic classification of 
an astronomical object, 
by explicitly combining 
existing knowledge
of the populations from which it might have been drawn
with the information contained in whatever measurements 
have been made of the source in question.
The next steps towards adapting this general approach to 
\hzq\ candidate selection are to examine the likelihood 
for photometric data (\sect{likelihood}) 
and to specialise to the specific case in which 
the source is assumed to be either a quasar or a star (\sect{pq}).

%%%%%%%%%%%%%%%%%%%%%%%%%%%%%%%%%%%%%%%%%%%%%%%%%%%%%%%%%%%%%%%%%%%%%%%%%%%%%%

\subsection{Photometric data}
\label{section:likelihood}

In optical and \nir\ astronomy,
even a low-quality spectrum is usually sufficient 
to establish the basic nature of the source with near total certainty,
obviating the need for the formal statistical approach described above.
Photometric measurements, however, are generally more ambiguous,
and some sort of Bayesian approach is required
to avoid making overly certain classifications 
(\eg\ \citealt{Mortlock_etal:2009b}).
For this reason
only photometric measurements are considered henceforth.

Given a model described by parameters $\parameters_\pop$,
the likelihood 
$\prob(\data | \parameters_\pop, \pop)$
of measuring photometric data $\data$
must include information on how the parameters relate to observables, 
as well as describing the 
stochastic aspects of the measurement process.
In the case of optical or \nir\ survey photometry,
the likelihood should account for a number of distinct effects:
the background uncertainty in the images;
the Poisson noise in the number of photons received from the source;
possibile inter-band noise correlations (\eg\ \citealt{Scranton_etal:2005});
and non-detections, including cases in which the 
background-subtracted counts are negative.
For the problem of assessing \hzq\ candidates 
it is reasonable to ignore some of these effects:
very few ambiguous candidates are more than a magnitude or two
brighter than the survey limits, so the source Poisson photon noise can be
neglected;
and the gain from including the inter-band noise correlations is negligible,
especially given the fact that the
correlations are often poorly known.
It is, however, vital to allow for non-detections, 
particular in the case of \hzqs\ which 
have negligible flux blueward of the redshifted \lya\ emission line.

Traditional logarithmic magnitudes \citep{Pogson:1856} 
cannot represent negative measured fluxes;
and, while the likelihood for non-detections
can be expressed in terms of
asinh magnitudes \citep{Lupton_etal:1999},
the resulting expressions are cumbersome (\apdx{likelihoodfaint}).
The most straightforward approach is to work in flux units 
(\ie\ calibrated and background-subtracted counts),
in which case the data vector is
$\data = \hat{\fluxes} 
  = \{ \hat{\flux}_1, \hat{\flux}_2, \ldots, \hat{\flux}_{\nband} \}$,
where $\hat{\flux}_\band$ is the reported flux in 
the $\band$'th of the $\nband$ bands.
Ignoring inter-band correlations, 
the likelihood can be separated into the form
\begin{equation}
\label{equation:likelihood_multi}
\prob[\hat{\fluxes} | \fluxes (\parameters_\pop)]
  = \prod_{\band = 1}^{\nband}
    \prob \left[ \hat{F}_\band | \flux(\vect{\parameters_\pop}) \right],
\end{equation}
where 
$F_{\pop,\band}(\parameters_\pop)$ is the true flux in band $\band$
of an object of type $\pop$ described by parameters $\parameters_\pop$.
(The explicit dependence of the true flux on $\parameters_\pop$
could be omitted, but it is retained here to emphasize
the fact that $\flux_\band$ is only ever an intermediate quantity.)

For sources within a few magnitudes of the survey limit
(which includes almost all the \hzq\ candidates)
the photometric errors are dominated by the 
uncertainties in the background subtraction, 
which is typically very 
well approximated as being additive and Gaussian in flux units.
However, many of the \hzq\ candidates under consideration
will be extreme outliers from the stellar locus,
and the frequency of such events in real data is almost
always higher than would be predicted by Gaussian statistics.
As all the probability calculations here are performed numerically
there would be no significant penalty for adopting a
more complicated noise distribution with stronger tails;
but the small number of outliers makes it difficult to assess what
distribution should be adopted.
Regardless of the specific form of the non-Gaussian tails 
of the photometric noise distribution, the net effect would
be to decrease $\pq$ due to the increased likelihood of stars
being scattered to have quasar-like colours.
Perhaps more importantly, the probabilities 
of most candidates would not be changed significantly:
given that only two classes are under consideration,
$\pq$ is determined primarily by the relative distance
of a source's measured colours 
from the quasar and star loci.
As such, the impact of inaccurate modelling of the photometric
errors on the resultant candidate samples -- and the relative ranking
of the candidates should be minimal.  
It is only in a few unusual cases 
that the 
relative likelihood of the measured photometry 
under the two different hypotheses is changed by increasing the 
tails of photometric noise distribution, 
and these tend to correspond to non-astronomical contaminants for which 
neither model is a good fit.

On balance it seems clearest to assume Gaussianity, 
for which the single-band likelihood is 
\[
\prob \left[ \hat{\flux}_\band | \flux(\vect{\parameters_\pop}) \right]
  = 
  \frac{1}{(2 \upi)^{1/2} \, \sigma_\band}
  \exp \left\{
    - \frac{1}{2} 
    \left[
      \frac{\hat{\flux}_\band - \flux_{\pop,\band}(\parameters_\pop)}
      {\sigma_\band}
    \right]^2
  \right\}, 
\]
\begin{equation}
\label{equation:likelihood}
\end{equation}
where
$\noise_\band$ is the 
the background uncertainty (in flux units).

If the photometric data are only given in terms of magnitudes then 
they can be converted into flux units, although some care is required
to ensure that the correct noise level is recovered.
The conversions for both logarithmic and asinh 
magnitudes are given in \apdx{likelihoodfaint}.

It is also possible that only upper limits are reported for sources 
which are undetected -- or, more accurately, were measured 
with a low signal--to--noise ratio -- 
in one or more bands.
With access to the raw data 
aperture fluxes can be measured for all undetected sources,
but in some cases this is not possible 
With only an upper limit it is impossible to
reconstruct the likelihood as given in \eq{likelihood},
as no information is retained about
how far below the detection limit the source's measured flux was.
In any band for which only an upper limit,
of $\flux_{\limiting,\band}$ $(\simeq 5 \sigma_\band$ in most cases), 
is given,
the likelihood is simply the probability that a source of true flux 
$\flux_{\pop,\band}(\parameters_\pop)$ would be
observed with a measured flux below
the stated detection limit.
Integrating over the unknown measured flux gives the 
likelihood of a non-detection as
\[
\prob \left[ \hat{F}_\band < \flux_{\limiting,\band} 
  | \flux(\vect{\parameters_\pop}) \right] 
  = 
  \frac{1}{2} \left\{
    1 + \erf\left[ 
      \frac{\flux_{\limiting,\band} - \flux_{\pop,\band}(\parameters_\pop)}
      {2^{1/2} \sigma_\band} \right] \right\},
\]
\begin{equation}
\label{equation:likelihood_upper}
\end{equation}
where $\erf(x) = 2 \, \pi^{-1/2} \int_0^x e^{-t^2} \, \diff t$ 
is the error function.

The likelihood for a source with 
detections in some bands and only upper limits in others
is a rather strange combination of 
probability densities (for the measurements)
and cumulative probabilities (for the upper limits).
However, this is not problematic in the context of model
comparison as 
the same combinations of differential and cumulative probabilities
appear in both the numerator and denominator of
\eq{pqsoweight}, and so cancel out appropriately when 
evaluating $\prob(\pop | \data, \detected)$.
The seamless handling of upper limits 
is one of the many benefits of the Bayesian approach to such 
problems.

The fact that upper limits can be self-consistently 
included in the classification formalism does not,
however,
change the fact that they represent a loss of information.
There are several ways in which this 
information loss could be characterised,
but the most relevant here is how 
$\prob(\pop | \data, \detected)$ is changed.
Given that the \stonlong\ of a non-detection is inevitably low,
it might seem that the effect of replacing a noisy
flux estimate with an upper limit would be minimal.
In the case of the 
\hzq\ candidates considered in \sect{example}, however, 
a `non-detection' of, 
\eg, $\hat{\flux}_\band \simeq 3 \sigma_\band$ 
is often sufficient to decisively reject a candidate.
A simple example demonstrating why this is the case is 
presented in \apdx{likelihoodfaint}.

While it is useful to be able to include upper limits,
flux estimates should be supplied and used if possible.
In particular,
flux estimates are obtained for all non-detections 
in the \sdssukidss\ quasar search described in \sect{example}
and hence 
the 
calculations of $\pq$ presented in
\sect{results} are based 
exclusively on \eq{likelihood}, not \eq{likelihood_upper}.

%%%%%%%%%%%%%%%%%%%%%%%%%%%%%%%%%%%%%%%%%%%%%%%%%%%%%%%%%%%%%%%%%%%%%%%%%%%%%%

\subsection{The probability that a source is a HZQ}
\label{section:pq}

The probabilistic classification formalism described above 
was developed specifically to 
assess the quality of the numerous 
superficially promising 
quasar candidates that are inevitably generated by a \hzq\ survey.
For a real point--source in such a sample only two 
possibilities are explicitly considered here:
it is a Galactic star (\ie\ $\pop = \str$); 
or it is a target \hzq\ (\ie\ $\pop = \qso$).
Thus the model space is reduced to $\models = \{ \qso, \str \}$ and 
\eq{pqsoweight} can be simplified to give the quasar probability as 
\begin{equation}
\label{equation:pq}
\pq = \prob(\qso | \data, \detected) 
  =
  \frac{\weight_\qso(\data, \detected)}
  {\weight_\qso(\data, \detected) + \weight_\str(\data, \detected)}.
\end{equation}

The notation used in \eq{pq} emphasizes the role of the data 
in calculating $\pq$, in the case of \hzqs, 
but the most important
single factor is the degree to which quasars are out-numbered by Galactic
stars (at the depths probed by current surveys, at least).
For this reason, $\pq \ll 1$ unless a source not only 
has the measured colours of a \hzq, but has sufficiently precise 
photometry that the data represent a statistically
significant deviation from the stellar locus.
An implication of these criteria is that it is almost impossible 
for faint sources close to a survey's detection limit to have 
high $\pq$ as their measured colours are inevitably imprecise --
almost all such sources with \hzq -like colours are 
better explained as being scattered stars.
In the space of possible multi-band measurements and errors 
there is at best only a small region for which the quasar probability
is significant -- if the survey is too shallow or does not cover 
the appropriate wavelengths
then $\pq$ will be low for all possible photometric measurements.
The Bayesian approach to candidate selection is,
in principle, as exacting as possible;
the next step is to see how it works in practice 
by applying it to a real data-set.

%%%%%%%%%%%%%%%%%%%%%%%%%%%%%%%%%%%%%%%%%%%%%%%%%%%%%%%%%%%%%%%%%%%%%%%%%%%%%%

\section{Searching for HZQs using UKIDSS and SDSS data}
\label{section:example}

The probabilistic approach to \hzq\ selection
described in \sect{prob} was developed
to prioritse the large number of candidates
generated by matching NIR UKIDSS sources to optical SDSS catalogues.
The generation of the cross-matched \sdssukidss\ sample
and the initial candidate selection are described in 
\sect{cand}.
Realistic models for the star and quasar populations
are developed in 
\sect{star} and \sect{quasar}, respectively.
The dependence of the quasar probability, $\pq$, 
on the measured photometry of an idealised source is explored 
in \sect{exampleprob}, 
and the possibilities for photometric redshift estimation 
are demontrated in \sect{zest}.
Finally, the probabilistic selection method is applied 
to the full \sdssukidss\ sample in \sect{results}.

%%%%%%%%%%%%%%%%%%%%%%%%%%%%%%%%%%%%%%%%%%%%%%%%%%%%%%%%%%%%%%%%%%%%%%%%%%%%%% 

\subsection{Initial candidate selection}
\label{section:cand}

The starting point for this search for $\redshift \ga 6$ quasars is
UKIDSS \citep{Lawrence_etal:2007},
a suite of five separate \nir\ surveys 
using the 
Wide Field Camera (WFCAM; \citealt{Casali_etal:2007})
on the 
United Kingdom Infrared Telescope (UKIRT).
A detailed technical description of the survey is
given by \cite{Dye_etal:2006},
although there have been a number of improvements in the
time since \citep{Warren_etal:2007}.
The most relevant of the projects is the 
Large Area Survey (LAS), which should eventually 
cover $\simm 3800 \unit{\sqdeg}$ in the UKIDSS
$Y$, $J$, $H$ and $K$ bands (defined in \citealt{Hewett_etal:2006}).
As of UKIDSS's Seventh Data Release (DR7),
on 2010 February 25,
the LAS had covered $\simm 1900 \unit{\sqdeg}$ 
in $Y$ and $J$ to
average depths 
of $\yv_\limiting = 20.0 \pm 0.1$ and
$\jv_\limiting = 19.5 \pm 0.2$ 
\citep{Dye_etal:2006,Warren_etal:2007}.
Querying the WFCAM Science Archive\footnote{The WSA is located at
{\tt{http://surveys.roe.ac.uk/wsa/}}.} (WSA;
\citealt{Hambly_etal:2008}) reveals 
that the DR7 LAS sample contains $\simm 2.2 \times 10^7$ catalogued sources 
that were detected in both $Y$ and $J$.
According to the \qlf\ of 
\cite{Jiang_etal:2008},
only \nexp\ \hzqs\ are expected with $\yv \leq 19.8$ 
in the DR7 LAS area; 
the problem, then, is to 
identify these few sources efficiently and reliably.

The first step in the filtering process is to match the \nir\ UKIDSS 
sample to the optical catalogues from the SDSS \citep{York_etal:2000}.
As of Data Release 7 (DR7; \citealt{Abazajian_etal:2009}),
the SDSS covers $\simm 1.2 \times 10^4 \unit{\sqdeg}$,
including almost the entire UKIDSS LAS area.
Observations were made 
in the custom 
\uprime, \gprime, \rprime, \iprime\ and \zprime\ 
filters \citep{Fukugita_etal:1996},
and photometry 
was obtained in all five bands for every detected source.
For point--sources the photometry was based on a model
of the point-spread function (PSF).
The SDSS main survey reaches single-scan depths of
$\iab_\limiting \simeq 22.5 \pm 0.2$ and $\zab_\limiting \simeq 20.8 \pm 0.2$,
and so sources close to the UKIDSS 
$Y$-band limit with $\imyval \ga 2.5$ and $\zmyval \ga 0.8$ are likely to be 
undetected by SDSS.
In the case of non-detections,
aperture photometry in the \iprime\ and \zprime\ bands 
was obtained from the SDSS images.
Aperture photometry was not obtained in the three bluest SDSS
bands, however, although potential quasar
candidates were rejected if they were detected 
in $\uprime$, $\gprime$ or $\rprime$.
More importantly, 
approximately $30 \percent$ of UKIDSS sources 
were observed more than once by SDSS, 
and in such cases 
the best flux estimates from the different scans
(\ie\ PSF-based if available or aperture otherwise) 
were combined using inverse-variance weighting.
The final result is a
combined \sdssukidss\ catalogue of sources with 
the best available survey photometry in the
\iprime, \zprime, $Y$ and $J$ bands 
(as well as $H$ and $K$, where available).

\begin{figure}
\includegraphics[width=\figwidth]{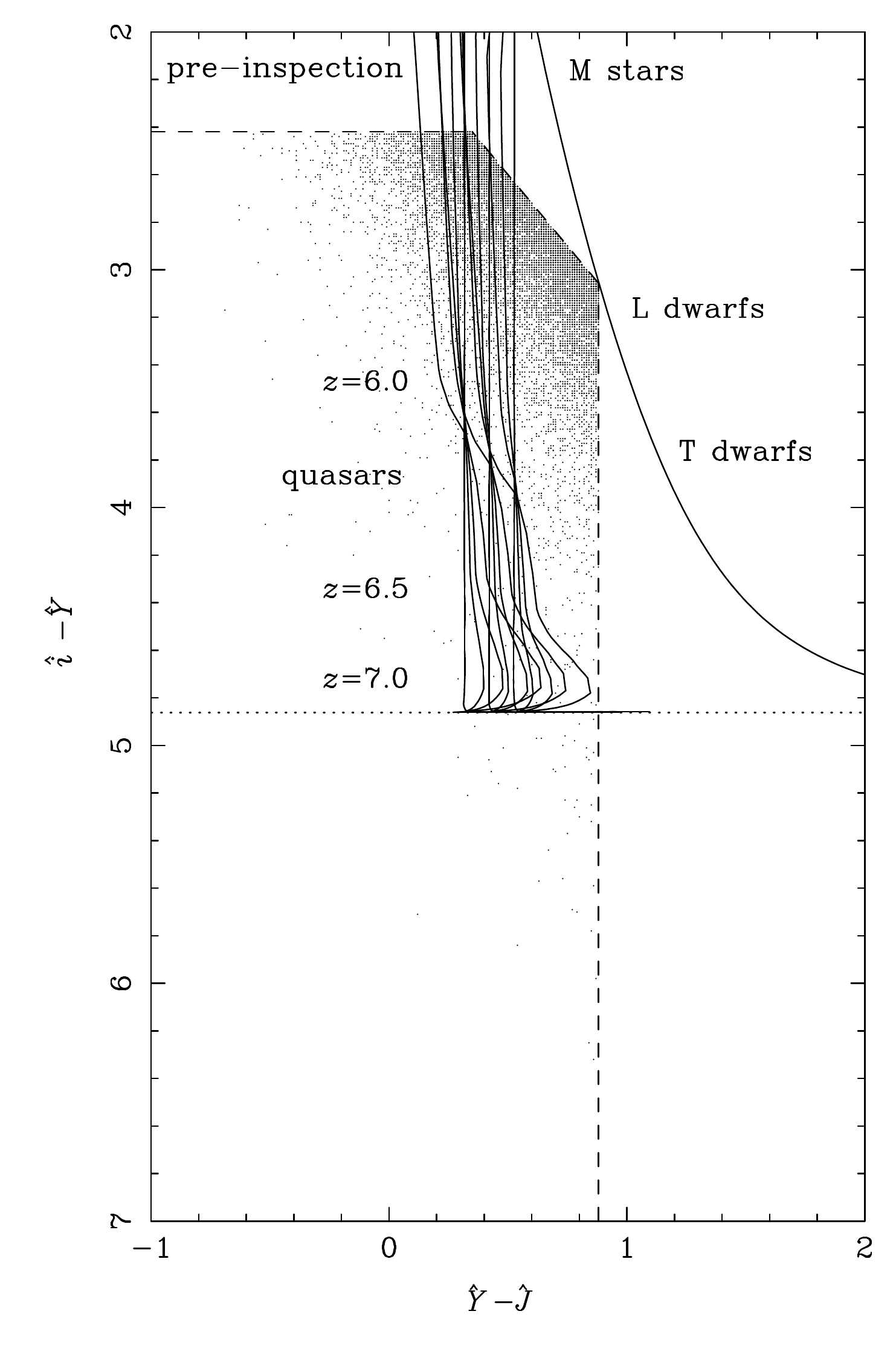}
\caption{Two-colour diagram showing the $\simm 1.1 \times 10^4$
UKIDSS DR7 LAS point--sources which, 
after being cross-matched to SDSS, 
have measured colours similar to those expected of \hzqs.
Also shown are 
the colours of
the stellar locus described in \sect{star}
and
the twelve quasars models described in \sect{quasar},
all calculated for a source that has $Y = 19.5$.
The dashed lines denote the initial pre-selection 
cuts that are applied before subsequent processing.
The horizontal dotted line shows the maximum 
$\imy$ value that a $Y = 19.5$ source could have in the absence of noise.}
\label{figure:ccobs}
\end{figure}

In the absence of photometric noise,
the target \hzqs\ are expected to
occupy a region of the \iprime, \zprime, $Y$ and $J$
\sdssukidss\ colour space that
is well separated from other astronomical sources 
(\cf\ \citealt{Hewett_etal:2006}).
The theoretical separation between \hzqs\ and cool stars 
in colour space is illustrated in 
\fig{ccobs}, which 
shows both the stellar locus (described in \sect{star})
and the model quasar tracks (described in \sect{quasar}).
The single dominant factor 
that ensures \hzqs\ have such distinct colours
is the near-complete absorption blueward 
of the \lya\ emission line due to intervening \hi.
In the redshift range 
$5.8 \la \redshift \la 6.4$ 
\lya\ is in the $\zprime$ band,
and so such quasars should be extremely red in $\imz$ (and $\imy$);
at higher redshifts ($6.6 \la \redshift \la 7.2$),
\lya\ is in the $Y$ band, leading to 
extremely red $\zmy$ or $\imy$ colours.  
By contrast, 
most Main Sequence stars 
are expected to have considerably bluer colours,
although the coolest M dwarfs have $\imyval \simeq 2$.
While L and T dwarfs have similar $\imy$ colours to \hzqs, 
they are expected to be 
significantly redder in $\ymj$ \citep{Hewett_etal:2006},
which is the reason that
the \hzq\ search only includes fields with 
observations in both $Y$ and $J$.

The vast majority of \sdssukidss\ sources can be 
rejected as \hzq\ candidates using a variety of 
automated cuts
(described more fully in \citealt{Mortlock_etal:2011b}):
sources with an unambigous, bright match in SDSS 
that gives $\imyobs \la 3$ are far bluer than the target \hzqs;
sources with red optical--NIR colours but with 
$\ymjobs \ga 0.9$ are almost certainly brown dwarfs;
most galaxies appear as extended sources in UKIDSS
(as characterised by, \eg, the UKIDSS {\tt{MergedClassStat}} statistic);
sources close to a bright star,
or that have required deblending in the SDSS processing,
or with significant UKIDSS error flags,
all have unreliable measured photometry;
sources with a significantly non-zero measured proper motion 
cannot be extra-Galactic.
Applying these cuts to the SDSS-matched 
UKIDSS DR7 LAS catalogue leaves 
$\simm 10^4$
apparently stationary, isolated point--sources 
that have the measured colours 
of \hzqs.
This sample is shown in \fig{ccobs},
along with with the inclusive colour cuts that were adopted.

How should the \hzq\ search proceed from here?
The ideal would be to
take spectra of all of the candidates, 
but doing so would require prohibitive observational resources --
\cite{Glikman_etal:2008}
needed 25 hour-long Keck observations to rule out
the most promising candidates from just
the $27.3 \unit{\sqdeg}$ covered by the 
UKIDSS Early Data Release (EDR; \citealt{Dye_etal:2006}).
Obtaining independent photometry of all the candidates is more feasible,
but still difficult due to limitations of telescope scheduling and the 
range of target right ascensions.
Even if the intention was to reobserve all candidates, 
some means of prioritizing the most promising is needed -- 
it is clear from \fig{ccobs} that a randomly selected 
candidate from this sample will almost certainly be a
scattered Galactic star.
It was this dilemma that led to the development of the 
Bayesian selection method described in \sect{prob}.
Before it can be implemented, however, 
models are needed for both the star and quasar populations.

%%%%%%%%%%%%%%%%%%%%%%%%%%%%%%%%%%%%%%%%%%%%%%%%%%%%%%%%%%%%%%%%%%%%%%%%%%%%%%

\subsection{The stellar population}
\label{section:star}

\begin{figure}
\includegraphics[width=\figwidth]{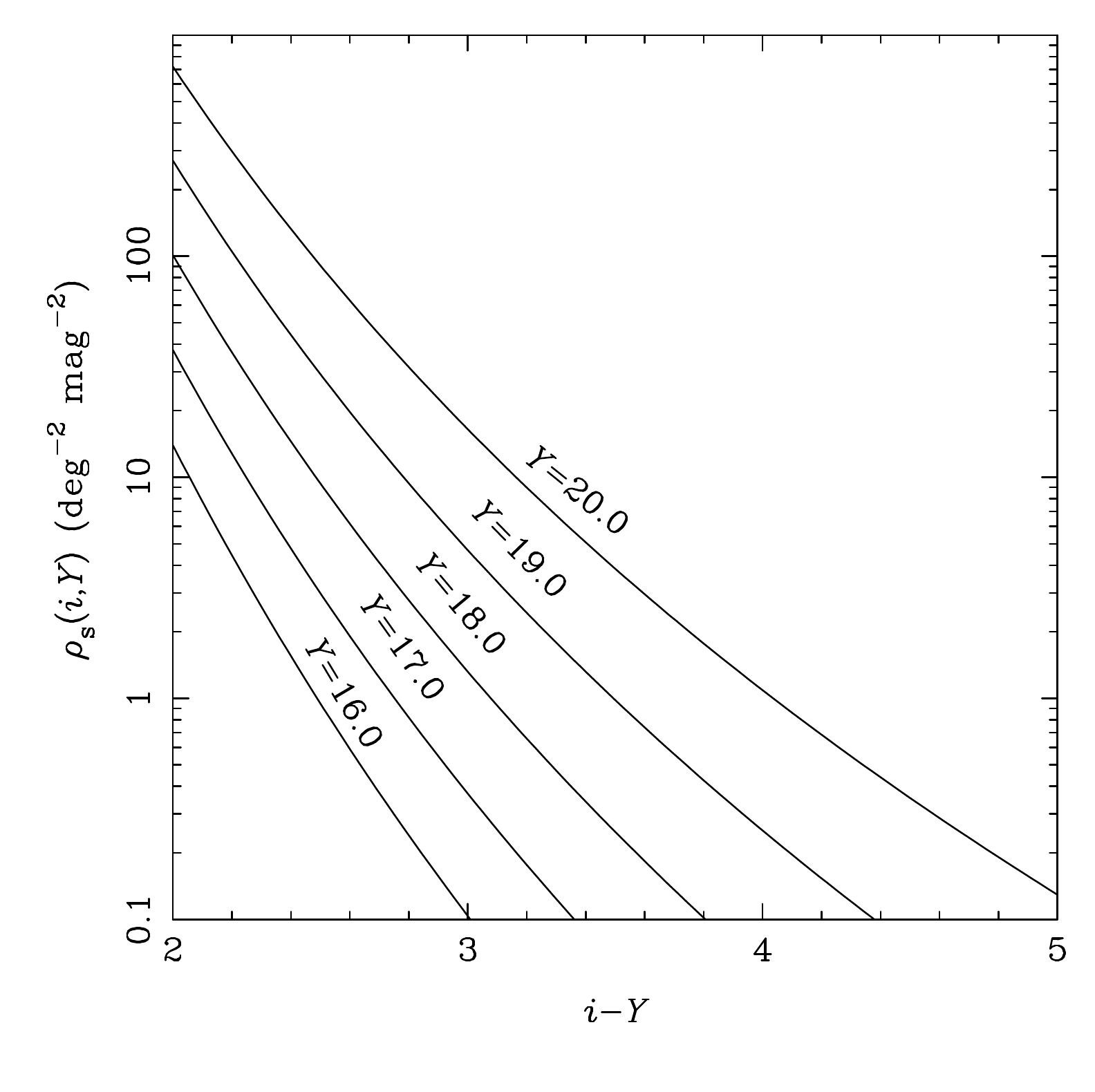}
\caption{The best-fit intrinsic distribution of \imy\ colours,
  $\density_\star(Y + \imy, Y)$
  for different values of $Y$ as labelled.}
\label{figure:iyhist}
\end{figure}

The \sdssukidss\ sources described in \sect{cand}
have the measured colours of \hzqs, 
but are predominantly Galactic stars
(as can be seen from \fig{ccobs}).
The detailed optical and NIR properties of the various possible
contaminants are described in detail by \cite{Hewett_etal:2006},
but it is critical here to establish likely make-up of the 
contaminating sources in detail.
Most Main Sequence stars are sufficiently hot that 
optical and \nir\ filters 
(and \iprime, \zprime, $Y$ and $J$ in particular)
only probe their Rayleigh--Jeans tails.
With $\imyval \simeq 1$, such stars
are so much bluer than the target \hzqs\
that they are not a significant contaminant.
Cooler M stars also have bluer
optical-\nir\ colours than the target \hzqs\ 
(\eg\ $\imyval \simeq 2$ as compared to $\imy \ga 3$ for the quasars),
but can be sufficiently faint in the $\iprime$ band that
a small fraction 
scatter to have $\imyobs \ga 3$.
Moreover, M stars have $\ymj \simeq 0.5$, much like the \hzqs,
and so the fact that their $Y$ and $J$ band photometry is typically 
fairly accurate actually increases the chance that some 
have the same observed colours as \hzqs\ in the 
UKIDSS and SDSS bands.
Conversely, the accurate $\ymj$ values measured for L and T
dwarfs 
(which can be just as red in optical--\nir\ colours as \hzqs)
combines with their lower numbers to ensure that they are not
the major source of contamination.
As even the faintest sources considered as possible \hzq\ candidates
have signal--to--noise ratios of $\ga 10$ in the $Y$ and $J$ bands,
the number of rare L or T dwarf with $\ymj \ga 1.0$ 
being measured to have $\ymjobs \simeq 0.4$ is 
considerably less than the number of more common M stars 
measured to have $\imyobs \ga 3$.
As candidate lists based on simple colour cuts are dominated by 
scattered cool M dwarfs (with some L and T dwarfs), 
an accurate model of the intrinsic distribution of M, L 
and T dwarfs in the \sdssukidss\ bands 
is necessary to calculate useful values of $\pq$.
In more visual terms, the only sources that need to be modelled
are those shown in \fig{ccobs} 
and the above arguments (along with the follow-up observations
described in \sect{results}) show that these are predominantly M dwarfs.

There are several possible approaches for modelling the stellar population. 
One option would be to fit the distribution of observed colours
or magnitudes (\cf\ \citealt{Richards_etal:2004,Bovy_etal:2010}), 
although this would not not correctly recover the tails of the 
(magnitude-dependent) noise distributions which are critical to 
this problem.
Under the approximation that most of the relevant sources are 
sufficiently faint to be dominated by background noise 
(\cf\ \sect{likelihood}), it might be possible to combine 
deconvolution techniques to
estimate the instrinsic distribution (\cf\ \citealt{Bovy_etal:2010}),
but this approach was not explored here.
The other extreme would be to develop a physical model for the 
local population of M, L and T dwarfs,
although the gain from this extra complication is minimal:
it is only important to describe the distribution of
measurable properties of these sources.

An intermediate approach to modelling the source population was adopted here.
The population of nearby M, L and T dwarfs 
was described by developing a parameterized 
function to describe their intrinsic 
(\ie\ not noise-convolved) 
distribution of magnitudes and colours.
The reason for modelling the intrinsic distribution 
is to be able to estimate the probability of stars scattering 
into sparsely populated regions of colour space.
While the core of the observed stellar distribution could be modelled 
empirically using the distribution of observed colours, such an approach
would not provide reliable estimates of the probability of a source
being scattered into the tails of the distribution 
(where the plausible \hzq\ candidates lie).
It is better to model separately the 
intrinsic population and the noise distribution (\sect{likelihood})
and to then convolve the two to provide the desired 
extreme scattering probability.

The adopted stellar population model has two
parameters, $\iab$ and $\yv$, both of which are observables.
In terms of the notation of \sect{prob}, the parameter space is 
defined by $\parameters_\str = \{ \iab, \yv \}$
and the number density of stars per unit solid angle is thus written as
$\density_\str(\parameters_\str) = \density_\str(\iab, \yv)$.
With $\imy$ serving as a proxy for stellar temperature,
the other colours (\ie\ $\zmy$ and $\ymj$) are,
to a sufficiently good approximation, functions of $\imy$.
The specific form of $\density_\str(\iab, \yv)$
was obtained 
by comparing the predicted distribution of observed photometry 
to a sample 
of well-measured \sdssukidss\ point--sources 
(with $15.0 \leq \hat{Y}\vega \leq 19.5$ and $\imyobs \geq 2.0$)
extracted from the WSA.
It was critical to ensure that the 
distribution describing the
intrinsic population was convolved with the 
correct photometric noise distribution
when comparing with the observed counts;
a full description of the fitting procedure is given in \apdx{starfit}.

Several different functional forms for 
$\density_\str(\iab, \yv)$ were investigated
before adopting
power-law number counts in combination with 
an exponential power-law colour distribution
for the reddest stars.
Taken together,
\[
\density_\str(\iab, \yv) = \density_\refval 10^{\alpha (Y - 18.0)}
\]
\begin{equation}
\label{equation:star}
\mbox{} 
\,\,\,\,\,\,\,
\times
  \Theta(\imy - 2)
  \frac{[\ln(\beta + \gamma Y)]^{1/\delta} 
    \, (\beta + \gamma Y)^{- (i\mbox{--}Y)^\delta}}
  {P\left[1/\delta, \ln(\beta + \gamma Y) \, 2^{\beta + \gamma Y}\right]}
  ,
\end{equation}
where $P(t, x) 
  = \int_x^\infty {x^\prime}^{t - 1} e^{-x^\prime}\, \diff x^\prime$ 
is the complementary incomplete gamma function
and the best-fit values of the free parameters 
are 
$\density_\refval = 20.9 \unit{mag}^{-2} \unit{deg}^{-2}$,
$\alpha = 0.45939287$,
$\beta = 551.74630495$,
$\gamma = -16.49969157$ and
$\delta = 0.04050890$. 
The best-fit distribution $\density_\str(\iab, Y)$
is shown as a function of $\imy$ for several values of 
$Y$ in \fig{iyhist}.

The model represents an average of the stellar population
over the range of Galactic latitudes, $b$, covered by the UKIDSS LAS.
The LAS was deliberately designed to avoid low-$b$ fields,
so the stellar density of the reddest stars 
(which are seen only to moderate distances)
varies by less than a factor of $\simm 2$ 
over the whole survey area.
Taking $\weight^\prime_\str(\data, \detected) 
  \simeq 2 \weight_\str(\data, \detected)$ 
in \eq{pq}, would decrease $\pq$ by a
factor of $\simm 5$ at most;
and, for the vast majority of sources 
which are decisively classified, 
$\pq$ remains essentially unchanged.
In only a small fraction of cases would a poor candidate be 
erroneously included for follow-up 
due to this effect.
For a survey covering a 
greater range of Galactic latitudes it 
would be important to include the $b$-dependence of 
$\density_\str(\iprime, Y)$, 
most simply by multiplying $\weight_\str(\data, \detected)$ 
by a $b$-dependent scaling,
the nature of which could be inferred from the survey data.

The predicted photometry in the other relevant bands 
(specifically $\zprime$ and $J$) 
is then given by the empirical colour relations
\begin{equation}
\label{equation:zmystar}
\zmy = 0.362 + 0.314 \, (\imy)
\end{equation}
and
\begin{equation}
\label{equation:ymjstar}
\ymj = 0.328 + 0.088 \, (\imy) + 0.0295 \, (\imy)^2.
\end{equation}

The two colour relationships 
in \eqs{zmystar} and (\ref{equation:ymjstar}),
together with the data, combine to give the likelihood 
(Eq.~\ref{equation:likelihood_multi}) as a function of $\iprime$ and $Y$.
Integrating the product of the likelihood and the stellar density
(given in Eq.~\ref{equation:star})
over these two parameters 
then gives the weighted evidence that a source is a star, 
$\weight_\str(\data, \detected)$, as defined in \eq{weight}.
To be more explicit,
specialising to the stellar case and the 
\sdssukidss\ filters allows 
\eq{weight} to be written as 
\begin{equation}
\label{equation:ws}
\weight_\str(\hatiprime, \hat{\zprime}, \hat{Y}, \hat{z}, \detected)
\end{equation}
\[
\mbox{}
  = \int_{-\infty}^\infty \int_{-\infty}^\infty
    \rho_\star(\iprime, Y)  \,
    \prob(\detected | \iprime, Y, \str) \,
    \prob(\hatiprime, \hat{\zprime}, \hat{Y}, \hat{z} | \iprime, Y, \str)
    \, \diff \iprime \, \diff Y.
\]
The colour relationships in \eqs{zmystar} and (\ref{equation:ymjstar})
combine with the $\iprime$ and $Y$ to give the predicted 
photometry in the four bands of interest,
from which the conversion to flux units allows the likelihood
to be written as a product of four Gaussians (\sect{likelihood}).
The detection probability is close to unity for the range
of $\iprime$ and $Y$ being considered, although 
in practice adopting detection cut-offs is the simplest way to 
ensure the integrals do not extend to such faint fluxes that 
confusion issues would become relevant.

%%%%%%%%%%%%%%%%%%%%%%%%%%%%%%%%%%%%%%%%%%%%%%%%%%%%%%%%%%%%%%%%%%%%%%%%%%%%%%

\subsection{The quasar population at high redshift}
\label{section:quasar}

UKIDSS is the first survey to have had a significant
chance of detecting $\redshift \simeq 7$ quasars,
so there are no empirical constraints 
on much of the target \hzq\ population.
This does not, however, mean that there is no information 
about the \hzq\ population beyond the current redshift limit 
--
it is perfectly reasonable to extrapolate from 
the results of $\redshift \simeq 6$ quasar surveys.
Indeed, one of the main principles of Bayesian reasoning is
that any available information should be applied if possible,
even if it is incomplete or imprecise.
A reasonable approximation to the correct prior 
(given by the actual, but unknown, \qlf\ in this case)
will result in final inferences that are superior to those
derived from any method which does not include any 
information about the likely numbers of $\redshift \simeq 7$ quasars.
The alternative approach is, in the case of rare object searches,
inevitably overly optimistic (\ie\ $\pq$ would be 
unreasonably high for large numbers of sources).  

The measured numbers of bright \hzqs\
(\eg\ \citealt{Fan_etal:2003,Jiang_etal:2008,Willott_etal:2010})
are consistent with a co-moving \qlf\footnote{The \qlf\ is 
defined such that the average number of quasars 
with absolute magnitudes between
$\mabsrest$ and $\mabsrest + \diff \mabsrest$ in a 
co-moving volume $\diff V_{\rmn{co}}$ at redshift $\redshift$ is 
$\lf_\qso(\mabsrest, \redshift) \, \diff \mabsrest \, \diff V_{\rmn{co}}$.}
given by a power-law of the form
\[
\lf_\qso(\mabsrest, \redshift) 
\]
\begin{equation}
\label{equation:qlf}
\mbox{}
  = 5.2\times 10^{-9}
     10^{0.84 (\mabsrest + 26.0) - 0.47 (\redshift - 6.0)} 
  \unit{mag}^{-1} \unit{Mpc}^{-3}.
\end{equation}
This parameterisation
combinines the magnitude dependence measured by
\cite{Fan_etal:2003} with the evolution model used by 
\cite{Jiang_etal:2008}\footnote{The \qlf\
parameters are reasonably well constrained by the data
with the exception of the evolution term.
The value found by 
\cite{Fan_etal:2001} over the 
redshift interval $3.6 \leq \redshift \leq 5.0$ was used,
although there is some evidence that the 
evolution at redshifts of $\redshift \ga 7$ might be 
stronger \citep{Mortlock_etal:2011b}.}.
\cite{Willott_etal:2010} 
found a significantly lower normalisation than \cite{Jiang_etal:2008}, 
and in principle the implied uncertainty should be included in
the calculation of $\pq$.  
However the resultant probabilities are unchanged in almost all cases,
and the higher normalisation was adopted to ensure a (marginally) more
inclusive candidate list.
(Moreover, the above values were already being used to select
UKDISS \hzq\ candidates before the results of \citealt{Willott_etal:2010}
were available.)

It would be possible to use $\mabsrest$ and $\redshift$ 
to parameterize the quasar
population, but it is more intuitive to convert 
$\mabsrest$ to the observable $Y$-band magnitude, 
so that the quasar population is,
like the stellar population defined in \sect{star},
characterised by an observable surface density.
Thus the quasar parameters used
are $\parameters_\pop = (Y, \redshift)$,
which leads to the population model
\begin{equation}
\label{equation:quasarsky}
\density_\qso(Y, \redshift) 
  = \frac{1}{4 \pi}
  \frac{\diff V_{\rmn{co}}}{\diff \redshift}
  \,
  \lf_\qso \left[Y - \dl(\redshift) - K_Y(\redshift) , \redshift \right],
\end{equation}
where
$\diff V_{\rmn{co}}$ is the co-moving volume of a
spherical shell of thickness $\diff \redshift$ at redshift $\redshift$, 
$\dl(\redshift)$ is the luminosity distance
and $K_Y(\redshift)$ is the $Y$-band quasar $K$-correction 
that converts the rest-frame AB magnitude at $0.1450 \unit{\micron}$ 
to the $Y$-band Vega magnitude.

The quasars' $K$-corrections are, in turn, evaluated using updated 
and expanded versions of the model quasar spectra 
developed by \cite{Hewett_etal:2006} and \cite{Maddox_etal:2008}.
In particular, 
they include a realistic model of the 
absorption blueward of the \lya\ emission line 
caused by the increased density of \hi\ 
above $z \simeq 5.8$, as measured by \cite{Fan_etal:2006b}.
The variety of the quasars' intrinsic properties is accounted for by
using colours from twelve different templates that span
four line-strengths and three continuum slopes.
As the main motivation of using multiple models is to ensure
appropriately high values of $\pq$ for 
the \hzqs\ with redder continuum slopes 
(that are closer to the stellar locus in $\ymj$ than the fiducual quasars),
the twelve models are weighted equally,
even though it would be more accurate to down-weight the less common 
templates.
Using a range of models also accounts for the colour variations 
that result from the combination of intrinsic quasar variability
and non-simultaneous measurements:
most UKIDSS observations took place several years 
after the SDSS observations of the same fields,
a time-scale on which 
\cite{Ivezic_etal:2004} found a typical variation of $\simm 0.15 \unit{mag}$.
When multiplied by the SDSS and UKIDSS filter profiles 
and integrated over wavelength,
the model spectra not only give $K_Y(\redshift)$,
but also the required optical-\nir\ colours.
All twelve colour tracks are shown in \fig{ccobs}.

The \hzq\ colour relationships described above,
together with the data, combine to give the likelihood
(Eq.~\ref{equation:likelihood_multi}) as a function of $\redshift$ and $Y$.
Integrating the product of the likelihood the quasar density
(given in Eq.~\ref{equation:quasarsky})
over these two parameters
then gives the weighted evidence defined in \eq{weight} as
\begin{equation}
\label{equation:wq}
\weight_\qso(\hatiprime, \hat{\zprime}, \hat{Y}, \hat{z}, \detected)
\end{equation}
\[
  = \int_0^\infty \!\! \int_{-\infty}^\infty
    \rho_\qso(Y, \redshift) \,
    \prob(\detected | Y, \redshift, \qso) \,
    \prob(\hatiprime, \hat{\zprime}, \hat{Y}, \hat{z} | Y, \redshift, \qso)
    \, \diff Y \, \diff \redshift.
\]
The quasar model loci give the 
predicted $\iprime$, $\zprime$, $Y$ and $J$ 
photometry,
from which the conversion to flux units allows the likelihood
to be written as a product of four Gaussians (\sect{likelihood}).
As with the integral over the stellar population (Eq.~\ref{equation:ws}), 
a simple detection cut-off is needed to 
ensure that the integral is not dominated 
by the numerous undetectable ultra-faint sources 
that are beyond the confusion limit in the relevant bands.

%%%%%%%%%%%%%%%%%%%%%%%%%%%%%%%%%%%%%%%%%%%%%%%%%%%%%%%%%%%%%%%%%%%%%%%%%%%%%%

\subsection{The probability that a source is a HZQ}
\label{section:exampleprob}

Having developed quantitative models for the 
stellar population (\sect{star})
and \hzq\ population (\sect{quasar}),
the measured $\iprime$, $\zprime$, $Y$ and $J$ band 
photometry of a source can then be used to calculate $\pq$ 
according to \eq{pqsoweight}.
The two-dimensional (weighted) evidence integrals over the 
quasar and star parameters 
are evaluated using simple numerical quadrature 
as this is faster than more general Monte Carlo 
techniques for a problem of such low dimensionality.
On a standard desktop computer one evaluation of $\pq$ 
takes between a tenth and a hundredth of a second, 
which is sufficiently fast that even the most
inclusive of candidate lists can be analysed.

The speed with which $\pq$ can be calculated 
also means that it is possible to explore 
how
$\pq$ depends on the measured photometry and the associated errors.
This is potentially important, 
as the reasons that some candidates have low
or high probabilities are not always obvious.
In this context it is useful to think of 
$\pq$ not as a quantity associated with candidates,
but as a function of the 
information that is particular to each source
(\ie\ the measured photometry and the associated uncertainties),
leading to 
$\pq = \pq(
    \hatiprime, \Delta \hatiprime,  
    \hat{\zprime}, \Delta \hat{\zprime},  
    \hat{Y}, \Delta \hat{Y},  
    \hat{J}, \Delta \hat{J}
  )$
for the \sdssukidss\ sample.
This function has too many parameters to explore comprehensively,
but many of its important features can be seen in 
two-dimensional projections.
Plotting $\pq$ in the space of measured colours 
also facilitates direct comparison with selection methods based 
on colour cuts,
which can be cast into a 
Bayesian form by treating them as if $\pq = 1$ for objects
satisfying the cuts and $\pq = 0$ otherwise.
In the regions of parameter space for which 
$\pq$ varies rapidly with the measured colours,
the Bayesian selection reduces to a cut-based method,
but with the important difference that the selection boundaries 
are defined objectively.

\begin{figure}
\includegraphics[width=\figwidth]{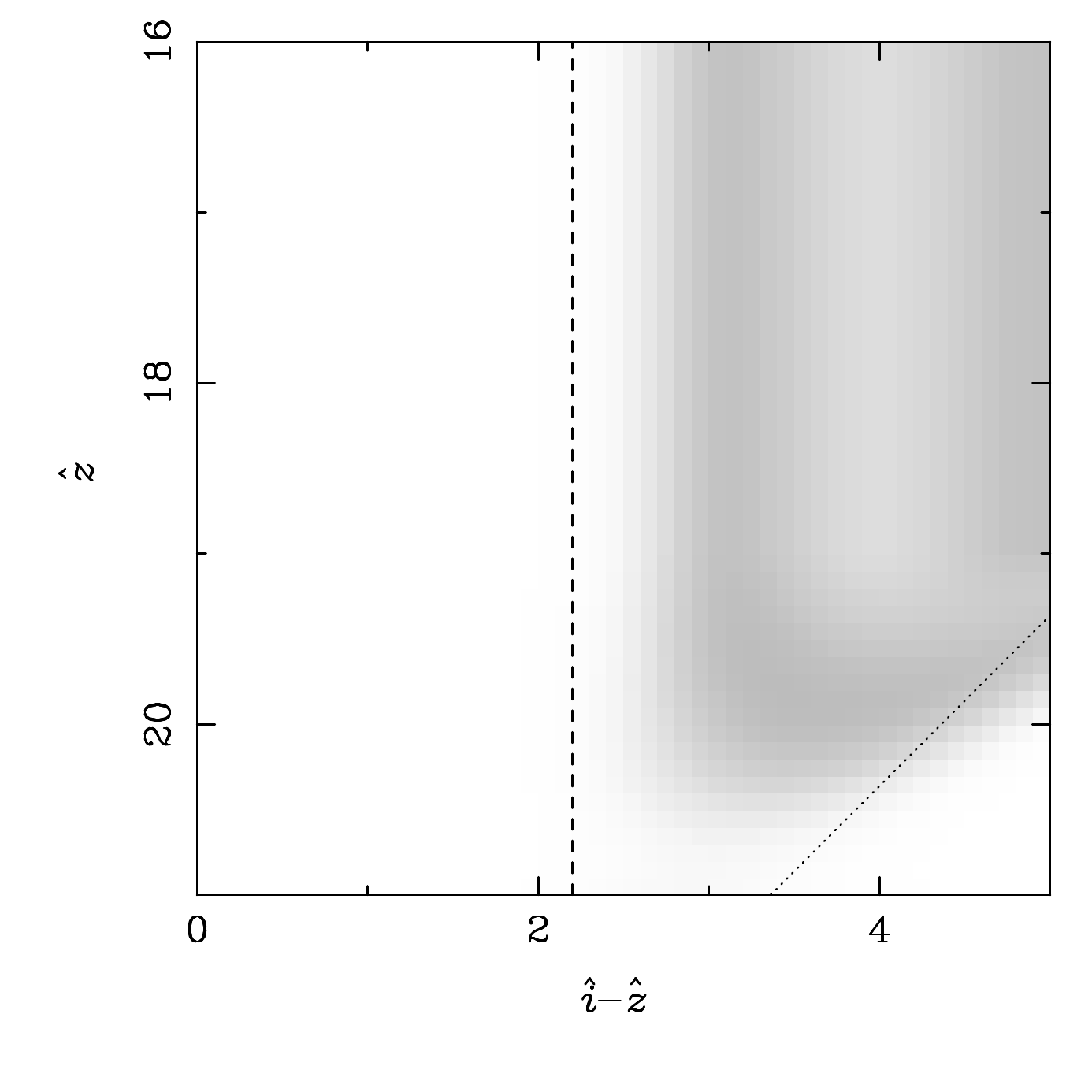}
\caption{The dependence of $\pq$ on
the measured \zprime\ band magnitude
for different value of the measured $\imz$ colour,
ranging from $\pq = 0$ (white) to $\pq = 1$ (grey).
The noise is as appropriate for
fiducial SDSS observations with
$\iprime_{\limiting} = 22.5$,
$\zprime_{\limiting} = 20.8$,
but the UKIDSS bands ($Y$ and $J$)
have been ignored.
The dashed line shows the initial SDSS \hzq\ selection
criterion defined in Fan et al.\ (2003)
and the dotted line shows the maximum 
theoretical value of \imz\ for a source with
$\zab = \hat{\zab}$ and zero flux in the \iprime\ band in the absence of noise.}
\label{figure:pq_colmag}
\end{figure}

The simplest non-trivial case is that of a two-band survey
for which each detected source can be treated as having
one measured magnitude and a single measured colour.
If the two bands are taken to be $\iprime$ and $\zprime$, 
this is a reasonable approximation to the SDSS \hzq\ survey, 
for which the initial selection criteria were 
$\hat{\zprime} < 20.2$ and $\imzobs  > 2.2$ 
\citep{Fan_etal:2001}.
Ignoring the $Y$ and $J$ bands and 
assuming the fiducial SDSS depths in $\iprime$ and $\zprime$
given in \sect{cand}, the variation of $\pq$ with 
$\hat{\zprime}$ and $\imzobs$ is shown in 
\fig{pq_colmag}.
The only sources which have $\pq \ga 0.1$ are those which have 
$\hat{\zprime} \la 20$ and $\hatiprime - \hat{\zprime} \ga 2.5$,
roughly corresponding to the region of parameter space selected by
\cite{Fan_etal:2001}.
The seemingly counter-intuitive result that the $\pq$
does not increase monotonically with $\imz$ 
is an artefact of the asinh magnitude system.

It is also noticeable from \fig{pq_colmag}
that no pair of (measured) $\hatiprime$
and $\hat{\zprime}$ values would result in $\pq \simeq 1$, 
a result which is independent of the depth of the observations.
This is because the sources which appear red in $\imz$
are not just scattered stars,
but also L and T dwarfs 
which actually have these red colours 
(and outnumber the target \hzqs).
The only way to generate a sample of candidates 
with higher $\pq$ is to obtain data in another band,
chosen such that quasars and the potential contaminants
have distinct colours.  
This can be done by follow-up 
(\eg\ in the $J$ band, as done by \citealt{Fan_etal:2001})
or by extending the wavelength coverage of the inital survey
(\eg\ UKIDSS $Y$-band imaging).
The choice between these two strategies is sometimes difficult, 
as adding an extra band to a survey costs area or depth,
whereas the number of follow-up observations required to complete
a two-band search is potentially prohibitive.
However in terms of this exploration of how $\pq$ depends on
the measured colours there is no distinction, 
as only the observational depths and the choice of bands is 
important, not the sequence of observations.

\begin{figure*}
\includegraphics[width=\thinfigwidth]{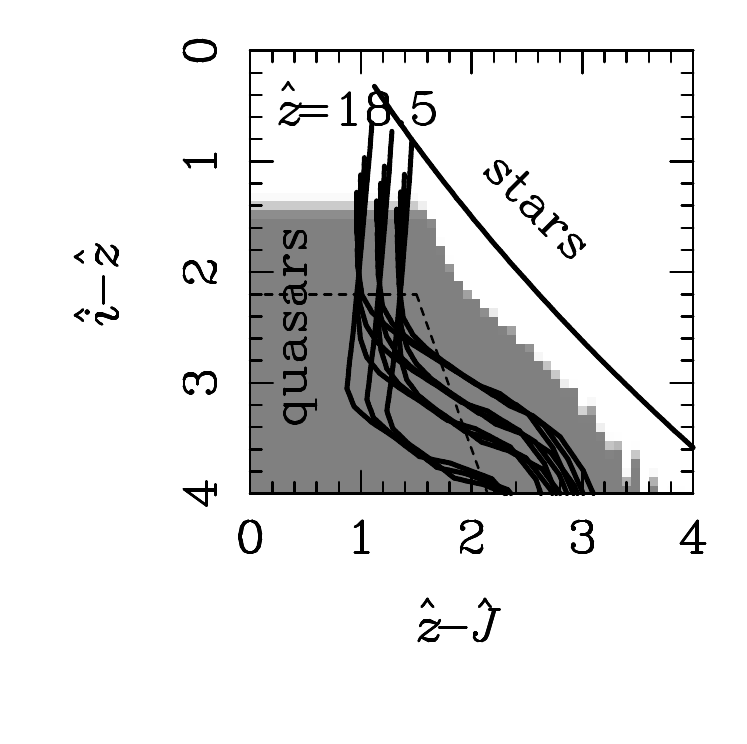}
\includegraphics[width=\thinfigwidth]{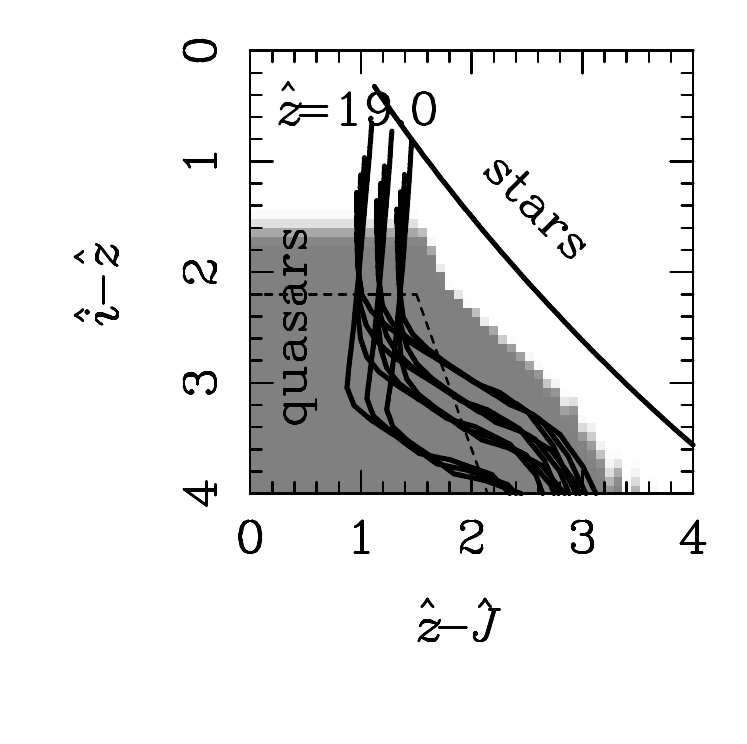}
\includegraphics[width=\thinfigwidth]{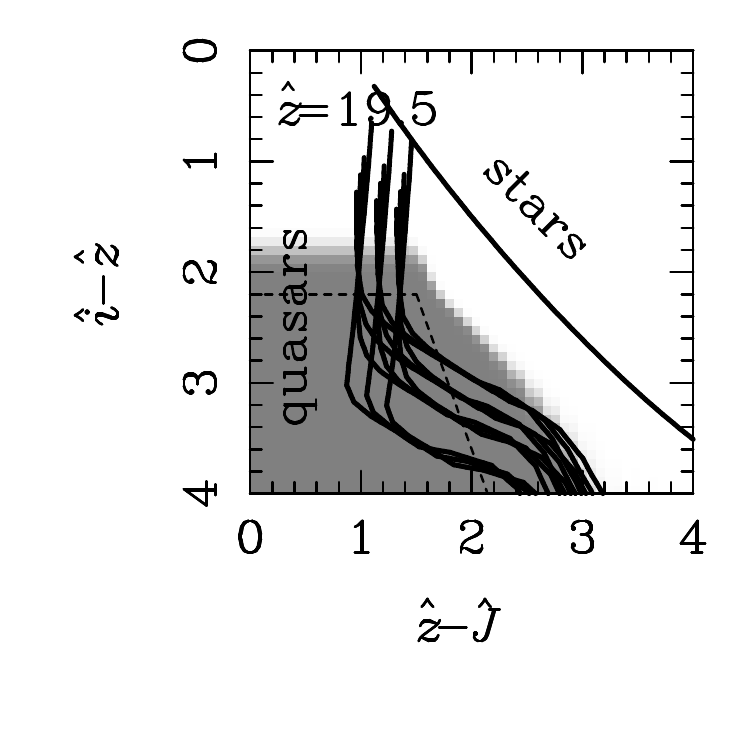}
\includegraphics[width=\thinfigwidth]{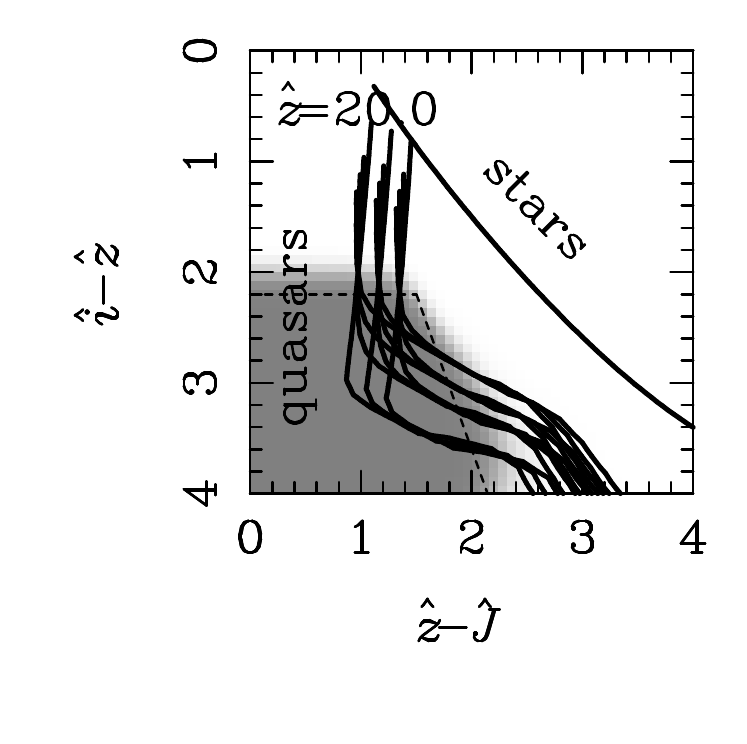}
\caption{The dependence of $\pq$ on
a source's observed $\imz$ and $\zmj$ colours,
ranging from $\pq = 0$ (white) to $\pq = 1$ (grey),
for
$\hat{\zprime} = 18.5$,
$\hat{\zprime} = 19.0$,
$\hat{\zprime} = 19.5$
and
$\hat{\zprime} = 20.0$, as labelled.
In all cases the noise is as appropriate for
fiducial SDSS and UKDISS observations with
$\iprime_{\limiting} = 22.5$,
$\zprime_{\limiting} = 20.8$
and
$J_{\limiting} = 22.5$,
but the $Y$ band has been ignored.
The quasar and star loci
are shown as solid curves 
and the dashed lines show the SDSS \hzq\ selection 
region defined in Fan et al.\ (2003).}
\label{figure:pq_colcol_y}
\end{figure*}

\begin{figure*}
\includegraphics[width=\thinfigwidth]{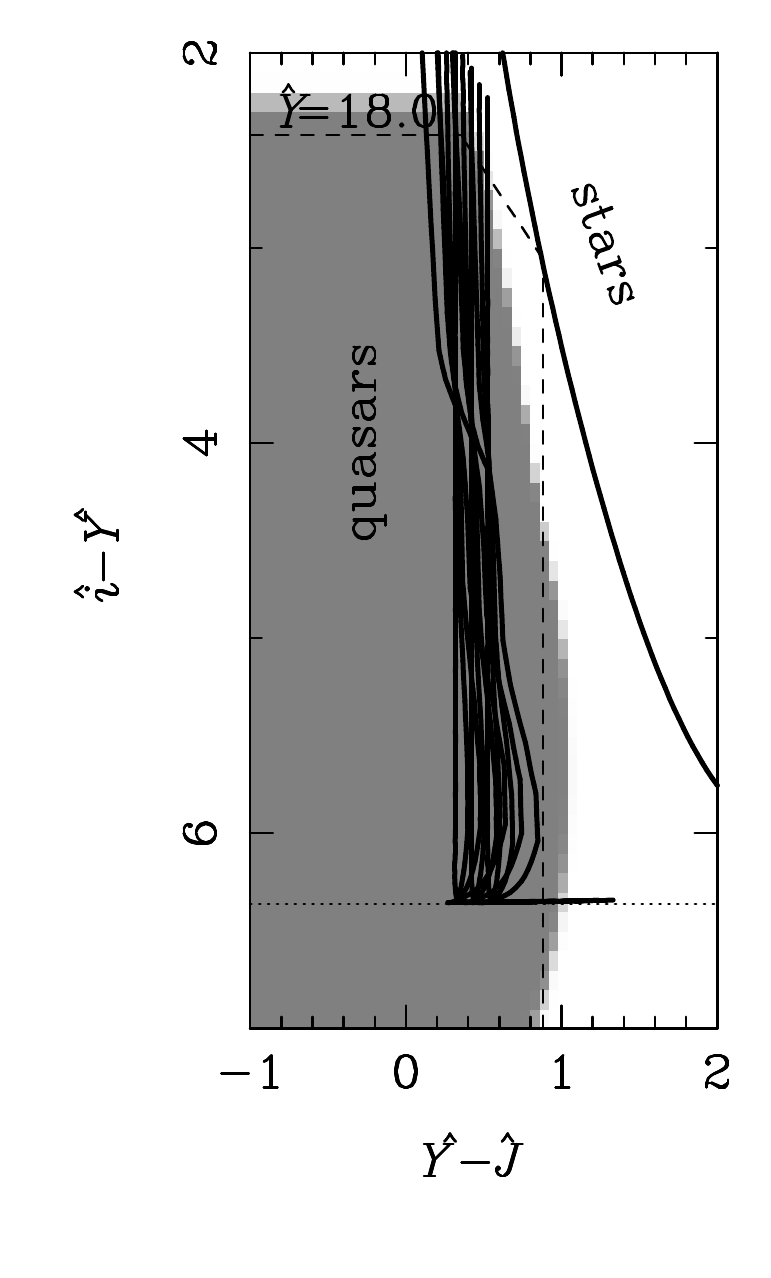}
\includegraphics[width=\thinfigwidth]{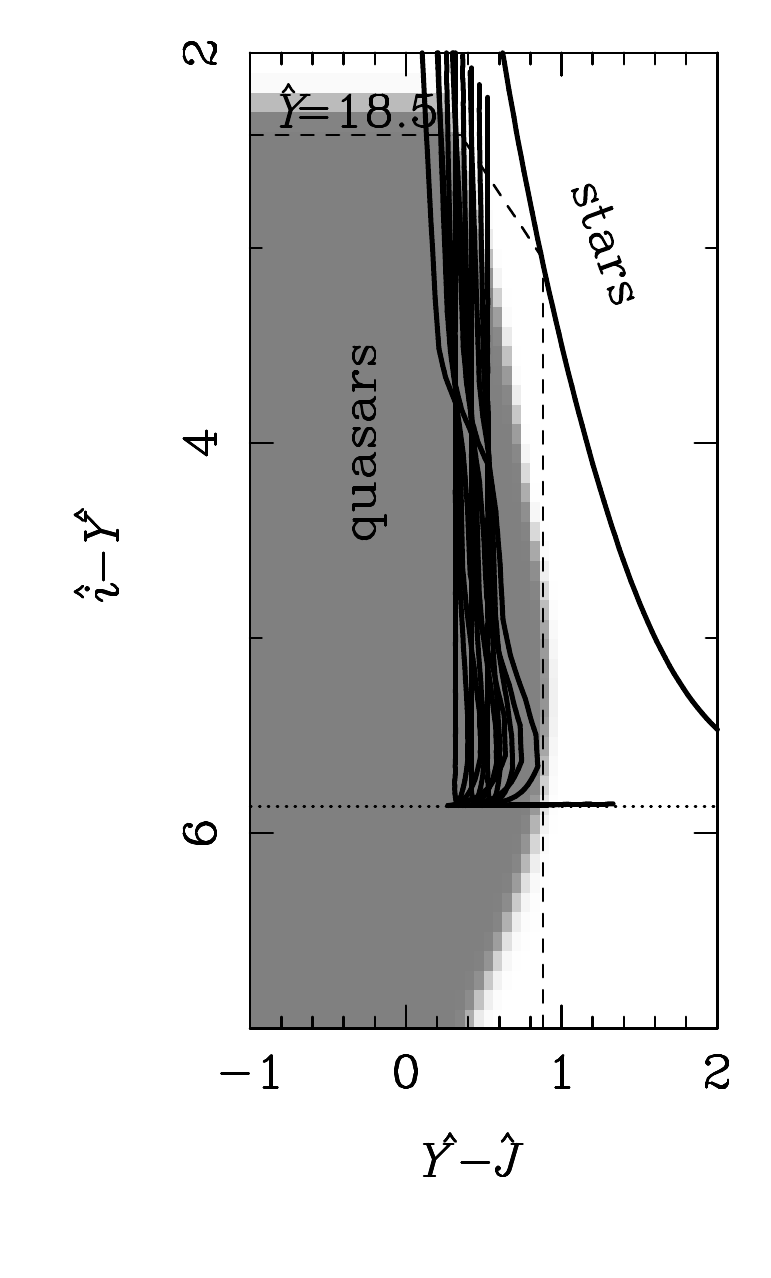}
\includegraphics[width=\thinfigwidth]{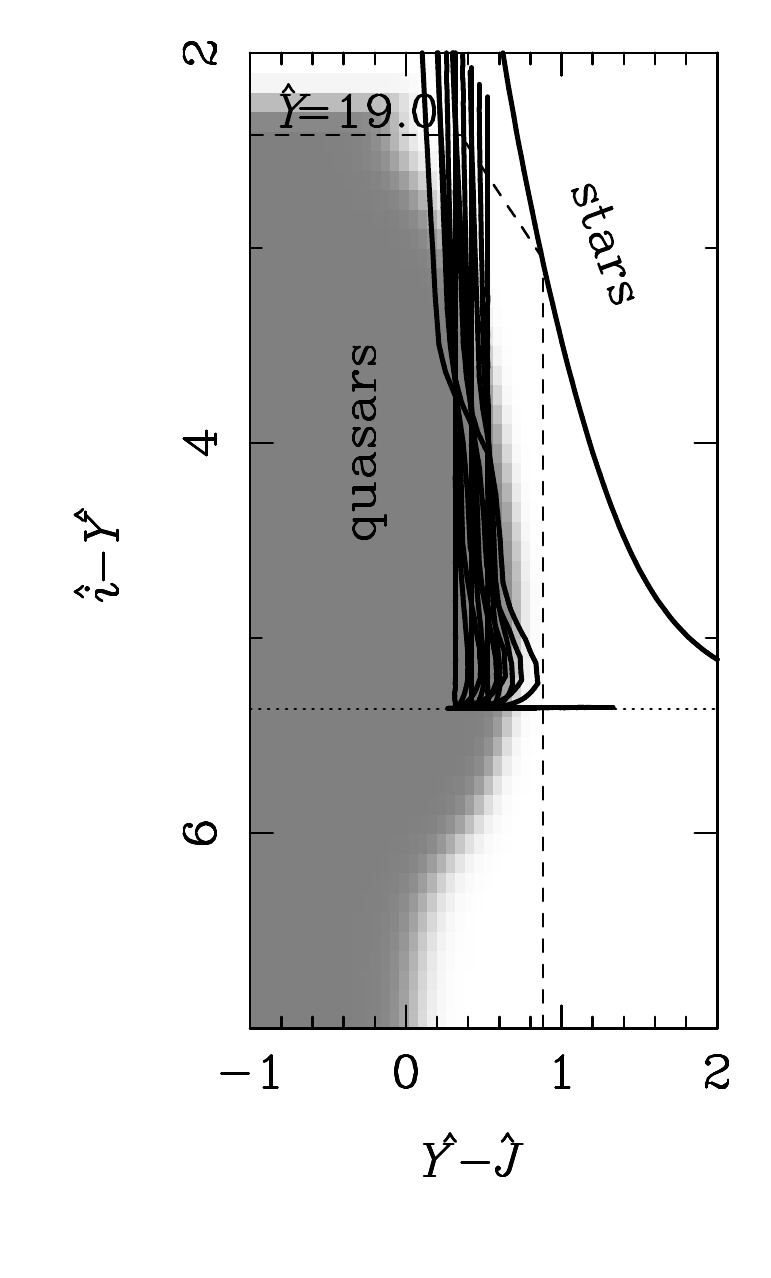}
\includegraphics[width=\thinfigwidth]{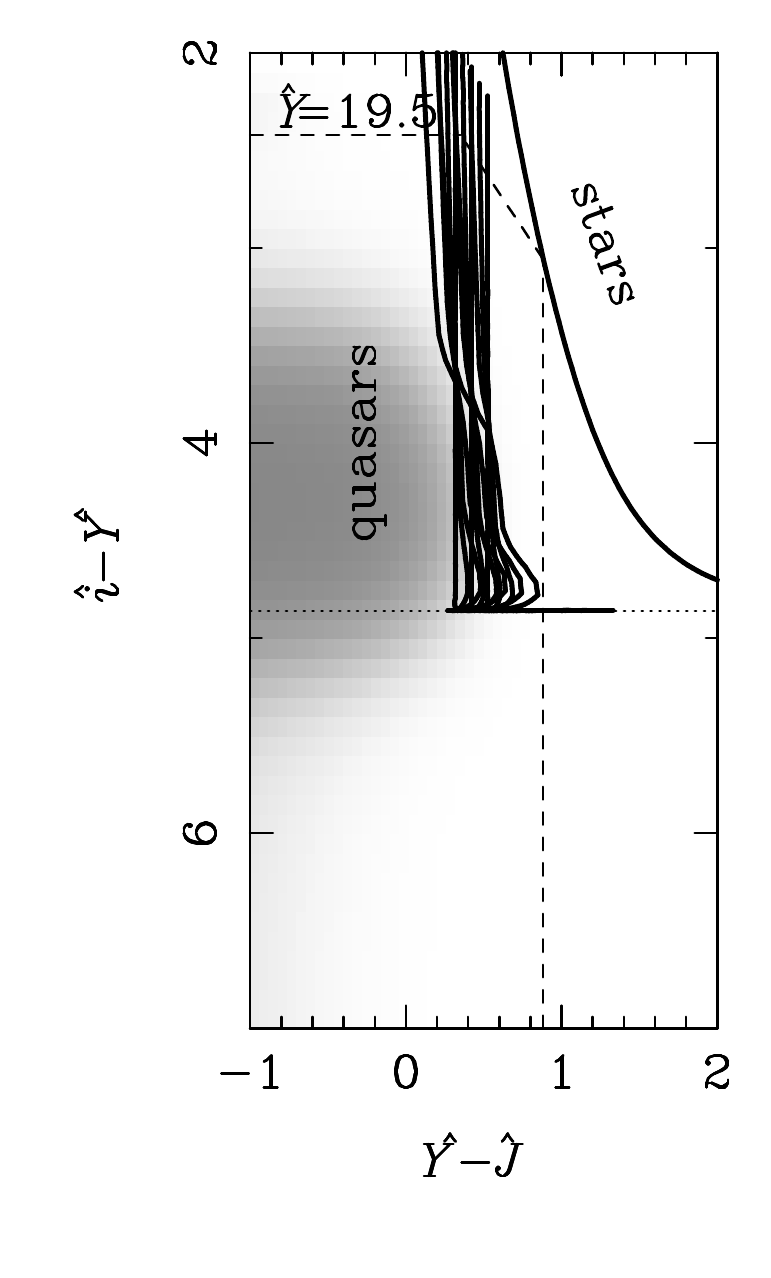}
\caption{The dependence of $\pq$ on
a source's observed $\imy$ and $\ymj$ colours,
ranging from $\pq = 0$ (white) to $\pq = 1$ (grey),
for
$\hat{Y} = 18.0$,
$\hat{Y} = 18.5$,
$\hat{Y} = 19.0$
and
$\hat{Y} = 19.5$, as labelled.
In all cases the noise is as appropriate for
fiducial SDSS and UKDISS observations with
$\iprime_{\limiting} = 22.5$,
$Y_{\limiting} = 20.2$
and
$J_{\limiting} = 22.5$,
but the \zprime\ band has been ignored.
The horizontal dotted line shows the maximum
theoretical value of \imy\ for a source with
$Y = \hat{Y}$ and zero flux in the \iprime band in the absence of noise;
this changes with $\hat{Y}$ due to the use of asinh magnitudes 
to represent the $\iprime$ band photometry.
The fiducial quasar and star loci
are shown as solid curves 
and the dashed lines show the basic pre-selection made 
to generate the \sdssukidss\ candidate sample.}
\label{figure:pq_colcol_z}
\end{figure*}

The above results imply that measurements in at least three bands 
are required to generate a sample of strong \hzq\ candidates,
and in particular that two appropriately chosen colours are needed.
Indeed, many \hzq\ searches have been based on pairs of colour cuts
(\eg\ \citealt{Warren_etal:1994,Fan_etal:2001}),
and this approach is compared directly with the Bayesian results
\figs{pq_colcol_y} and \ref{figure:pq_colcol_z}.
In both cases the depths in the included bands 
are chosen to match the fiducual UKIDSS and SDSS values 
given in \sect{cand},
but data in the missing band ($Y$ and $\zprime$, respectively) is ignored.
\Fig{pq_colcol_y} approximates the 
optical SDSS search of \cite{Fan_etal:2001} 
(but cannot mimic the 
CHFQS described by \citealt{Willott_etal:2007} because the star and
quasar models do not go sufficiently deep),
whereas \fig{pq_colcol_z} represents one of the 
obvious selection options from the matched UKIDSS-SDSS data.
As expected, 
$\pq$ has a similar colour-dependence,
being low near the stellar locus and higher where the quasars 
are expected to be found.
There is also a strong correspondence between the region
of high $\pq$ and the specific selection region defined by 
\cite{Fan_etal:2003}, 
particularly close to the the \zab -band selection cut at $\hat{\zab} = 20.1$.
There are, however, significant systematic differences between
the \cite{Fan_etal:2003} selection region and the the high-$\pq$ region.
The most obvious difference is that
$\pq$ also varies with magnitude for a given set of measured colours.
The most important aspect of the magnitude-dependence
is the decrease in the size of the high-$\pq$ region 
close to the detection limit in the reference band 
(\ie\ as $\zab \rightarrow 20.8$ or $Y \rightarrow 20.2$).
For sources well above the detection limit(s) the photometric errors 
are sufficiently small that there is only a minimal chance of such bright
stars being measured with \hzq -like colours.
But for fainter sources close to the detection limit the effective 
width of the observed stellar locus is greatly increased
and, in the example shown in \fig{pq_colcol_y}, 
encompasses the \hzq\ locus.

\begin{figure}
\includegraphics[width=\figwidth]{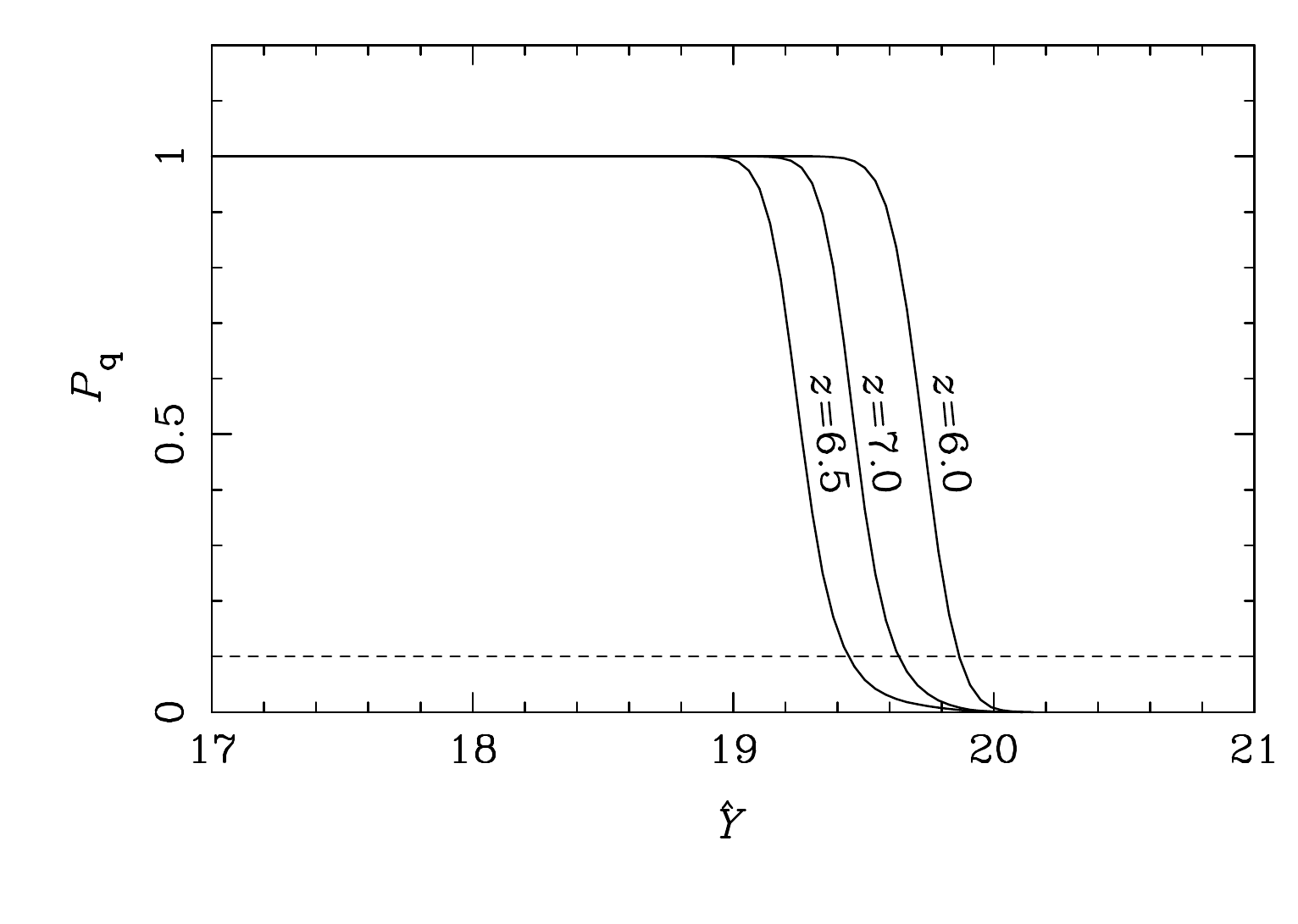}
\caption{The dependence of $\pq$ on
a source's observed $Y$-band magnitude
given it has the observed 
$\imy$, $\zmy$ and $\ymj$ colours of a \hzq\ 
with a redshift of $\redshift = 6.0$,
$\redshift = 6.5$ 
and 
$\redshift = 7.0$, as labelled.
The dashed horizontal line shows the cut of $\pq = 0.1$ used
in selecting \sdssukidss\ candidates for follow-up 
and hence gives the approximate 
completeness limit of the \hzq\ survey at different
redshifts.}
\label{figure:pq_mapp}
\end{figure}

The somewhat counter-intuitive consequence is that 
a sample of sources with (nearly) identical measured colours 
can include both near-certain \hzqs\ and 
obviously uninteresting scattered stars.
The dependence of $\pq$ on reference magnitude is shown in
\fig{pq_mapp} for sources with the measured colours of 
quasars with 
redshifts of $\redshift = 6.0$, $\redshift = 6.5$ and $\redshift = 7.0$.
\hzqs\ over the whole redshift range of interest have $\pq \simeq 1$ 
for $\hat{Y} \la 19$, after which $\pq$ falls fairly sharply 
due to the greater numbers of stars scattered to have quasar-like colours.
For each of the three redshift values
the source's measured colours are constant across the plot,
and so it remains perfectly consistent with being a \hzq;
the change comes about as the observed stellar locus is broader
for higher $\hat{Y}$, and for $\hat{Y} \ga 19.5$ effectively
covers the quasar loci.  
The redshift-dependence of the effective depth comes about
due to the small variations of the \hzqs' $\ymj$ colour as 
various emission lines appear in different filters.
As can be seen in \fig{ccobs},
the expected $\ymj$ colours of 
\hzqs\ increases from $\ymj \simeq 0.4$ at a redshift of 
$\redshift \simeq 6.0$ to $\sim 0.6$ at a redshift of $\redshift \simeq 6.8$,
bringing them closer to the stellar locus.
As a result the maximum depth at which they remain well-separated from the 
observationally-broadened stellar locus is decreased and 
the effective depth (defined as the $Y$ magnitude at which 
$\pq = 0.5$) decreases from $\hat{Y} \simeq 19.7$ at a redshift of
$\redshift \simeq 6.0$ to $\hat{Y} \simeq 19.3$ at a redshift of 
$\redshift = 6.5$.
At higher redshifts, however, the small increase in $\ymj$ is 
much less important than the large increase in the 
\hzqs' expected $\imy$ values.  As a result the effective depth 
at a redshift of $\redshift \simeq 7.0$ has increased to
$\hat{Y} \simeq 19.5$.
One implication of these various subtle effects
is that the selection function of the \sdssukidss\ \hzq\ search 
given in \cite{Mortlock_etal:2011b} is 
more strongly redshift-dependent than the 
SDSS \citep{Fan_etal:2003}
or CFHQS \citep{Willott_etal:2010} selection functions.

\begin{figure*}
\includegraphics[width=\thinfigwidth]{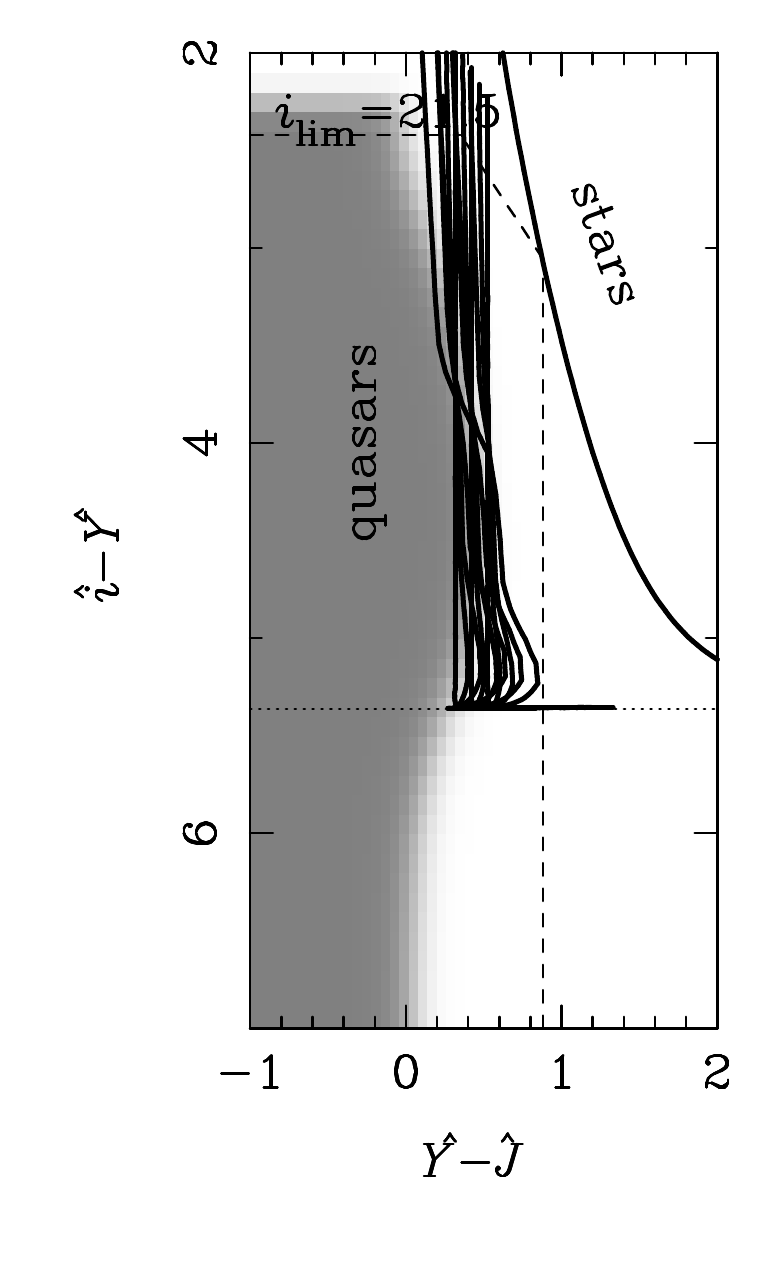}
\includegraphics[width=\thinfigwidth]{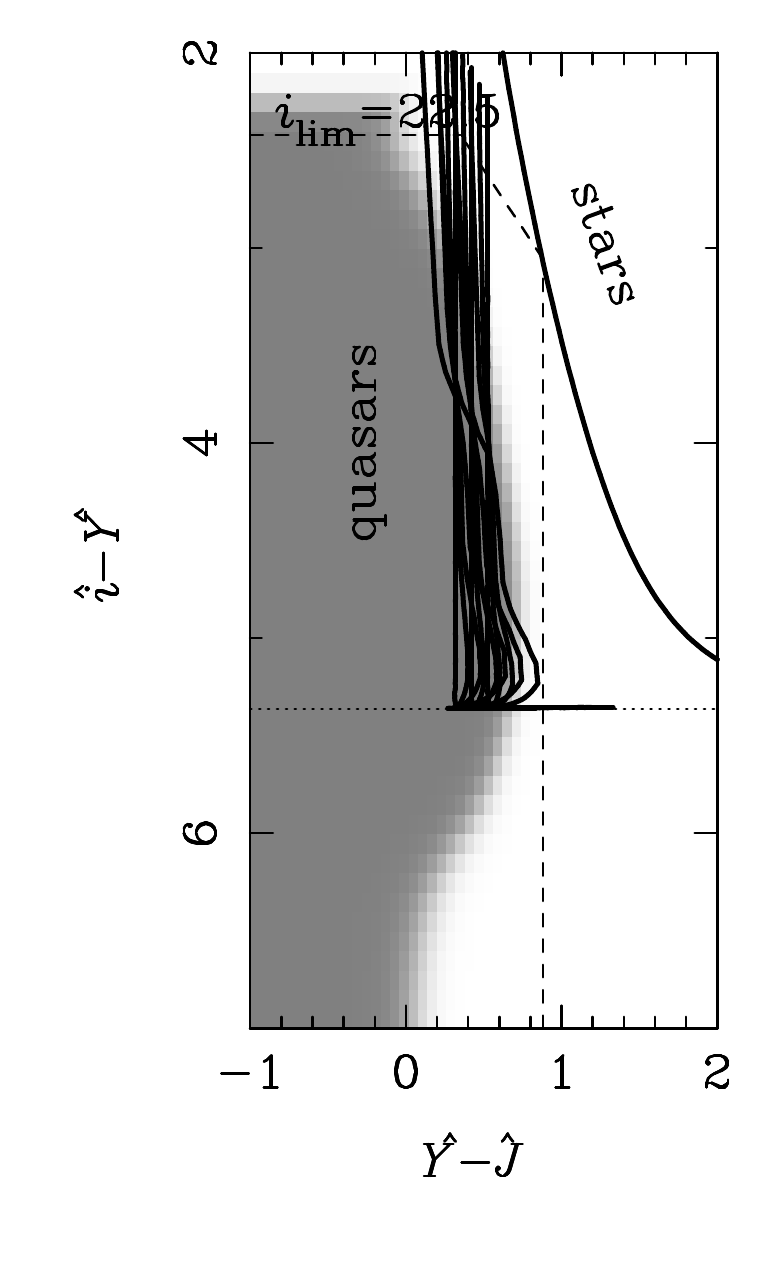}
\includegraphics[width=\thinfigwidth]{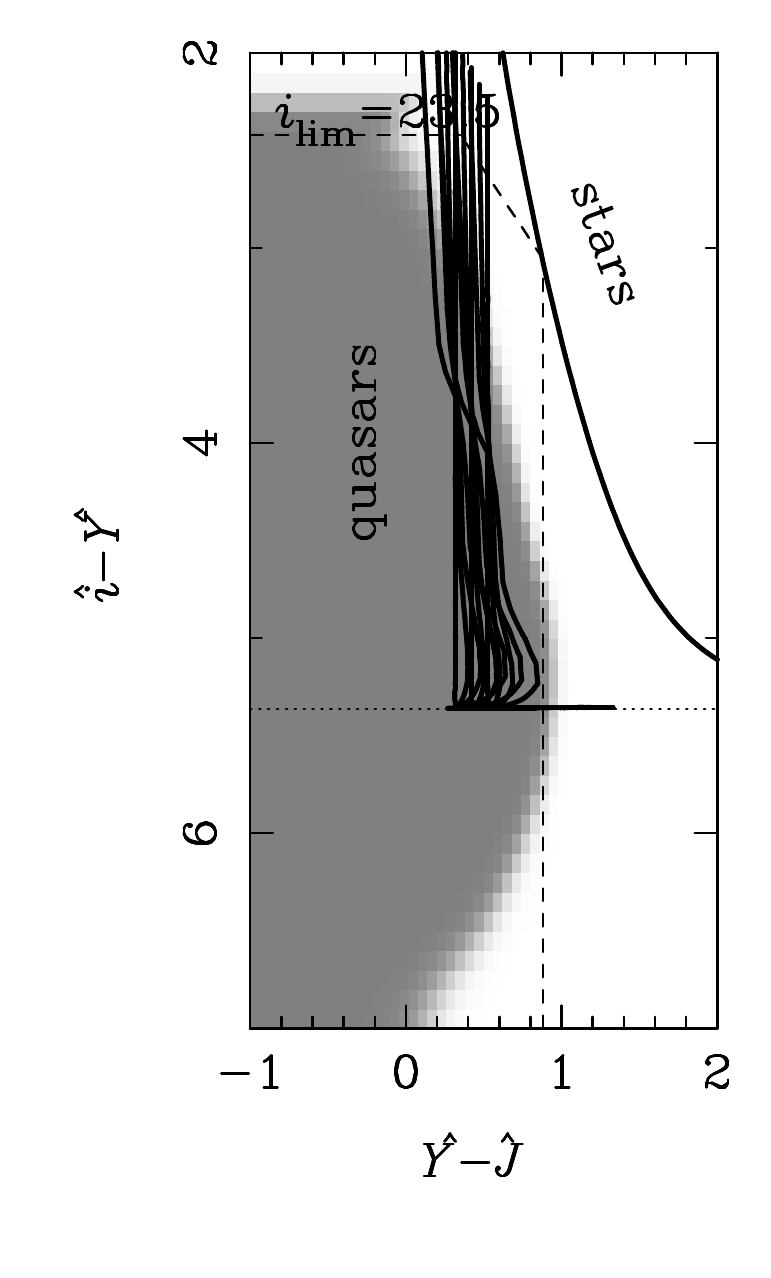}
\includegraphics[width=\thinfigwidth]{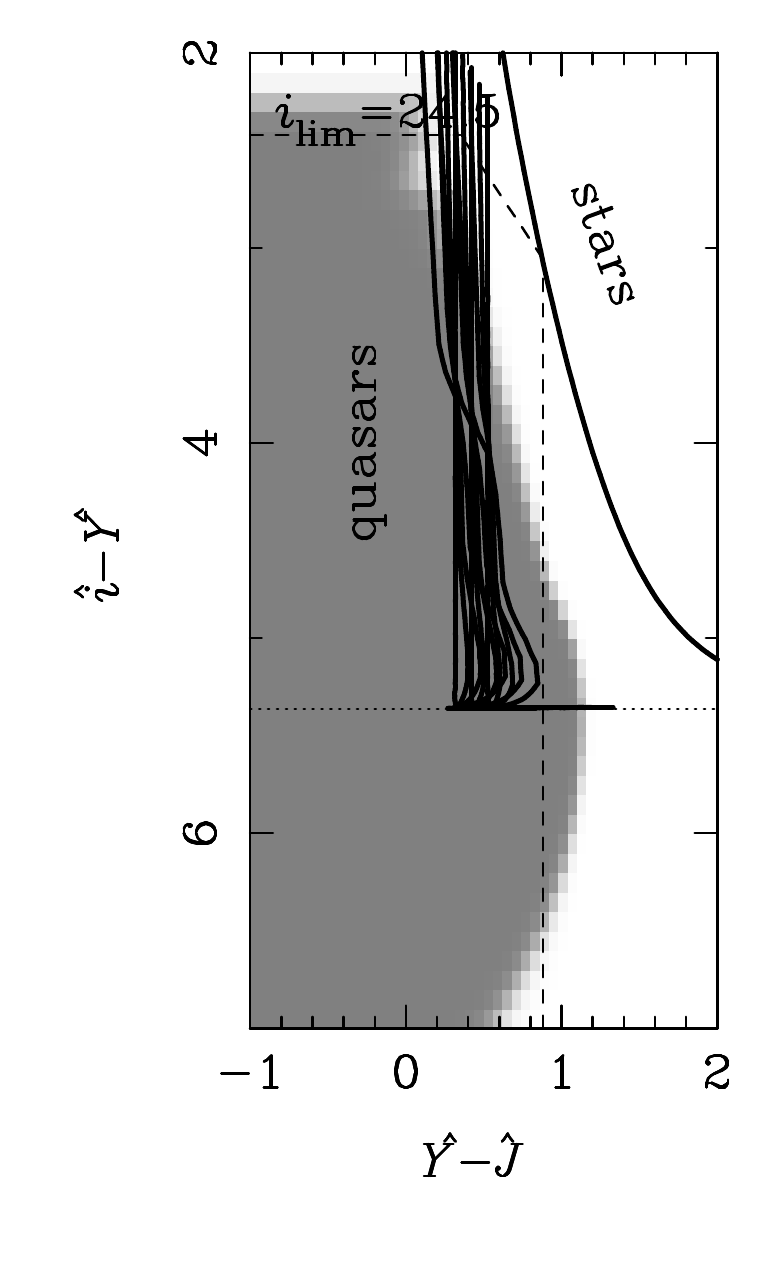}
\caption{The dependence of $\pq$ on
a source's observed $\imy$ and $\ymj$ colours,
ranging from $\pq = 0$ (white) to $\pq = 1$ (grey),
for
$\iprime_\limiting = 21.5$,
$\iprime_\limiting = 22.5$,
$\iprime_\limiting = 23.5$
and
$\iprime_\limiting = 24.5$, as labelled.
In all cases the noise is as appropriate for
fiducial SDSS and UKDISS observations with
$\iprime_{\limiting} = 22.5$
and
$J_{\limiting} = 22.5$,
but the $\zprime$ band has been ignored.
The source has a measured 
$Y$ band magnitude of $\hat{Y} = 19.5$ 
and measured $\iprime$ and $J$ band magnitudes as implied by the 
observed colours. 
The horizontal dotted line shows the maximum
theoretical value of \imy\ for a source with
$Y = \hat{Y}$ and zero flux in the \iprime band in the absence of noise;
this changes with $\hat{Y}$ due to the use of asinh magnitudes
to represent the $\iprime$ band photometry.
The quasar and star loci
are shown as solid curves 
and 
the dashed lines show the basic pre-selection made
to generate the \sdssukidss\ candidate sample.}
\label{figure:pq_depth}
\end{figure*}

Thus far the emphasis has been on the variation of
$\pq$ 
with the properties of a source,
but it is also revealing to investigate how $\pq$ 
depends on survey depth.
\Fig{pq_depth} shows how $\pq$ varies with $\iprime$ band 
depth (assuming a fiducial magnitude of $\hat{Y} = 19.5$).
As expected, extra depth in the $\iprime$ band increases the 
confidence with which $\redshift \simeq 6$ quasars can be identified.
It is possible to push closer to the stellar locus 
(\ie\ redder in $\ymj$) with confidence, and also possible to
find fainter \hzqs\ (\ie\ going deeper in $Y$).
Increasing the depth in the $Y$ or $J$ bands 
is not nearly as useful, because the $\ymj$ colour is already 
sufficiently well measured for a $\hat{Y}\vega \simeq 19$ source.
However extra depth in all three bands would allow a deeper survey,
and hence greater numbers of \hzqs, albeit of lower 
intrinsic luminosity.
This variation also shows the importance of calculating $\pq$
using the measured noise levels in each field of a survey,
rather than just using generic survey-wide depths.

In almost all the above examples 
the transitions between regions
of high and low $\pq$ are quite sharp,
with no large areas of uncertainty.
The transition scale is set
by the photometric errors,
although for a given observation and reference magnitude
(as is the case here)
the errors vary with colour.
This also explains why the transition is more gradual 
in regions of colour space which are expanded by the 
decreased variation of colour with flux,
which is the case for red $\imzobs$ or $\imyobs$ here.
The one case of an obviously gradual transition
is shown the right panel of \fig{pq_colcol_y},
in which
the source is sufficiently faint that the measurement
uncertainties in all
the relevant bands are $\ga 0.2 \unit{mag}$.

%%%%%%%%%%%%%%%%%%%%%%%%%%%%%%%%%%%%%%%%%%%%%%%%%%%%%%%%%%%%%%%%%%%%%%%%%%%%%%

\subsection{Photometric redshift estimation}
\label{section:zest}

\begin{figure*}
\includegraphics[width=\thinfigwidth]{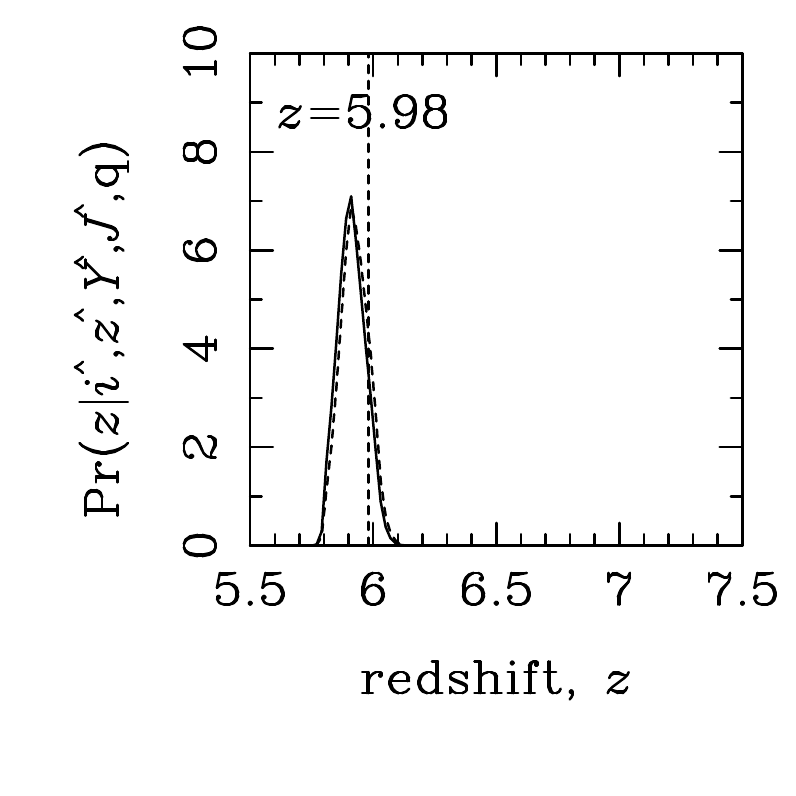}
\includegraphics[width=\thinfigwidth]{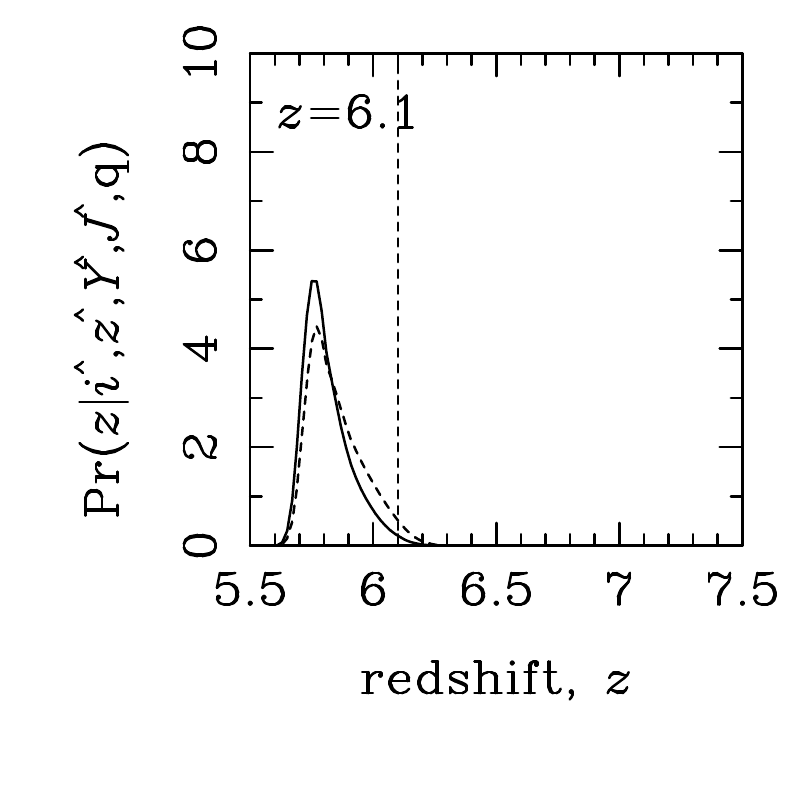}
\includegraphics[width=\thinfigwidth]{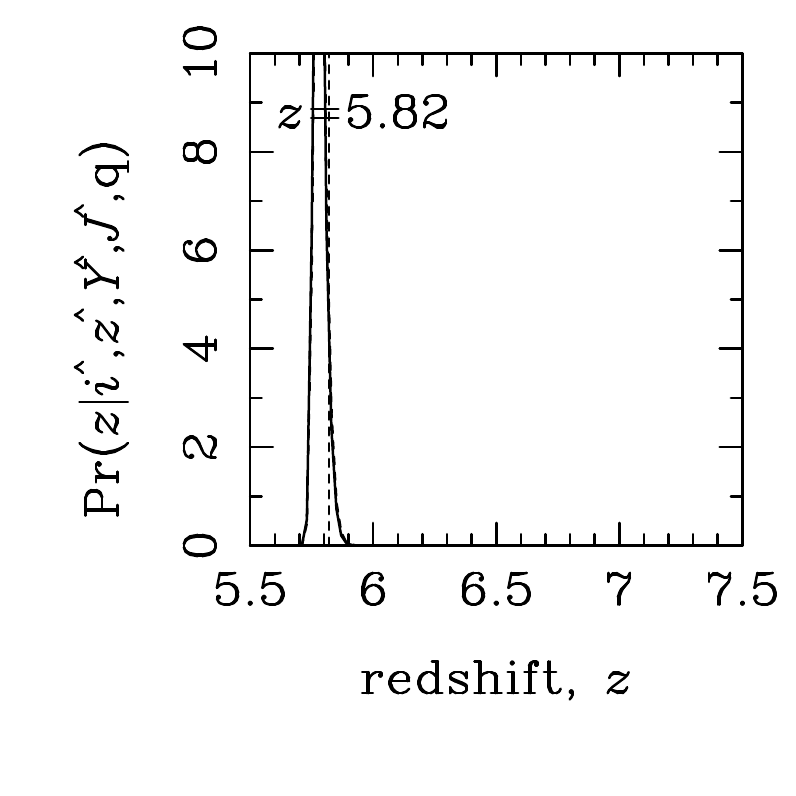}
\includegraphics[width=\thinfigwidth]{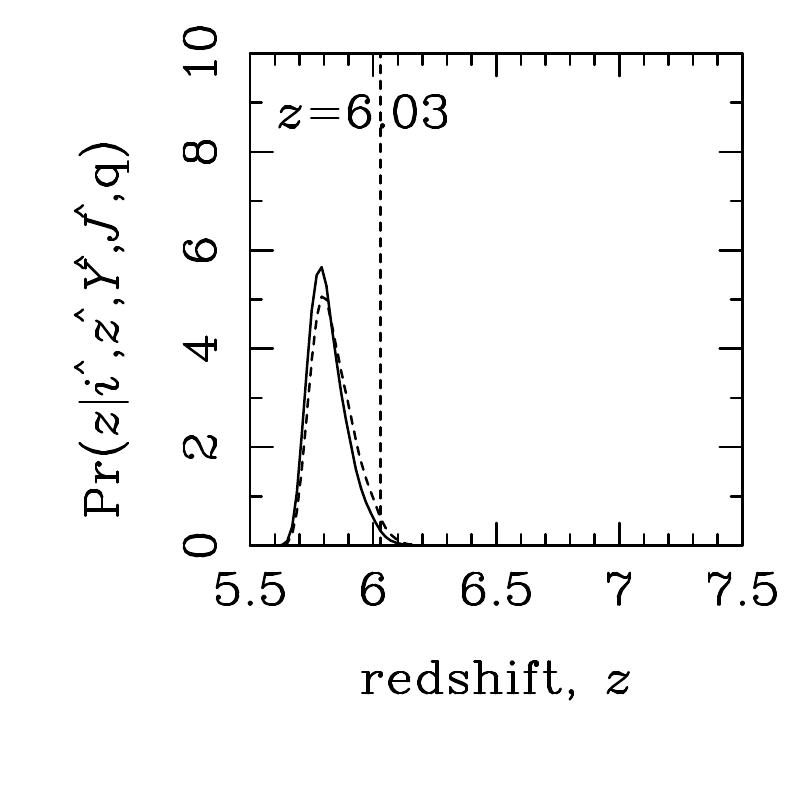}
\includegraphics[width=\thinfigwidth]{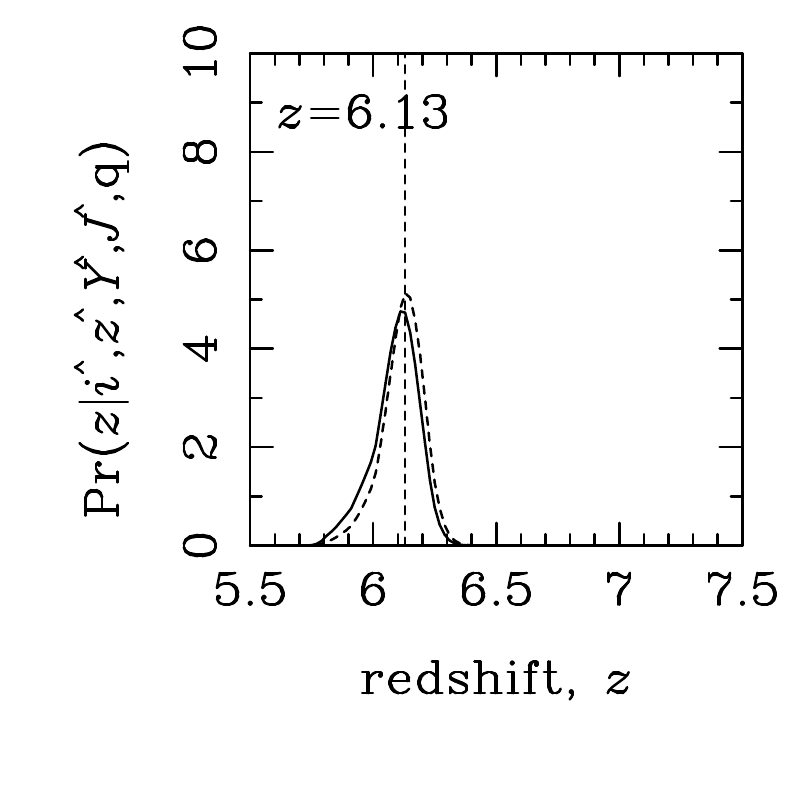}
\includegraphics[width=\thinfigwidth]{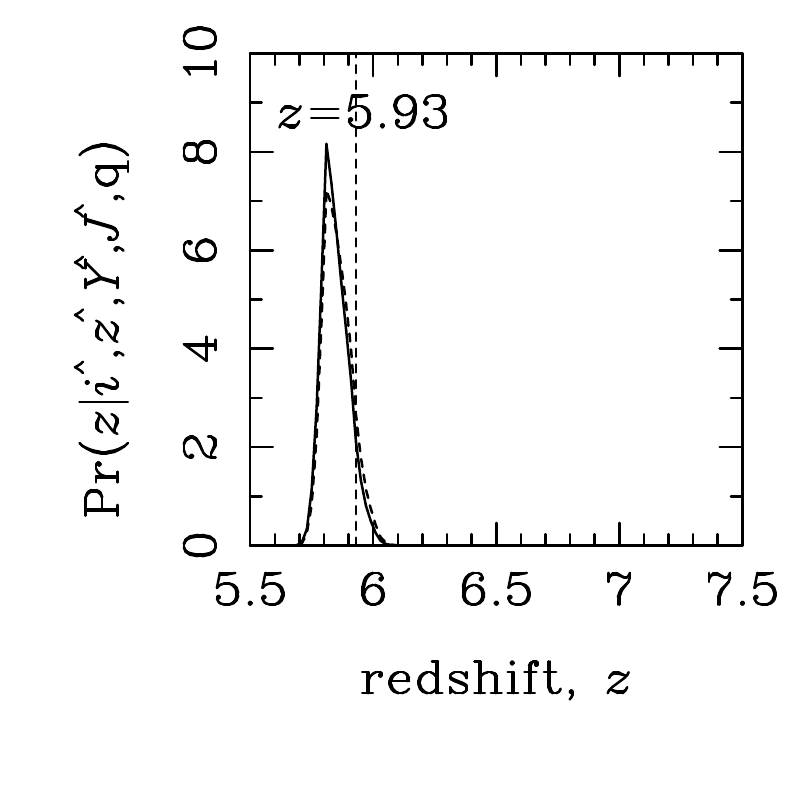}
\includegraphics[width=\thinfigwidth]{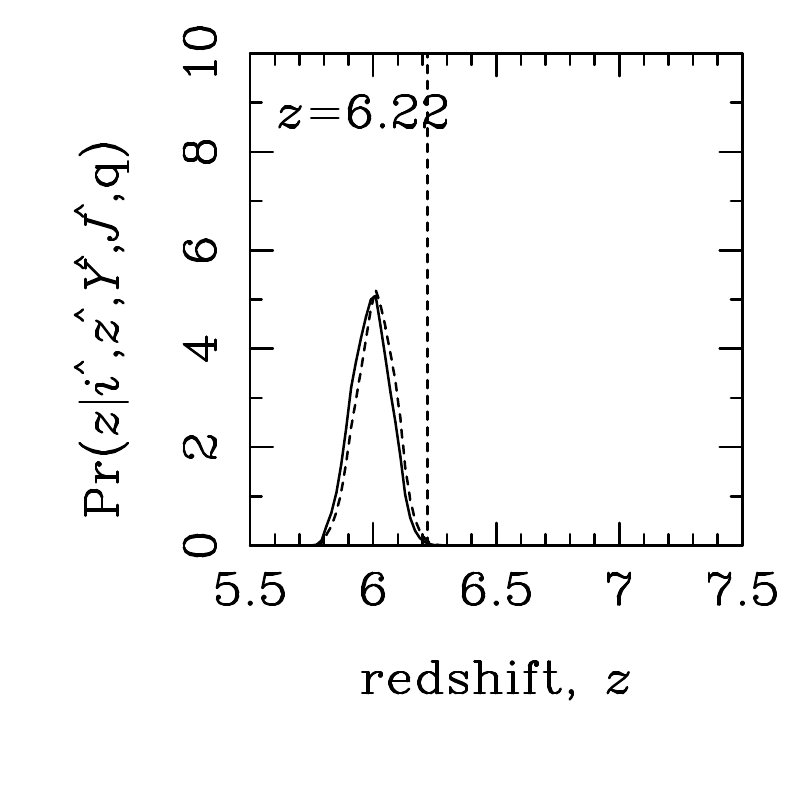}
\includegraphics[width=\thinfigwidth]{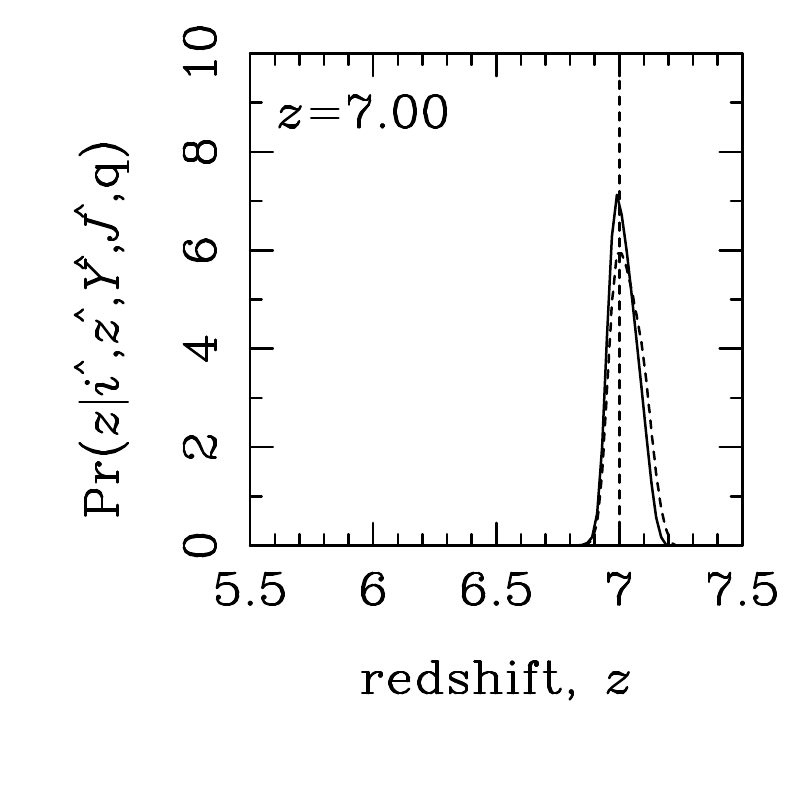}
\caption{Posterior distributions of the redshifts of the \nqso\ 
  \hzqs\ in UKIDSS DR7 
  inferred from the \sdssukidss\ survey photometry.  
  The eighth panel shows the results from a simulated 
  $\redshift = 7$ quasar with 
  $\hatiprime = 24.35 \pm 0.50$,
  $\hat{z} = 22.10 \pm 0.20$,
  $\hat{Y} = 19.50 \pm 0.020$
  and
  $\hat{J} = 19.06 \pm 0.030$.
  Results for both a realistc (solid curve)
  and uniform (dashed curve) prior in the quasar's redshift,
  $\redshift$, and $Y$-band magnitude are shown.  
  The vertical lines show the spectroscopic redshift 
  in the case of the \nqso\ real sources and the simulated 
  redshift in the case of the synthetic $\redshift = 7$ quasar.}
\label{figure:pz}
\end{figure*}

The most important information that can be extracted from
the measurements of a candidate is the probability that it is a \hzq;
but if the source is assumed to be a quasar then
the photometry can also be used to estimate its likely redshift.
Given that a spectrum is necessary to confirm any candidate
as a quasar, there is little long-term utility of such
photometric redshift estimates but
they can useful in prioritizing follow-up observations of
\hzq\ candidates.
The essenetial logic that it is most efficient to
pursue high-$\pq$ objects first is somewhat modified by the
fact that there is a particular premium
on finding the most distant sources
(\ie\ \hzqs\ with $\redshift \ga 6.5$ in the case of UKIDSS).
It would probably make more sense to follow-up
a $\zprime$-band drop-out $\redshift \ga 6.5$ quasar candidate with
$\pq \simeq 0.1$
than a more secure candidate with $\pq \simeq 0.9$ that was
well-detected in the $\zprime$ band
(a fact which might have contributed to the high $\pq$ in the first place).

The calculation is
analagous to that used in 
Bayesian photometric redshift estimation of galaxies
(\eg\ \citealt{Benitez:2000}) 
but with the important difference that quasar spectra exhibit far less
variety than those of galaxies, 
so that it is far easier to obtain reliable estimates.
The posterior distribution of a (putative) quasar's redshift,
given photometric measurements
$\hatiprime$, $\hat{\zprime}$, $\hat{Y}$ and $\hat{J}$, 
is calculated by modifying \eq{wq} to marginalise only over the 
unknown true $Y$-band magnitude. 
That gives 
\[
\prob(\redshift | \hatiprime, \hat{\zprime}, \hat{Y}, \hat{z}, 
  \qso, \detected)
\]
\begin{equation}
\label{equation:pz}
\mbox{}
  = 
  \frac{
  \int_{-\infty}^\infty
    \rho_\qso(Y, \redshift) \,
    \prob(\detected | Y, \redshift, \qso) \,
    \prob(\hatiprime, \hat{\zprime}, \hat{Y}, \hat{z} | Y, \redshift, \qso)
    \, \diff Y
  }
  {
    \weight_\qso(\hatiprime, \hat{\zprime}, \hat{Y}, \hat{z})
  },
\end{equation}
where $\rho_\qso(Y, \redshift)$ is given in \eq{quasarsky},
the appropriate form of the likelihood, 
$\prob(\hatiprime, \hat{\zprime}, \hat{Y}, \hat{z} | Y, \redshift, \qso)$,
is discussed in \sect{likelihood}
and the denominator
$\weight_\qso(\hatiprime, \hat{\zprime}, \hat{Y}, \hat{z})$
ensures the probability is correctly normalised.
(The posterior distribution in any of the model parameters 
-- for either quasars or stars -- 
could similarly evaluated by 
modifying Eq.~\ref{equation:weight} to marginalise over nuisance variables.)

The results of evaluating \eq{pz} 
for the \nqso\ \hzqs\ discovered in UKIDSS 
are shown in \fig{pz}.
In each case the posterior distributions inferred from the 
\sdssukidss\ survey photometry 
(under both realistic and fiducial uniform priors 
in $\redshift$ and $Y$) are compared to the 
spectroscopically-measured redshift.
The similarity of the distributions for the two different priors 
indicate that the detailed knowledge of the \hzq\ population 
is not important:
the \sdssukidss\ photometric data
(in combination with the model of the quasars' colours)
are sufficient to give largely prior-independent redshift estimates that 
have uncerainties of $\Delta \redshift \simeq 0.1$.
Moreover, these inferences are broadly verified by 
the subsequent spectroscopic redshift measurements, 
although there is a suggestion that the 
photometric estimates are systematically low.
The basic success of this method is unsurprising -- 
from \fig{ccobs} it is clear that a good measurement 
even of just
$\imy$ would be sufficient to obtain a reasonable redshift estimate 
-- although the full precision depends on all the 
available photometry.

In the absence of any discoveries of \hzqs\ with redshifts
of $\ga 6.5$ in UKIDSS DR7, 
the photometric redshifts constraints obtained for a simulated 
$\redshift = 7.0$ quasar are shown
the bottom-right panel of \fig{pz}.
Assuming that the 
extrapolation of the spectroscopic \hzq\ models described in 
\sect{quasar} are valid,
the photometric redshift constraints from \sdssukidss\ data 
are accurate to $\Delta \redshift \simeq 0.1$ over the entire
range $5.8 \la \redshift \la 7.2$.
Any $\redshift \ga 6.5$ candidates
should thus be readily identifiable from the 
survey photometry and can
prioritsed in the follow-up process.

%%%%%%%%%%%%%%%%%%%%%%%%%%%%%%%%%%%%%%%%%%%%%%%%%%%%%%%%%%%%%%%%%%%%%%%%%%%%%%

\subsection{Results}
\label{section:results}

\begin{figure}
\includegraphics[width=\figwidth]{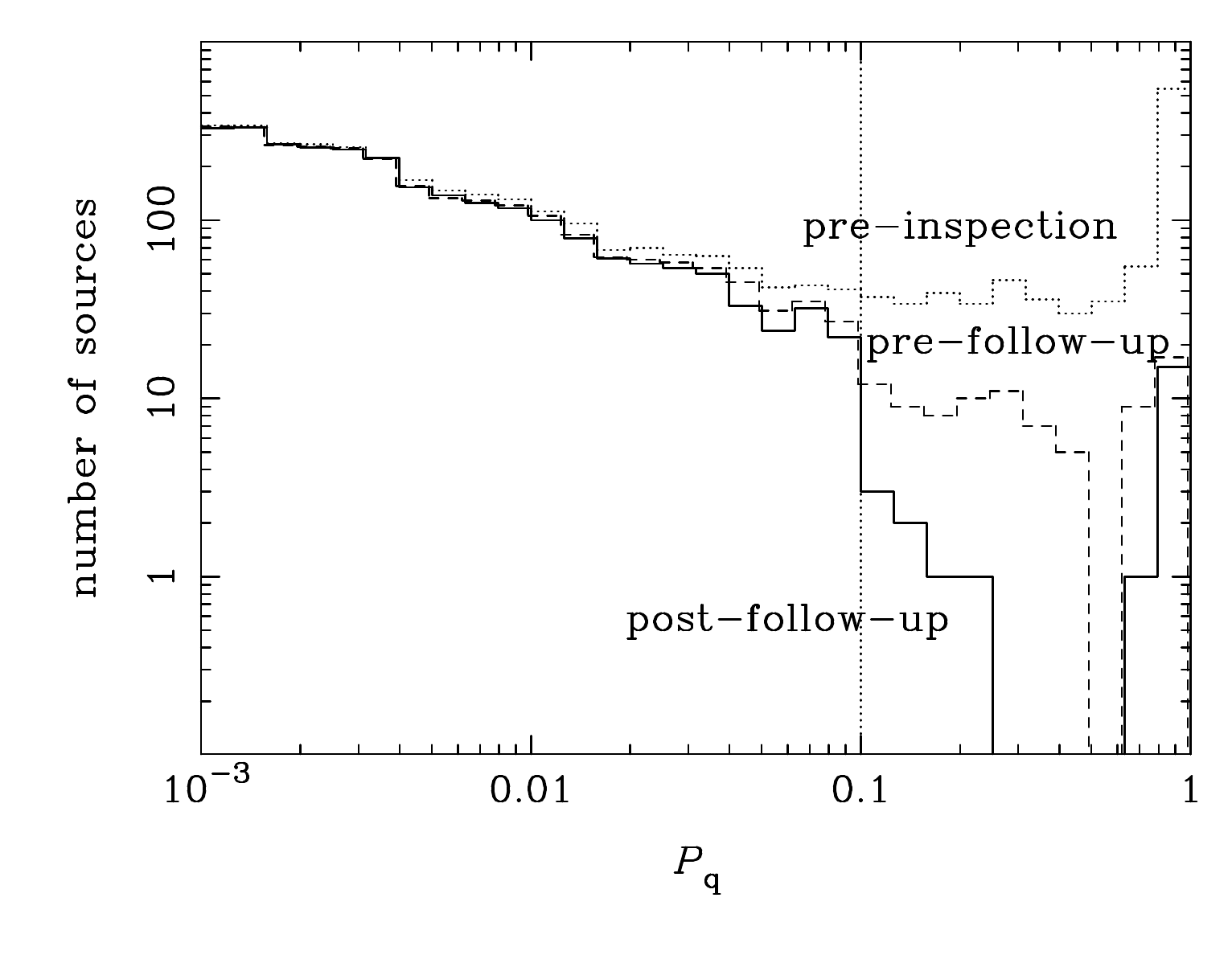}
\caption{Histogram of the
  quasar probabilities, $\pq$,
  of the most promising stationary point--sources
  in UKIDSS DR7. 
  The dotted line shows all sources that passed the automatic filtering
  described in \sect{cand}.
  The solid line shows the results after follow-up photometry.
  All the sources with $\pq \geq 0.1$ have been
  inspected visually, whereas only a random fraction of sources with
  $\pq < 0.1$ have been inspected (in the earlier stages of the
  project, before the selection criteria were finalised),
  and so still include many contaminants. 
  There are a small number of candidates with $0.1 \leq \pq \la 0.3$
  that are yet to be reobserved.}
\label{figure:pqdist}
\end{figure}

\begin{figure*}
\includegraphics[width=\figwidth]{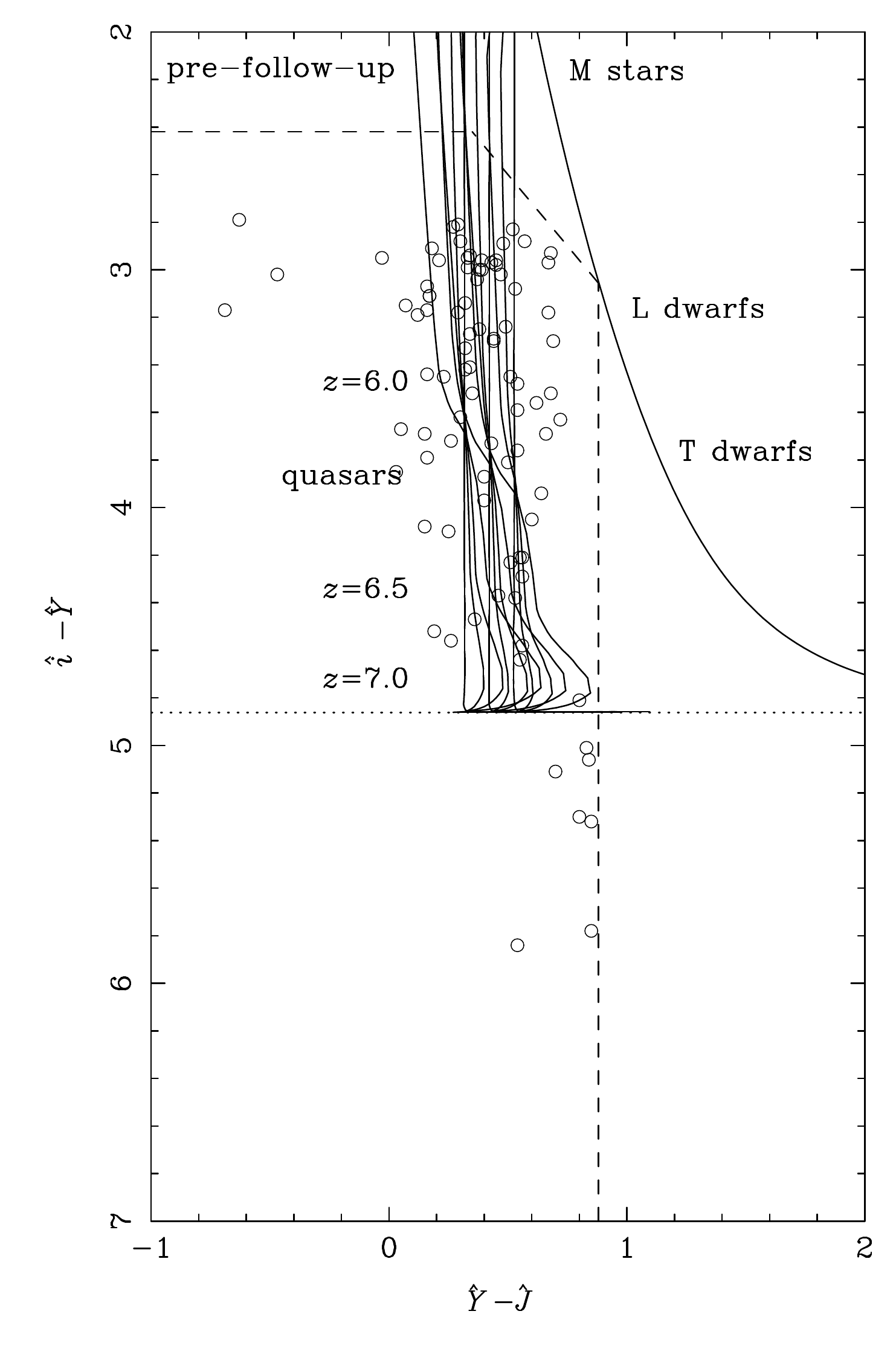}
\includegraphics[width=\figwidth]{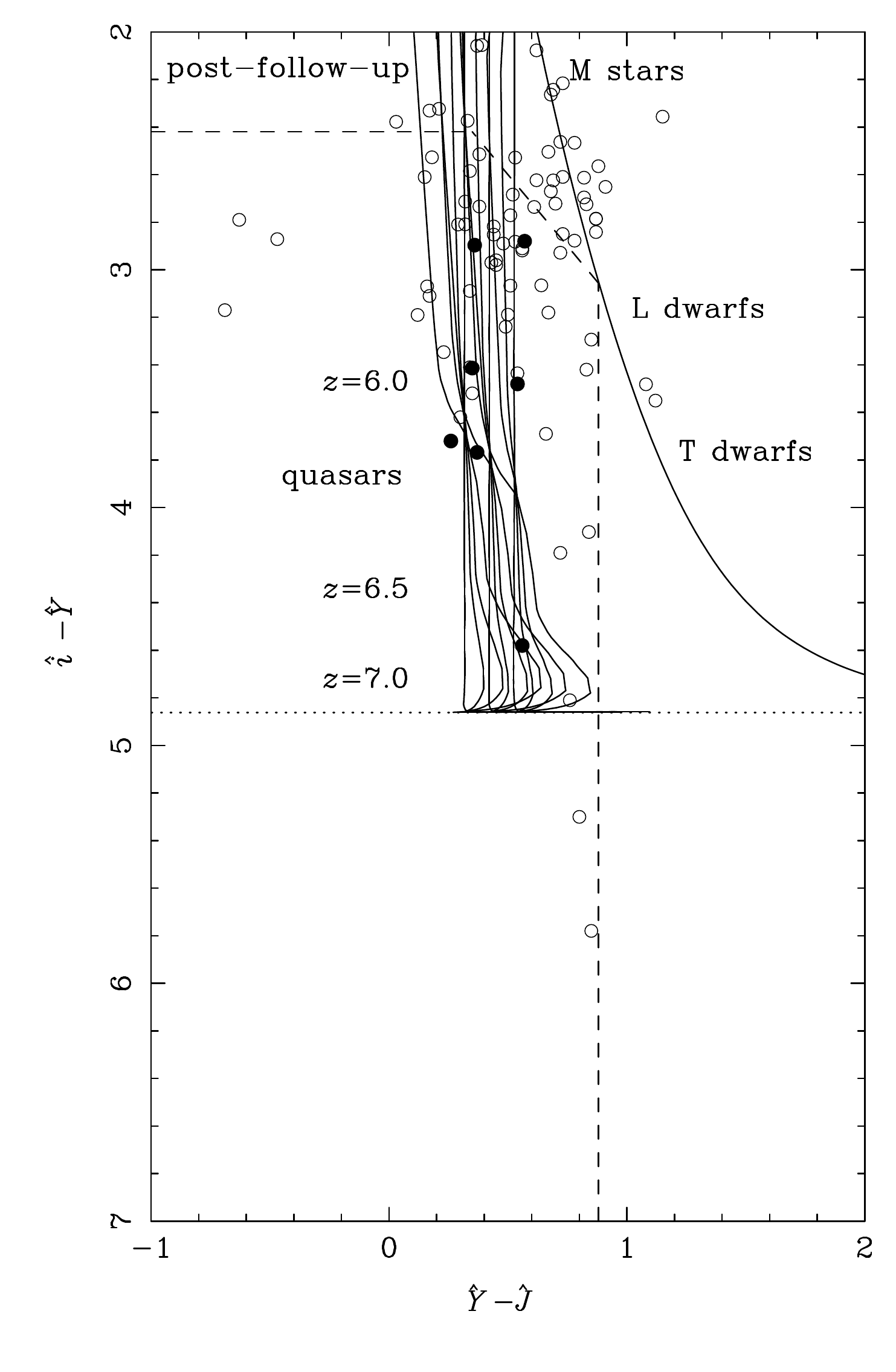}
\caption{Two-colour diagrams showing the 
UKIDSS DR7 LAS point--sources which,
after being cross-matched to SDSS,
have $\pq \geq 0.1$ from the \sdssukidss\ survey photometry.
The colours and from the survey photometry
are shown in the left panel; 
the results of follow-up photometry are shown in the right panel
(in which the open symbols are rejected candidates and 
the solid symbols are the \nqso\ confirmed \hzqs\ in the UKIDSS DR7 sample).
A single follow-up measurement 
(generally in the $\iprime$ band) was sufficient to reject most candidates,
so the some colours (most often $\ymj$) in the two panels remain the same;
also, there are a few candidates which are yet to be reobserved and so 
appear at the same position in both panels 
(as do the previously known \hzqs).
Also shown are the colours of
the twelve quasar models described in \sect{quasar}
and the stellar locus described in \sect{star},
in both cases calculated for a source that has $Y = 19.5$.
The dashed lines denote the initial pre-selection
cuts that are applied before subsequent processing
and the horizontal dotted line shows the maximum
$\imy$ value that a $Y = 19.5$ source could have in the absence of noise.}
\label{figure:cccand}
\end{figure*}

Having understood how the photometric measurements 
of a source combine with models of the star and quasar populations
to give $\pq$ (\sect{exampleprob}), 
the Bayesian selection method could be applied with confidence to the 
\sdssukidss\ sample of 
$\simm 10^4$ candidate \hzqs\ described in \sect{cand}.
The value of $\pq$ was calculated for each source,
and the high-probability tail of the resultant distribution 
is shown as the dotted histogram in \fig{pqdist}.
The main result was that the vast majority of sources 
with \hzq -like colours 
had $\pq \ll 1$ and were not considered further.
Given the low number of \hzqs\ expected in the sample,
the rejection of most candidates was both desireable and predictable, 
and immediately rendered the follow-up task far more manageable.

Applying a cut of $\pq \geq 0.1$
left just \ninitial\ promising candidates in UKIDSS DR7,
a completely automated reduction by a factor of 
$\simm 2.5 \times 10^4$ from the initial \sdssukidss\ sample.
Having extracted essentially all the useful information
from the photometry, 
the next step was to inspect the 
SDSS and UKIDSS images of the remaining 
candidates\footnote{During the the development of the
selection algorithm a large number of candidates with 
$\pq \geq 0.1$ were also investigated. 
Even though this effort was subsequently revealed to be 
unnecessary, the fact that all of these low-probability 
candidates were rejected was an important check on the whole process.}.
Most of the candidates were, unsurprisingly, revealed to be spurious,
with a large variety of explanations
(see \citealt{Mortlock_etal:2011b} for more details):
faint galaxies with supernovae that were
present in the UKIDSS observations but absent
when the SDSS observations were made;
Solar System asteroids that were observed by UKIDSS close to
turn-around\footnote{Asteroids
were most problematic in the early stages of the UKIDSS survey,
before (generally non-contemperaneous) $H$ and $K$ band observations were 
made in most fields.  As detailed 
further in \cite{Mortlock_etal:2011b},
any sources which were observed in the $Y$ and $J$ bands at 
the elongation expected of Main Belt asteroids at turn-around 
but which were completely absent in
the $\iprime$, $\zprime$, $H$ and $K$ bands 
were not considered further.};
UKIDSS cross--talk or persistence \citep{Dye_etal:2006}; 
and various data artefacts that resulted in obviously incorrect
photometry.
In theory, all such contaminants could be included as
additional models (along with stars and quasars) 
in the Bayesian classification scheme;
but, aside from the difficulties
in specifying the likely numbers and properties
of these various contaminants, 
these contaminants are sufficiently rare
that their identifcation does not consume
significant resources.

Of the \ninitial\ UKIDSS DR7 objects with $\pq \geq 0.1$, 
only \nreal\ were
confirmed as real, stationary, astronomical sources with no obvious 
data problems that might result in erroneous photometry.
\Fig{cccand} shows the colours of these candidates 
both from the intial \sdssukidss\ survey photometry (left panel)
and then updated after follow-up observations (right panel).
The distribution of these (real) candidates' quasar probabilities 
is shown as the dashed histogram in \fig{pqdist}.
The distribution of $\pq$ values 
can be used as a guide to calibrate the probability calculation:
if the models of the measurement process and the two populations 
were completely accurate then the sum of $\pq$ over all the candidates
would be approximately equal to the expected number of \hzqs\ in the sample.
That is not the case here: 
whereas only \nexp\ \hzqs\ are expected,
the initial probability sum is $\simm 60$
and the sum after follow-up observations 
is $\simm 40$.
The most plausible reason for these discrepancies is the use of 
a Gaussian likelihood (see \sect{likelihood}) rather than a distribution
with heavier tails to account better for outliers.  
The fact that the sum even after follow-up is still so high indicates that
the probabilities even for poor candidates with $\pq \la 0.01$ are 
systematically high.
It is, however, the realtive probabilities (\ie\ their rankings)
that of primary importance 
-- the real reason for exploring a Bayesian selection algorithm 
in the first place was not so much to determine the 
probability that the candidates are \hzqs,
but to answer the
distinct question of which 
of the identified candidates were the most likely to be quasars.

Having ranked the candidates, 
the first stage of the follow-up campaign was to see whether
any of the red \sdssukidss\ sources had been catalogued previously.
Cross-matching with SDSS revealed that 
\nqsoprev\ of the highest-ranked objects were known \hzqs\
(SDSS~J083643.9$+$005453.2 
  at redshift $\redshift = 5.82$, \citealt{Fan_etal:2001};
SDSS~J141111.3$+$121737.3 
  at redshift $\redshift = 5.93$, \citealt{Fan_etal:2004};
and 
SDSS~J162331.8$+$311200.5 
  at redshift $\redshift = 6.22$, \citealt{Fan_etal:2004}).
All \nqsoprev\ were very strong candidates with $\pq \geq 0.99$
and so clearly would have been discovered by subsequent 
observations had they been required.
The remaining candidates
were all queued for follow-up
photometric observations at 
The Liverpool Telescope ($\iprime$ filter), 
The Isaac Newton Telscope ($\iprime$ and $\zprime$ filters) 
or UKIRT ($Y$ and $J$ filters).
The follow-up images were generally deeper than the 
SDSS and UKIDSS survey observations, 
but even more important than an increase in photometric precision
was that these new measurements were independent of the 
candidate selection process.
Whereas the initial selection 
could be thought of as a method 
of identifying stars for
which the SDSS measurements are faint in \iprime\ or 
in which the UKIDSS data are bright in $Y$,
the follow-up photometry should be unbiased.
Every time a measurement was made the quasar probability
was recalculated with the new photometry;
a candidate was discarded if $\pq \la 0.01$ at any stage.
Many candidates were rejected after just one follow-up observation,
in most cases because they were revealed to be significantly 
brighter in the $\iprime$ band than indicated by the initial
survey photometry.
These candidates can be seen with $\imyobs \simeq 2.5$ in \fig{cccand};
as expected, their observed colours are much more like those 
of the reddest M dwarfs.
(Further follow-up observations in the $Y$ band
would probably reveal that the true $\imy$ values 
are bluer still -- as the initial sample was 
selected to be red in $\imy$,
the initial $Y$-band photometry of the candidates
tends to be biased bright, just as their initial $\iprime$-band 
photometry is biased faint.)
It is also striking that a number of the rejected candidates 
still have the observed $\imy$ and $\ymj$ colours of \hzqs, 
again illustrating the strong role that the Bayesian priors 
(particularly the relative numbers of stars and quasars) 
play in this process.

If a candidate still had $\pq \simeq 1$ after 
reobservation in at least the $\iprime$, $Y$ and $J$ bands then
spectroscopic observations were obtained.
As detailed in \cite{Mortlock_etal:2011b},
spectra were taken of 
\nspec\ UKIDSS DR7 candidates:
\nqsonew\ were confirmed as new \hzqs\footnote{The 
first new \hzq\
discovered in UKIDSS, \firstqso\ \citep{Venemans_etal:2007}, was 
identified before this probabilistic method had been developed,
and was also subsequently revealed to be a broad absorption line
(BAL) object \citep{Mortlock_etal:2009a}.
As BALs are not explicitly included in the quasar model described
in \sect{quasar} they will not be found reliably by the
selection method implemented here.}
(\citealt{Mortlock_etal:2009a,Venemans_etal:2011,Patel_etal:2011}).
At the end of this follow-up process 
every source with $\pq \geq 0.1$ from the \sdssukidss\ survey
photometry should either be convincingly rejected or 
spectroscopically confirmed as a \hzq.
This separation into sources with $\pq \simeq 0$ and 
confirmed \hzqs\ with $\pq = 1$ is a natural result of 
obtaining more information about those candidates that were initially 
promising but fundamentally ambiguous. 
The separation is not readily apparent in \fig{cccand}
as neither the $\zprime$-band photometry nor the photometric errors 
are shown (and because a few candidates are yet to be followed up
and so not yet decisively classified),
but it is well illustrated by the solid histogram in \fig{pqdist}.

The choice of a rigorous probability cut-off (\ie\ $\pq \geq 0.1$ here) 
is not necessary to maximize the yield from the survey 
in terms of \hzqs\ numbers but is needed to
evaluate the selection function (\ie\ completeness) of the final sample.
Indeed, the primary selection criterion for a \hzq\ of given 
intrinsic properties is that it has $\pq \geq 0.1$ in the \sdssukidss\
survey photometry.
The selection probability is not, however, equal to $\pq$;
rather it given by the fraction of \hzqs\ that, when observed 
with the appropriate noise levels in the \sdssukidss\ bands, 
would have $\pq \geq 0.1$.
The selection probability is evaluated in \cite{Mortlock_etal:2011b} as 
part of the estimation of the \hzq\ luminosity function from the
UKIDSS DR7 data.

%%%%%%%%%%%%%%%%%%%%%%%%%%%%%%%%%%%%%%%%%%%%%%%%%%%%%%%%%%%%%%%%%%%%%%%%%%%%%%

\section{Conclusions}
\label{section:conc}

A probabilistic approach to quasar selection can,
at least in principle, 
make use of all the relevant information about a candidate \hzq:
knowledge of the quasar and star populations;
the stochastic nature of the measurement process;
and, of course, the data obtained for the source in question.
Having derived a general Bayesian formalism for \hzq\ selection,
this approach was applied to real quasar candidates taken from the 
cross-matched UKIDSS and SDSS data-sets.
The $\simm 1900 \unit{\sqdeg}$ surveyed as of UKIDSS DR7
contained only \nreal\ real astronomical sources with 
quasar probabilities of $\pq \geq 0.1$,
and most of these candidates were quickly rejected with follow-up 
photometry (a single \iprime -band image sufficing in many cases).
Aside from \nqsoprev\ known \hzqs, 
this follow-up process left just \nspec\ strong photometric candidates,
of which \nqsonew\ were redshift $\redshift \simeq 6$ quasars.
If a cut-based approach had been adopted then 
follow-up observations would have been required 
for $\simm 10^3$ candidates.
Not only was the Bayesian selection method very efficient, 
but it was also entirely quantitive and objective, 
as needed to estimate the high-redshift \qlf\ from the sample
(see \citealt{Mortlock_etal:2011b}).

It is also possible to combine this model-selection approach
with parameter estimation.  In the context of \hzqs\ it is
clear that the highest redshift objects are the most important,
in which case $\pq$ could be combined with an estimate of the putative
quasar's redshift to rank potentialy record-breaking \hzqs\ above
those at redshifts which have already been explored.
Comparing the resultant photometric \hzq\ redshift estimates of
the \nqso\ confirmed quasars with their spectroscopic redshifts
confirms that the photometry is sufficiently imformative about
a quasar's redshift that it can be used 
to prioritize candidates in the follow-up process.

The Bayesian \hzq\ selection method described here will 
continue to be used 
in the analysis of subsequent UKIDSS data releases,
and may also be applied to data from future \nir\ surveys 
such as 
Pan--STARRS \citep{Kaiser:2002} and
VISTA \citep{Emerson_etal:2004}.
The utility of -- and need for -- such a complicated approach
to \hzq\ selection depends on the survey details:
the bands used and the depths reached 
determine the degree to which the target \hzqs\ are
separated in colour space from the contaminating stellar population(s).
In almost all cases, however, the applications of these Bayesian 
selection methods would represent a step 
closer to extracting
as much science as possible from the available data by
investigating fainter sources.

A corollary of the survey-dependent nature of Bayesian 
quasar selection is that 
the expected distribution of candidate probabilities could be
a useful tool in survey design.  
This was particular apparent by the degree to which the 
high-$\pq$ region of the SDSS data space matched the 
\hzq\ colour cuts adopted by \cite{Fan_etal:2001}.
Given that the trade-off between
broader filters (giving a higher \stonlong) and narrower
filters (giving better-defined colours) can only be assessed 
properly in the context of the expected source populations,
the separation of the probability distributions for simulated sources
of different types 
would be a powerful diagnostic.

Nonetheless, the principles behind the \hzq\ selection method
presented here are completely generic,
and could be usefully adapted to any astronomical 
classification problem in which the 
available data on the sources of interest 
do not permit decisive classifications to be made.
The price is the need to model the relevant source populations,
but the pay-off in the case of a search
rare objects is a massive reduction in the amount of 
follow-up observations required to extract the few 
unusual sources of interest.

%%%%%%%%%%%%%%%%%%%%%%%%%%%%%%%%%%%%%%%%%%%%%%%%%%%%%%%%%%%%%%%%%%%%%%%%%%%%%%

\section*{Acknowledgments}

Xiahoui Fan,
Sebastian Jester
and 
Gordon Richards
all provided invaluable insights into the subtleties of quasar selection.

MP acknowledges support from the University of London's Perren Fund.
PCH, RGM and BPV acknowledge support from the STFC-funded
Galaxy Formation And Evolution programme at the Institute of
Astronomy.

This work is based in part on data obtained during the UKIDSS project.
Many thanks to the staff at UKIRT, the Cambridge 
Astronomical Survey Unit, and the Wide Field Astronomy Unit, Edinburgh, 
for their work in implementing UKIDSS.

Funding for the SDSS and SDSS-II has been provided by the 
Alfred P.\ Sloan Foundation, the Participating Institutions, the 
National Science Foundation, the U.S.\ Department of Energy, 
the National Aeronautics and Space Administration, 
the Japanese Monbukagakusho, the Max Planck Society, 
and the Higher Education Funding Council for England. 
The SDSS Web Site is {\tt{http://www.sdss.org/}}.

The SDSS is managed by the Astrophysical Research Consortium for the 
Participating Institutions. 
The Participating Institutions are the 
American Museum of Natural History,
Astrophysical Institute Potsdam,
University of Basel,
University of Cambridge,
Case Western Reserve University,
University of Chicago,
Drexel University,
Fermilab,
the Institute for Advanced Study,
the Japan Participation Group,
Johns Hopkins University,
the Joint Institute for Nuclear Astrophysics,
the Kavli Institute for Particle Astrophysics and Cosmology,
the Korean Scientist Group,
the Chinese Academy of Sciences (LAMOST),
Los Alamos National Laboratory,
the Max-Planck-Institute for Astronomy (MPIA),
the Max-Planck-Institute for Astrophysics (MPA),
New Mexico State University,
Ohio State University,
University of Pittsburgh,
University of Portsmouth,
Princeton University,
the United States Naval Observatory,
and the University of Washington.

The Liverpool Telescope is operated on the island of La Palma by
Liverpool John Moores University in the Spanish Observatorio del Roque
de los Muchachos of the Instituto de Astrofisica de Canarias with
financial support from the UK Science and Technology Facilities
Council.

%%%%%%%%%%%%%%%%%%%%%%%%%%%%%%%%%%%%%%%%%%%%%%%%%%%%%%%%%%%%%%%%%%%%%%%%%%%%%%

\bibliographystyle{mn2e.bst}
\bibliography{references}

\begin{thebibliography}{}

\bibitem[\protect\citeauthoryear{{Abazajian} et~al.,}{{Abazajian}
  et~al.}{2009}]{Abazajian_etal:2009}
{Abazajian} K.~N.,  et~al., 2009, \apjs, 182, 543

\bibitem[\protect\citeauthoryear{{Bailer-Jones}, {Smith}, {Tiede}, {Sordo} \&
  {Vallenari}}{{Bailer-Jones} et~al.}{2008}]{BailerJones_etal:2008}
{Bailer-Jones} C.~A.~L.,  {Smith} K.~W.,  {Tiede} C.,  {Sordo} R.,
  {Vallenari} A.,  2008, \mnras, 391, 1838

\bibitem[\protect\citeauthoryear{{Barkana} \& {Loeb}}{{Barkana} \&
  {Loeb}}{2001}]{Barkana_Loeb:2001}
{Barkana} R.,  {Loeb} A.,  2001, \physrep, 349, 125

\bibitem[\protect\citeauthoryear{{Becker}, {Fan}, {White} \& et al.}{{Becker}
  et~al.}{2001}]{Becker_etal:2001}
{Becker} R.~H.,  {Fan} X.,  {White} R.~L.,    et al. 2001, AJ, 122, 2850

\bibitem[\protect\citeauthoryear{{Ben{\'{\i}}tez}}{{Ben{\'{\i}}tez}}{2000}]{Be%
nitez:2000}
{Ben{\'{\i}}tez} N.,  2000, \apj, 536, 571

\bibitem[\protect\citeauthoryear{{Bovy} et~al.,}{{Bovy}
  et~al.}{2010}]{Bovy_etal:2010}
{Bovy} J.,  et~al., 2010, ApJ, submitted

\bibitem[\protect\citeauthoryear{{Casali} et~al.,}{{Casali}
  et~al.}{2007}]{Casali_etal:2007}
{Casali} M.,  et~al., 2007, \aap, 467, 777

\bibitem[\protect\citeauthoryear{{Dunkley} et~al.,}{{Dunkley}
  et~al.}{2009}]{Dunkley_etal:2009}
{Dunkley} J.,  et~al., 2009, \apjs, 180, 306

\bibitem[\protect\citeauthoryear{{Dye} et~al.,}{{Dye}
  et~al.}{2006}]{Dye_etal:2006}
{Dye} S.,  et~al., 2006, \mnras, 372, 1227

\bibitem[\protect\citeauthoryear{{Emerson}, {Sutherland}, {McPherson}, {Craig},
  {Dalton} \& {Ward}}{{Emerson} et~al.}{2004}]{Emerson_etal:2004}
{Emerson} J.~P.,  {Sutherland} W.~J.,  {McPherson} A.~M.,  {Craig} S.~C.,
  {Dalton} G.~B.,    {Ward} A.~K.,  2004, The Messenger, 117, 27

\bibitem[\protect\citeauthoryear{{Fan} et~al.,}{{Fan}
  et~al.}{2006}]{Fan_etal:2006b}
{Fan} X.,  et~al., 2006, AJ, 132, 117

\bibitem[\protect\citeauthoryear{{Fan}, {Hennawi}, {Richards} \& et al.}{{Fan}
  et~al.}{2004}]{Fan_etal:2004}
{Fan} X.,  {Hennawi} J.~F.,  {Richards} G.~T.,    et al. 2004, AJ, 128, 515

\bibitem[\protect\citeauthoryear{{Fan}, {Narayanan}, {Lupton} \& et al.}{{Fan}
  et~al.}{2001}]{Fan_etal:2001}
{Fan} X.,  {Narayanan} V.~K.,  {Lupton} R.~H.,    et al. 2001, AJ, 122, 2833

\bibitem[\protect\citeauthoryear{{Fan}, {Narayanan}, {Strauss}, {White},
  {Becker}, {Pentericci} \& {Rix}}{{Fan} et~al.}{2002}]{Fan_etal:2002}
{Fan} X.,  {Narayanan} V.~K.,  {Strauss} M.~A.,  {White} R.~L.,  {Becker}
  R.~H.,  {Pentericci} L.,    {Rix} H.-W.,  2002, \aj, 123, 1247

\bibitem[\protect\citeauthoryear{{Fan}, {Strauss}, {Schneider} \& et al.}{{Fan}
  et~al.}{2003}]{Fan_etal:2003}
{Fan} X.,  {Strauss} M.~A.,  {Schneider} D.~P.,    et al. 2003, AJ, 125, 1649

\bibitem[\protect\citeauthoryear{{Fukugita}, {Ichikawa}, {Gunn}, {Doi},
  {Shimasaku} \& {Schneider}}{{Fukugita} et~al.}{1996}]{Fukugita_etal:1996}
{Fukugita} M.,  {Ichikawa} T.,  {Gunn} J.~E.,  {Doi} M.,  {Shimasaku} K.,
  {Schneider} D.~P.,  1996, AJ, 111, 1748

\bibitem[\protect\citeauthoryear{{Gehrels}, {Ramirez-Ruiz} \& {Fox}}{{Gehrels}
  et~al.}{2009}]{Gehrels_etal:2009}
{Gehrels} N.,  {Ramirez-Ruiz} E.,    {Fox} D.~B.,  2009, \araa, 47, 567

\bibitem[\protect\citeauthoryear{{Glikman}, {Eigenbrod}, {Djorgovski},
  {Meylan}, {Thompson}, {Mahabal} \& {Courbin}}{{Glikman}
  et~al.}{2008}]{Glikman_etal:2008}
{Glikman} E.,  {Eigenbrod} A.,  {Djorgovski} S.~G.,  {Meylan} G.,  {Thompson}
  D.,  {Mahabal} A.,    {Courbin} F.,  2008, \aj, 136, 954

\bibitem[\protect\citeauthoryear{{Gunn} \& {Peterson}}{{Gunn} \&
  {Peterson}}{1965}]{Gunn_Peterson:1965}
{Gunn} J.~E.,  {Peterson} B.~A.,  1965, ApJ, 142, 1633

\bibitem[\protect\citeauthoryear{{Hambly} et~al.,}{{Hambly}
  et~al.}{2008}]{Hambly_etal:2008}
{Hambly} N.~C.,  et~al., 2008, \mnras, 384, 637

\bibitem[\protect\citeauthoryear{{Hazard}, {Mackey} \& {Shimmins}}{{Hazard}
  et~al.}{1963}]{Hazard_etal:1963}
{Hazard} C.,  {Mackey} M.~B.,    {Shimmins} A.~J.,  1963, \nat, 197, 1037

\bibitem[\protect\citeauthoryear{{Hewett}, {Foltz} \& {Chaffee}}{{Hewett}
  et~al.}{1995}]{Hewett_etal:1995}
{Hewett} P.~C.,  {Foltz} C.~B.,    {Chaffee} F.~H.,  1995, AJ, 109, 1498

\bibitem[\protect\citeauthoryear{{Hewett}, {Warren}, {Leggett} \&
  {Hodgkin}}{{Hewett} et~al.}{2006}]{Hewett_etal:2006}
{Hewett} P.~C.,  {Warren} S.~J.,  {Leggett} S.~K.,    {Hodgkin} S.~T.,  2006,
  MNRAS, 367, 454

\bibitem[\protect\citeauthoryear{{Ivezi{\'c}} et~al.,}{{Ivezi{\'c}}
  et~al.}{2004}]{Ivezic_etal:2004}
{Ivezi{\'c}} {\v Z}.,  et~al., 2004, IAUS, 222, 525

\bibitem[\protect\citeauthoryear{{Jaynes}}{{Jaynes}}{2003}]{Jaynes:2003}
{Jaynes} E.~T.,  2003, {Probability Theory}.
Cambridge University Press, Cambridge, UK

\bibitem[\protect\citeauthoryear{{Jiang} et~al.,}{{Jiang}
  et~al.}{2008}]{Jiang_etal:2008}
{Jiang} L.,  et~al., 2008, \aj, 135, 1057

\bibitem[\protect\citeauthoryear{{Jiang}, {Fan}, {Bian}, {Annis}, {Chiu},
  {Jester}, {Lin}, {Lupton}, {Richards}, {Strauss}, {Malanushenko},
  {Malanushenko} \& {Schneider}}{{Jiang} et~al.}{2009}]{Jiang_etal:2009}
{Jiang} L.,  {Fan} X.,  {Bian} F.,  {Annis} J.,  {Chiu} K.,  {Jester} S.,
  {Lin} H.,  {Lupton} R.~H.,  {Richards} G.~T.,  {Strauss} M.~A.,
  {Malanushenko} V.,  {Malanushenko} E.,    {Schneider} D.~P.,  2009, \aj, 138,
  305

\bibitem[\protect\citeauthoryear{{Kaiser} et~al.,}{{Kaiser}
  et~al.}{2002}]{Kaiser:2002}
{Kaiser} N.,  et~al., 2002, in {Tyson} J.~A.,  {Wolff} S.,  eds, Society of
  Photo-Optical Instrumentation Engineers (SPIE) Conference Series Vol.~4836 of
  Society of Photo-Optical Instrumentation Engineers (SPIE) Conference Series.
p.~154

\bibitem[\protect\citeauthoryear{{Lawrence} et~al.,}{{Lawrence}
  et~al.}{2007}]{Lawrence_etal:2007}
{Lawrence} A.,  et~al., 2007, \mnras, 379, 1599

\bibitem[\protect\citeauthoryear{{Lupton}, {Gunn} \& {Szalay}}{{Lupton}
  et~al.}{1999}]{Lupton_etal:1999}
{Lupton} R.~H.,  {Gunn} J.~E.,    {Szalay} A.~S.,  1999, AJ, 118, 1406

\bibitem[\protect\citeauthoryear{{Maddox}, {Hewett}, {Warren} \&
  {Croom}}{{Maddox} et~al.}{2008}]{Maddox_etal:2008}
{Maddox} N.,  {Hewett} P.~C.,  {Warren} S.~J.,    {Croom} S.~M.,  2008, \mnras,
  386, 1605

\bibitem[\protect\citeauthoryear{{Mortlock}}{{Mortlock}}{2009}]{Mortlock:2009}
{Mortlock} D.~J.,  2009, in {Hobson} M.~P.,  {Jaffe} A.~H.,  {Liddle} A.,
  {Mukherjee} P.,   {Parkinson} D.,  eds, Bayesian Methods In Cosmology
  {Cambridge University Press}, p.~193

\bibitem[\protect\citeauthoryear{{Mortlock} et~al.,}{{Mortlock}
  et~al.}{2009}]{Mortlock_etal:2009a}
{Mortlock} D.~J.,  et~al., 2009, \aap, 505, 97

\bibitem[\protect\citeauthoryear{{Mortlock}, {Patel}, {Warren}, {Venemans},
  {Hewett}, {McMahon}, {Simpson} \& {Sharp}}{{Mortlock}
  et~al.}{2011}]{Mortlock_etal:2011b}
{Mortlock} D.~J.,  {Patel} M.,  {Warren} S.~J.,  {Venemans} B.~P.,  {Hewett}
  P.~C.,  {McMahon} R.~G.,  {Simpson} C.~J.,    {Sharp} R.~G.,  2011, MNRAS, in
  preparation

\bibitem[\protect\citeauthoryear{{Mortlock}, {Peiris} \&
  {Ivezi{\'c}}}{{Mortlock} et~al.}{2009}]{Mortlock_etal:2009b}
{Mortlock} D.~J.,  {Peiris} H.~V.,    {Ivezi{\'c}} {\v Z}.,  2009, \mnras, 399,
  699

\bibitem[\protect\citeauthoryear{{Patel} et~al.,}{{Patel}
  et~al.}{2011}]{Patel_etal:2011}
{Patel} M.,  et~al., 2011, MNRAS, in preparation

\bibitem[\protect\citeauthoryear{{Pogson}}{{Pogson}}{1856}]{Pogson:1856}
{Pogson} N.,  1856, \mnras, 17, 12

\bibitem[\protect\citeauthoryear{{Press}, {Teukolsky}, {Vetterling} \&
  {Flannery}}{{Press} et~al.}{2007}]{Press_etal:2007}
{Press} W.~H.,  {Teukolsky} S.~A.,  {Vetterling} W.~T.,    {Flannery} B.~P.,
  2007, {Numerical Recipes: The Art Of Scientific Computing (3rd Ed.)}.
Cambridge University Press

\bibitem[\protect\citeauthoryear{{Richards} et~al.,}{{Richards}
  et~al.}{2004}]{Richards_etal:2004}
{Richards} G.~T.,  et~al., 2004, ApJS, 155, 257

\bibitem[\protect\citeauthoryear{{Richards} et~al.,}{{Richards}
  et~al.}{2009}]{Richards_etal:2009}
{Richards} G.~T.,  et~al., 2009, \aj, 137, 3884

\bibitem[\protect\citeauthoryear{{Schmidt}}{{Schmidt}}{1963}]{Schmidt:1963}
{Schmidt} M.,  1963, Nature, 197, 1040

\bibitem[\protect\citeauthoryear{{Schmidt}, {Schneider} \& {Gunn}}{{Schmidt}
  et~al.}{1995}]{Schmidt_etal:1995}
{Schmidt} M.,  {Schneider} D.~P.,    {Gunn} J.~E.,  1995, \aj, 110, 68

\bibitem[\protect\citeauthoryear{{Schneider}}{{Schneider}}{1999}]{Schneider:19%
99}
{Schneider} D.~P.,  1999, in {Holt} S.,  {Smith} E.,  eds, After the Dark Ages:
  When Galaxies were Young (the Universe at $2 < z < 5$) Vol.~470 of American
  Institute of Physics Conference Series, {Surveys for High-redshift Quasars}.
p.~233

\bibitem[\protect\citeauthoryear{{Scranton}, {Connolly}, {Szalay}, {Lupton},
  {Johnston}, {Budavari}, {Brinkman} \& {Fukugita}}{{Scranton}
  et~al.}{2005}]{Scranton_etal:2005}
{Scranton} R.,  {Connolly} A.~J.,  {Szalay} A.~S.,  {Lupton} R.~H.,  {Johnston}
  D.,  {Budavari} T.,  {Brinkman} J.,    {Fukugita} M.,  2005, \aj, submitted

\bibitem[\protect\citeauthoryear{{Sivia} \& {Skilling}}{{Sivia} \&
  {Skilling}}{2006}]{Sivia_Skilling:2006}
{Sivia} D.~S.,  {Skilling} J.,  2006, {Data Analysis: A Bayesian Tutorial}.
Oxford University Press, Oxford, UK

\bibitem[\protect\citeauthoryear{{Skrutskie} et~al.,}{{Skrutskie}
  et~al.}{2006}]{Skrutskie_etal:2006}
{Skrutskie} M.~F.,  et~al., 2006, \aj, 131, 1163

\bibitem[\protect\citeauthoryear{{Stark}, {Ellis}, {Chiu}, {Ouchi} \&
  {Bunker}}{{Stark} et~al.}{2010}]{Stark_etal:2010}
{Stark} D.~P.,  {Ellis} R.~S.,  {Chiu} K.,  {Ouchi} M.,    {Bunker} A.,  2010,
  \mnras, 408, 1628

\bibitem[\protect\citeauthoryear{{Venemans} et~al.,}{{Venemans}
  et~al.}{2011}]{Venemans_etal:2011}
{Venemans} B.~P.,  et~al., 2011, MNRAS, in preparation

\bibitem[\protect\citeauthoryear{{Venemans}, {McMahon}, {Warren},
  {Gonzalez-Solares}, {Hewett}, {Mortlock}, {Dye} \& {Sharp}}{{Venemans}
  et~al.}{2007}]{Venemans_etal:2007}
{Venemans} B.~P.,  {McMahon} R.~G.,  {Warren} S.~J.,  {Gonzalez-Solares} E.~A.,
   {Hewett} P.~C.,  {Mortlock} D.~J.,  {Dye} S.,    {Sharp} R.~G.,  2007,
  MNRAS, 376, L76

\bibitem[\protect\citeauthoryear{{Warren} et~al.,}{{Warren}
  et~al.}{2007}]{Warren_etal:2007}
{Warren} S.~J.,  et~al., 2007, \mnras, 375, 213

\bibitem[\protect\citeauthoryear{{Warren}, {Hewett} \& {Osmer}}{{Warren}
  et~al.}{1994}]{Warren_etal:1994}
{Warren} S.~J.,  {Hewett} P.~C.,    {Osmer} P.~S.,  1994, ApJ, 421, 412

\bibitem[\protect\citeauthoryear{{Willott} et~al.,}{{Willott}
  et~al.}{2007}]{Willott_etal:2007}
{Willott} C.~J.,  et~al., 2007, \aj, 134, 2435

\bibitem[\protect\citeauthoryear{{Willott} et~al.,}{{Willott}
  et~al.}{2010}]{Willott_etal:2010}
{Willott} C.~J.,  et~al., 2010, \aj, 139, 906

\bibitem[\protect\citeauthoryear{{Wyithe}}{{Wyithe}}{2008}]{Wyithe:2008}
{Wyithe} J.~S.~B.,  2008, \mnras, 387, 469

\bibitem[\protect\citeauthoryear{{York} et~al.,}{{York}
  et~al.}{2000}]{York_etal:2000}
{York} D.~G.,  et~al., 2000, \aj, 120, 1579

\end{thebibliography}

%%%%%%%%%%%%%%%%%%%%%%%%%%%%%%%%%%%%%%%%%%%%%%%%%%%%%%%%%%%%%%%%%%%%%%%%%%%%%%

\appendix 

\begin{figure}
\includegraphics[width=\figwidth]{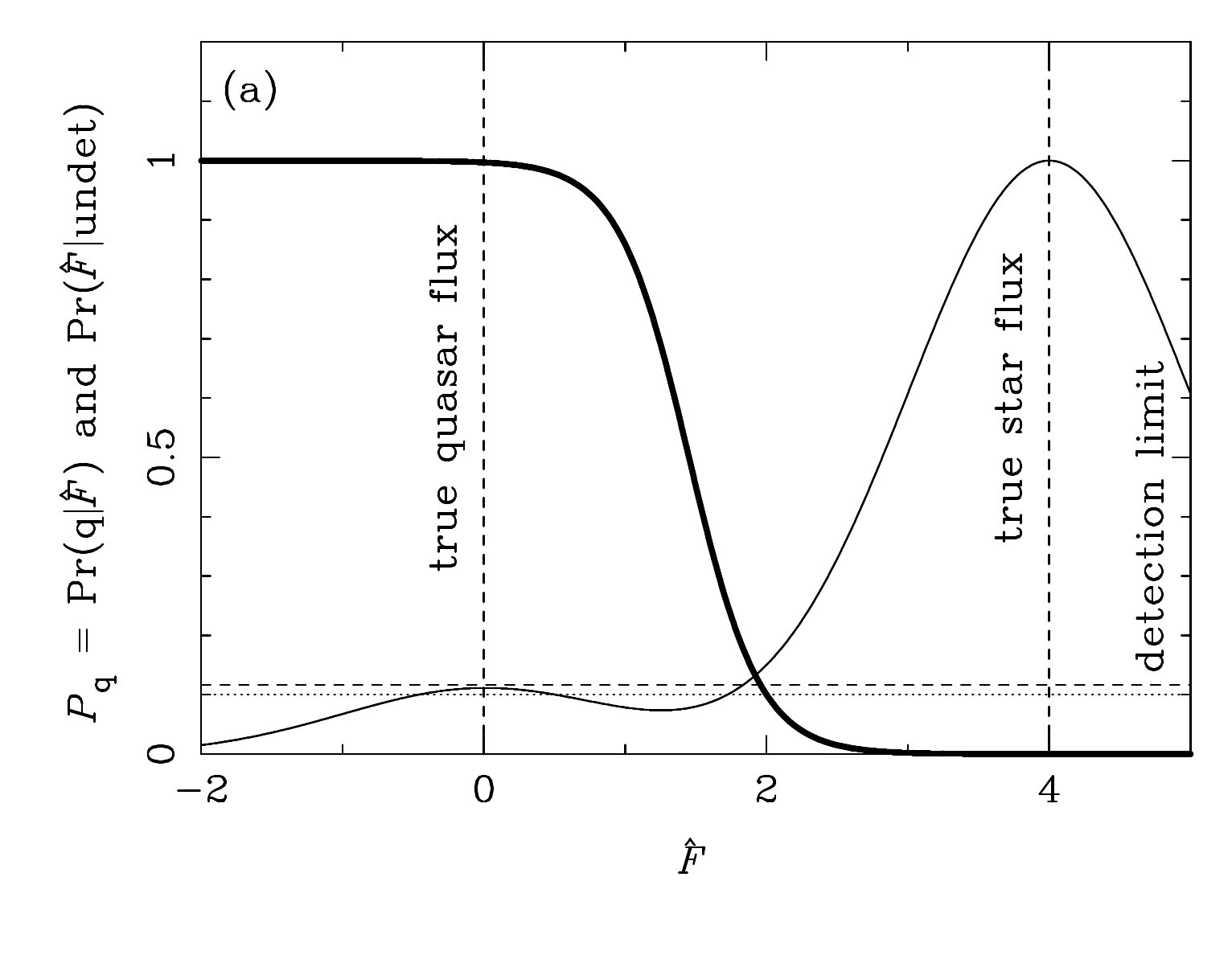}
\caption{The probabilty that a source is a quasar, $\pq$,
  shown as a function of the estimated flux, $\hat{\flux}$,
  in a band blueward of the \lya\ break.
  Three values of the probability are shown:
  the prior probability implied by the relative numbers of stars
  and quasars (dotted horizontal line);
  the posterior that would have been obtained if
  only the upper limit that $\hat{\flux} \leq 5 \sigma$ was
  available (dashed horizontal line);
  and the posterior obtained from the
  actual flux measurement (thick solid curve).
  Also shown is the (unnormalised) distribution of measured fluxes
  that arises from the relative numbers of stars and quasars
  and their true flux values (thin solid curve).}
\label{figure:nondet}
\end{figure}

%%%%%%%%%%%%%%%%%%%%%%%%%%%%%%%%%%%%%%%%%%%%%%%%%%%%%%%%%%%%%%%%%%%%%%%%%%%%%%

\section{The likelihood for a single photometric measurement}
\label{section:likelihoodfaint}

Some care must be taken in formulating the likelihood 
for photometric data.
As argued in \sect{likelihood},
there are large number of potential complications 
(\eg\ 
the Poisson fluctuations in the number of source photons received;
the influence of nearby sources;
cosmic rays, bad pixels, \etc, that cause 
the noise distribution to have non-Gaussian tails;
inter-band noise correlations),
but in most cases it is unnecessary to model all these effects.
The focus here is on a standard optical or \nir\ observation
of a faint source for which there is a non-negligible chance
that the background-subtracted counts are negative.
For such faint sources\footnote{The 
Gaussian approximation is also appropriate
for bright sources
from which the number of source photons expected
in the observation is sufficiently high
that the Poisson distribution of received photons is
sufficiently narrow.},
the likelihood is very well approximated as a Gaussian 
in flux units (\cf\ Eq.~\ref{equation:likelihood}).

In practice, however,
optical and NIR photometric data are seldom reported in the form 
of the estimated flux and the associated error,
and some of the limitations of the common alternative representations 
are described here.
The most fundamental problem is the use of upper limits 
in the case of non-detections, 
which discards potentially vital information (\sect{upperlimits}).
The use of magnitudes 
to report the measurements of faint sources 
can be anywhere from awkward to incorrect
(as shown in \sect{fluxunits}),
and so formulae for transforming 
magnitude data into flux units are given in \sect{formulae}.

%%%%%%%%%%%%%%%%%%%%%%%%%%%%%%%%%%%%%%%%%%%%%%%%%%%%%%%%%%%%%%%%%%%%%%%%%%%%%%

\subsection{Non-detections and upper limits}
\label{section:upperlimits}

A source is commonly considered to have been detected
if, in a particular obervation, it has a 
measured flux of $\hat{\flux} \ga \stonmin \, \sigma$, 
where $\sigma$ is the effective background uncertainty in the image
(expressed in flux units) and 
$\ston \geq \stonmin \simeq 5 \sigma$
is the minimum signal--to--noise ratio required for a source
to be considered as `detected'.
Having made a detection,
it is standard to report both $\hat{\flux}$ and $\sigma$ 
(albeit possibly transformed into magnitudes; see \sect{fluxunits}).
For most faint sources 
(\eg\ with $\hat{\flux} \la 10 \, \stonmin \, \sigma$)
all the significant information in the measurement is encoded in
$\hat{\flux}$ and $\sigma$.
If, however,
$\hat{\flux} < \stonmin \sigma$ 
(\ie, the source is undetected),
then it is rare that both $\hat{\flux}$ and $\sigma$ are reported.
Transforming instead to, \eg,
a 3-$\sigma$ upper limit of $\hat{\flux} + 3 \sigma$ 
is a common (if pointless) option,
but it at least there is no loss of information 
(provided the value of $\sigma$ is also retained).
However, either giving the above upper limit alone 
or, worse, reporting only that a source's measured flux 
was below $\flux_\limiting = \stonmin \, \sigma$,
does discard some information.
The question addressed here is:
If just $\flux_\limiting$ is reported
(instead of $\hat{\flux}$ and $\sigma$ together),
is the information loss significant?

In statistical terms,
the difference manifests itself in the likelihood.
Given a flux measurement, the likelihood is,
as in \eq{likelihood},
\begin{equation}
\label{equation:lik_flux_apdx}
\prob(\hat{\flux} | \flux) = \frac{1}{(2 \pi)^{1/2} \, \sigma}
  \exp \left[ - \frac{1}{2} 
    \left( 
      \frac{\hat{\flux} - \flux}{\sigma}
    \right)^2 \right],
\end{equation}
where $\flux$ is the true flux of the source.
With access only an upper limit,
the likelihood is the probability
that the background-subtracted counts are below the detection 
threshold, which is given by
\begin{equation}
\label{equation:likelihood_upper_apdx}
\prob( \hat{\flux} < \flux_{\limiting} | \flux )
  = 
  \frac{1}{2} 
    \left[
    1 + \erf\left( 
      \frac{\flux_\limiting - \flux}
      {2^{1/2} \sigma} \right)
    \right],
\end{equation}
where $\erf(x) = 2 \, \pi^{-1/2} \int_0^x e^{-t^2} \, \diff t$
is the error function.

What would be the result of adopting 
an upper limit in place of a low \stonlong\ flux measurement?
As the first is a probability density and the second a 
(cumulative) probability,
it is not particularly meaningful to compare the likelihoods directly.
Instead, it is more appropriate to explore 
whether the resultant inferences differ significantly,
which can be done by performing 
a simplified 
version of the model selection problem discussed in \sect{prob}.
Assuming a 
surface density $\Sigma_\str$ of stars, 
all of which have a flux of $\flux_\str$ in the band in question, 
and a surface density $\Sigma_\qso$ of quasars,
all of which have a flux of $\flux_\qso$ in this band,
then the probability that a source is a quasar, $\pq$,
is given by inserting either \eq{lik_flux_apdx}
or \eq{likelihood_upper_apdx} 
into \eq{pq}.
If flux measurements are used then
\begin{equation}
\pq = \frac{\Sigma_\qso 
    \exp\left[- \frac{1}{2}\left(\frac{\hat{\flux} - \flux_\qso}
    {\sigma}\right)^2\right]}
    {\Sigma_\qso 
    \exp\left[- \frac{1}{2}\left(\frac{\hat{\flux} - \flux_\qso}
    {\sigma}\right)^2\right]
  + \Sigma_\str 
    \exp\left[- \frac{1}{2}\left(\frac{\hat{\flux} - \flux_\str}
    {\sigma}\right)^2\right]};
\end{equation}
if only upper limits are available then 
\begin{equation}
\pq = \frac{
  \Sigma_\qso \left[1 + \erf\left( \frac{\flux_\limiting - \flux_\qso}
      {2^{1/2} \sigma} \right) \right]}
  {\Sigma_\qso \left[1 + \erf\left( \frac{\flux_\limiting - \flux_\qso} 
      {2^{1/2} \sigma} \right) \right]
   + \Sigma_\str \left[1 + \erf\left( \frac{\flux_\limiting - \flux_\str} 
     {2^{1/2} \sigma} \right) \right]
   }.
\end{equation}
These two expressions for $\pq$ are compared in \fig{nondet}, 
under the assumption that $\Sigma_\str = 10 \Sigma_\qso$,
$\flux_\str = 4 \sigma$ and $\flux_\qso = 0$.
The quasar probability inferred from the upper limit
is independent of $\hat{\flux}$
(provided it is lower than the detection limit) and,
in this case, happens to be similar
to the prior probability given by the relative 
numbers of stars and quasars.
Using the formal flux estimate, however, 
the quasar hypothesis is decisively rejected if $\hat{\flux} \ga 3 \sigma$.
Moreover, for the priors and model fluxes chosen here,
the majority of measurements would be in this regime,
because most of the undetected sources would be stars
for which the measured flux was only just below
the detection threshold.
This predicted distribution of measured fluxes is also 
shown in \fig{nondet}.

In this simplified example the use of an upper limit in place of a formal
flux estimate results in unnecessary extra
uncertainty in the classification of most `undetected' sources.
The equivalent calculation for the
\sdssukidss\ sample
is, of course, more complicated,
but the above result still holds:
the majority of candidates which 
are not detected in the $\iprime$ and $\zprime$ bands
have $\hat{\flux} \ga 3 \sigma$, 
sufficient to reject them as possible \hzqs\ with great confidence.
A broader implication of the above arguments is that 
measurements and uncertainties should always be reported in
preference to upper limits.

%%%%%%%%%%%%%%%%%%%%%%%%%%%%%%%%%%%%%%%%%%%%%%%%%%%%%%%%%%%%%%%%%%%%%%%%%%%%%%

\subsection{The use of magnitudes to represent photometric data}
\label{section:fluxunits}

Photometric data are usually reported in terms of 
either logarthmic or asinh magnitudes.
Given an estimated magnitude and its associated error it is 
then natural to adopt a Gaussian likelihood based on these 
values
(or equivalently, to construct a least-squares estimate in 
terms of the appropriately weighted magnitude differences).
Gaussians in magnitude and flux units are obviously not 
equivalent, 
but are the differences sufficient to 
result in significantly changed inferences?

The starting point to answering this question is the basic
formulae that relate magnitude to flux 
(\sect{logmag} and \ref{section:asinhmag}).
Using these definitions it is then possible to
transform a Gaussian distribution from magnitude units to 
flux units to see how the resultant inferences differ (\sect{gausslik}).

%%%%%%%%%%%%%%%%%%%%%%%%%%%%%%%%%%%%%%%%%%%%%%%%%%%%%%%%%%%%%%%%%%%%%%%%%%%%%%

\subsubsection{Logarithmic magnitudes}
\label{section:logmag}

The traditional logarithmic magnitude
corresponding to (positive) flux $\flux$ is given by 
\citep{Pogson:1856} as
\begin{equation}
\label{equation:logmag}
m = - \frac{5}{2 \ln(10)} 
  \ln \left( \frac{\flux}{\flux_0} \right),
\end{equation}
where $\flux_0$ is the zero--point flux 
for which $m = 0$ by construction.
This formula can be inverted to give
\begin{equation}
\label{equation:logmaginv}
\flux = \flux_0 
  \exp\left[{- \frac{2 \ln(10)}{5} m} \right].
\end{equation}
Differentiating \eq{logmag} gives the Jacobian used
to convert probability densities as
\begin{equation}
\label{equation:dmdf_log}
\left| \frac{\diff m}{\diff \flux} \right|
  = \frac{5}{2 \ln(10) \flux}.
\end{equation}

%%%%%%%%%%%%%%%%%%%%%%%%%%%%%%%%%%%%%%%%%%%%%%%%%%%%%%%%%%%%%%%%%%%%%%%%%%%%%%

\subsubsection{Asinh magnitudes}
\label{section:asinhmag}

The asinh magnitude scheme was introduced by \cite{Lupton_etal:1999}
to overcome the inability of the logarithmic magnitudes to represent
negative flux estimates, 
while retaining the familiar behaviour for high fluxes.
Asinh magnitudes are defined by 
\begin{equation}
\label{equation:asinhmag}
m = m_0 - \frac{5}{2 \ln(10)} 
    \asinh\left( \frac{1}{2} 
      \frac{\flux}{10^{- 2 m_0 / 5} \flux_0} \right) ,
\end{equation}
where $m_0$ is,
in the limit of high $m_0$,
the zero--point asinh magnitude corresponding to zero flux.
As
$\lim_{x \rightarrow \infty} \asinh(x/2) = \ln(x)$, 
\eq{asinhmag} approaches \eq{logmag} 
for $\flux \gg 10^{- 2 m_0 / 5} \flux_0$;
but for $|\flux| \la 10^{- 2 m_0 / 5} \flux_0$ 
the asinh magnitude is proportional to flux.
Inverting \eq{asinhmag} 
gives the flux as
\begin{equation}
\label{equation:asinhmaginv}
\flux = 2 \times 10^{- 2 m_0 / 5} \flux_0
  \sinh\left[ \frac{2 \ln(10)}{5} (m_0 - m) \right] .
\end{equation}
Differentiating \eq{asinhmag} then gives the Jacobian needed to 
transform variables as 
\begin{equation}
\label{equation:dmdf_asinh}
\left| \frac{\diff m}{\diff \flux} \right| 
  = \frac{5}{2 \ln(10)}  
    \frac{1}
    {\left[
      \flux^2 + 4 (10^{- 2 m_0 / 5} \flux_0)^2  
    \right]^{1/2}}.
\end{equation}

%%%%%%%%%%%%%%%%%%%%%%%%%%%%%%%%%%%%%%%%%%%%%%%%%%%%%%%%%%%%%%%%%%%%%%%%%%%%%%

\subsubsection{Gaussian likelihoods in magnitude units}
\label{section:gausslik}

\begin{figure*}
\includegraphics[width=\midfigwidth]{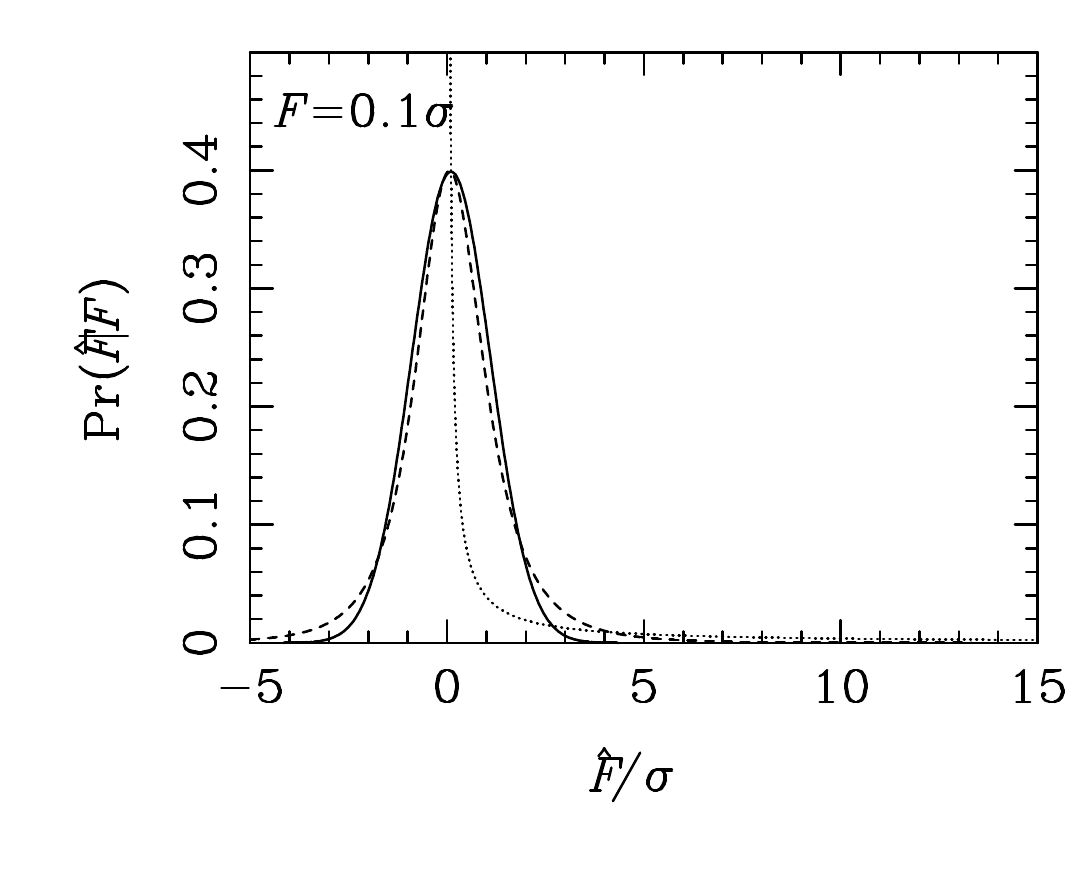}
\includegraphics[width=\midfigwidth]{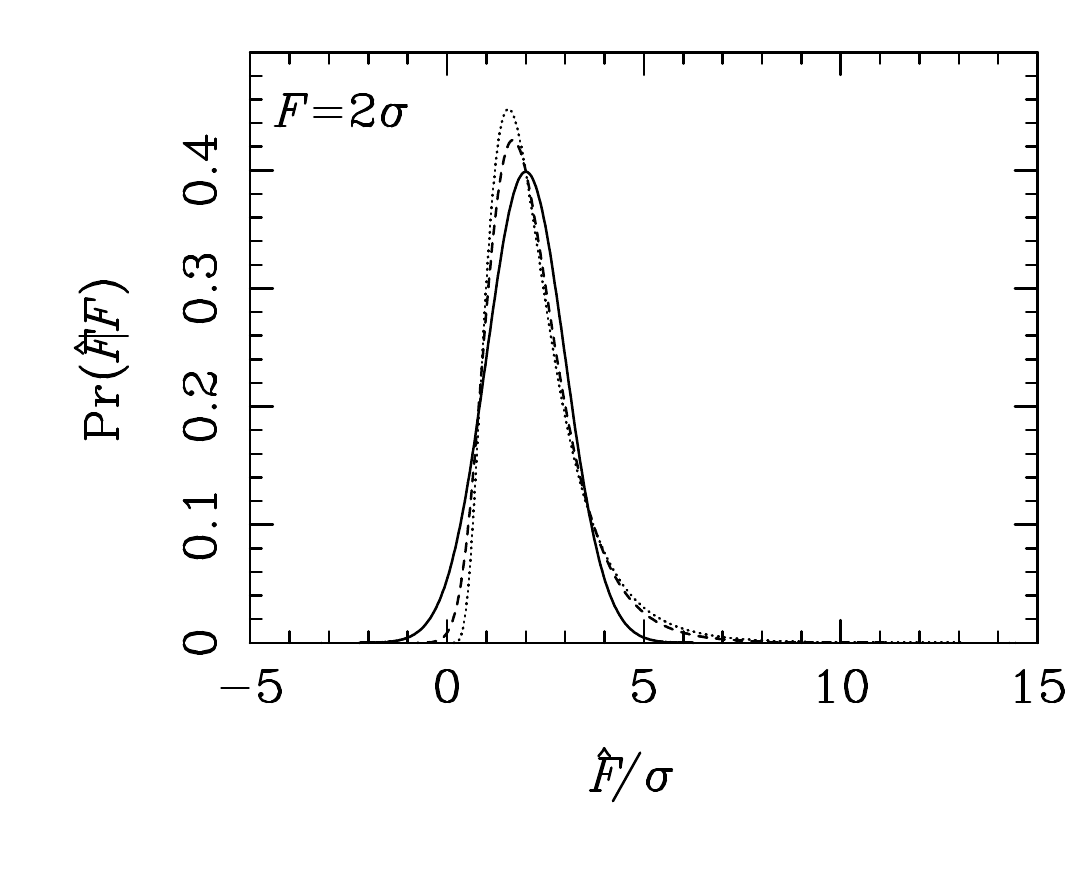}
\includegraphics[width=\midfigwidth]{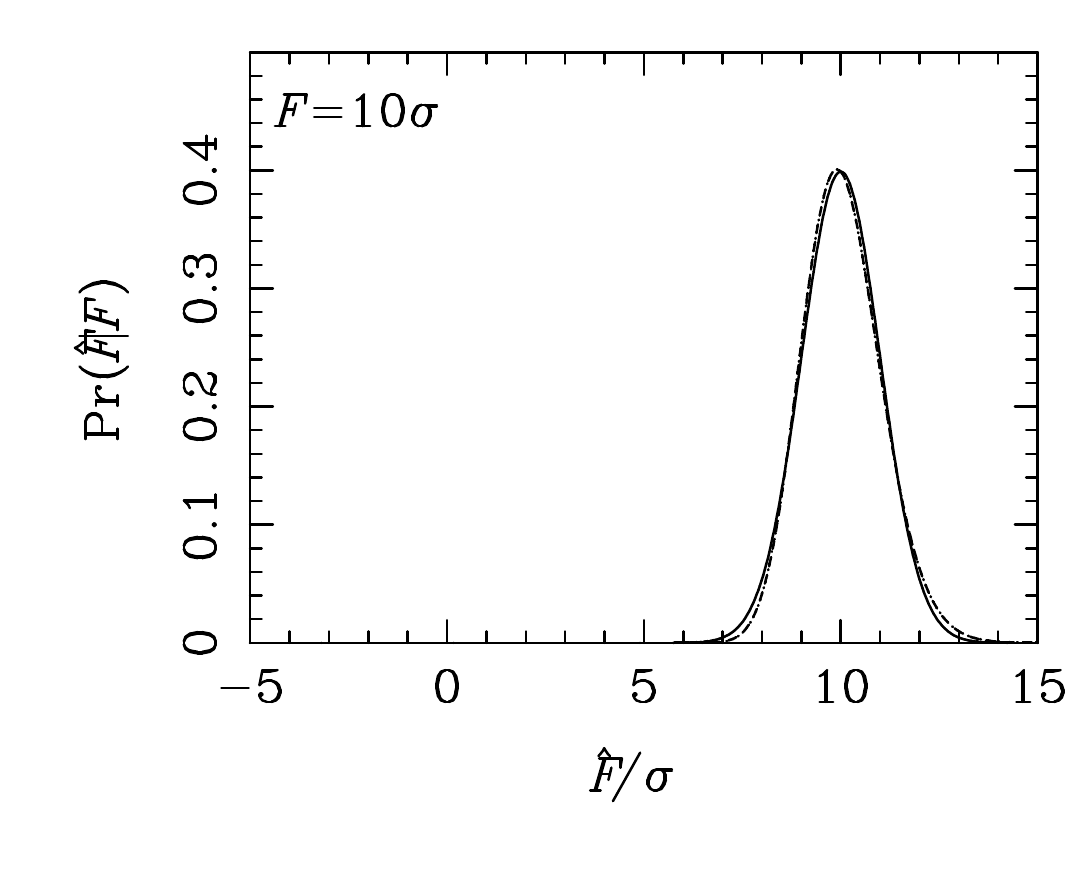}
\caption{Gaussian likelihoods in flux (solid curves),
  logarithmic magnitudes (dotted curves)
  and asinh magnitudes (dashed curves)
  transformed into flux units
  for sources with $\flux = 0.1 \sigma$, $\flux = 2.0 \sigma$
  and $\flux = 10.0 \sigma$, as labelled,
  where $\sigma$ is the background noise in flux units.
  The zero--point asinh magnitude was chosen so that its
  relationship with $\sigma$ matches
  typical SDSS $\zprime$-band observations.}
\label{figure:likelihood}
\end{figure*}

\begin{figure}
\includegraphics[width=\figwidth]{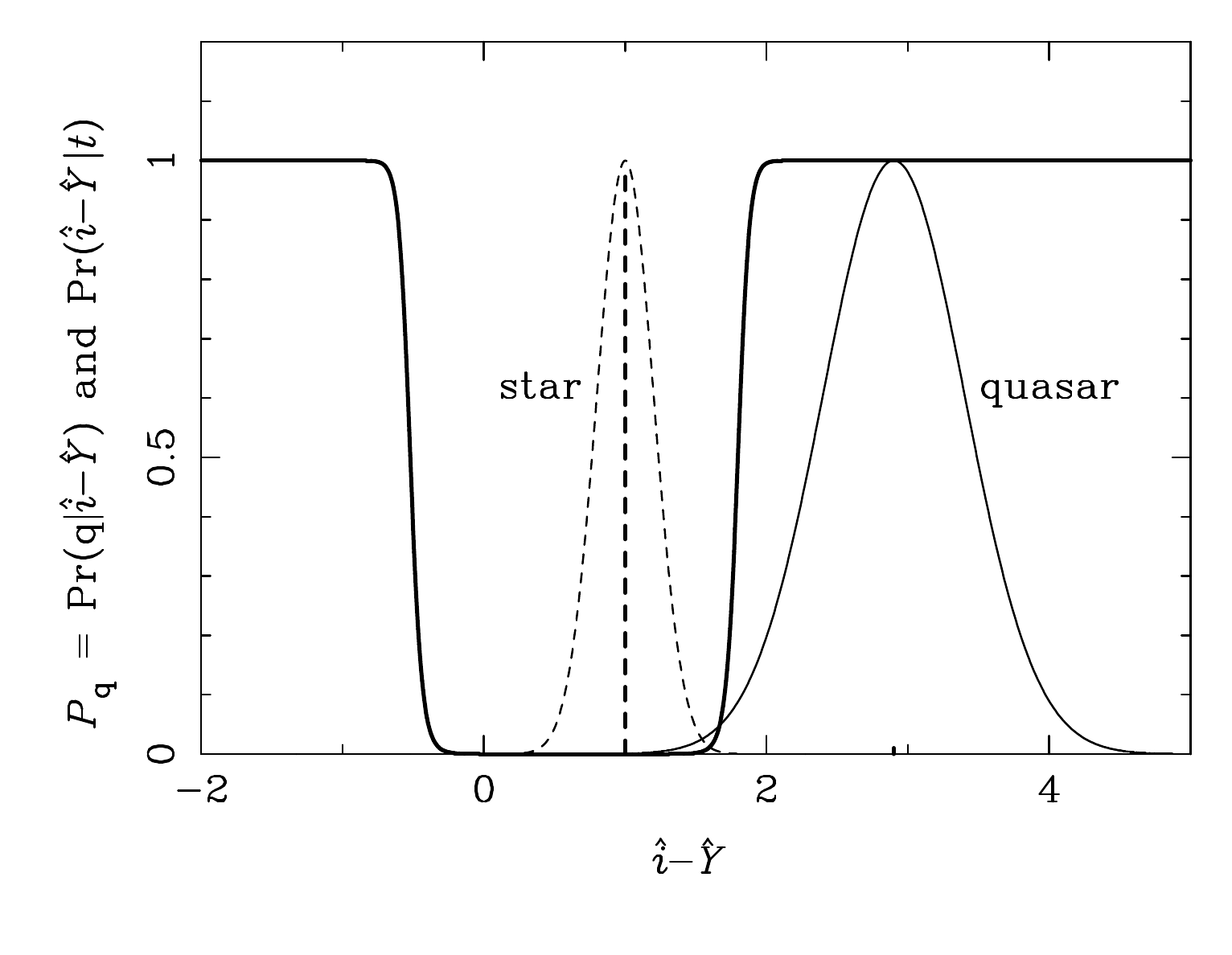}
\caption{The probability that a source is a quasar, $\pq$,
  shown as a function of the observed $\imy$ colour of the source.
  The thin lines show the (arbitrarily normalised) likelihoods
  of the star and the quasar, which are assumed to be Gaussian in
  magnitude units; the thick vertical lines give the relative
  normalisation of the two populations.
  The nonsensical rise of $\pq$ to unity for $\hat{i} - \hat{Y} \la 0$
  is purely an artefact of the seemingly innocent assumption that
  the likelihood is Gaussian in magnitude units.}
\label{figure:modelcomp}
\end{figure}

Given an estimated magnitude, $\hat{m}$, and an error
$\Delta \hat{m}$, it is common to assume that the likelihood
is Gaussian in magnitude units, and hence given by 
\begin{equation}
\label{equation:lik_mag}
\prob(\hat{m} | m) = \frac{1}{(2 \pi)^{1/2} \, \Delta \hat{m}}
  \exp \left[ - \frac{1}{2} 
    \left( \frac{\hat{m} - m}{\Delta \hat{m}} \right)^2 \right],
\end{equation}
where 
$m$ is the true magnitude of the source in this band.
Changing variables to flux units, the
implied likelihood would then be
\[
\prob(\hat{\flux} | \flux)
\]
\begin{equation}
\mbox{}
  = \frac{1}{(2 \pi)^{1/2} \, \Delta \hat{m}}
  \exp \left\{ - \frac{1}{2} 
    \left[ \frac{m(\hat{\flux}) - m(\flux)}{\Delta \hat{m}} \right]^2 
    \right\} 
  \left| \frac{\diff m}{\diff \flux} \right|,
\end{equation}
where 
$m(\flux)$ 
and 
$| \diff m / \diff \flux|$ are given in 
\eqs{logmag} and (\ref{equation:dmdf_log}) for logarithmic magnitudes and 
\eqs{asinhmag} and (\ref{equation:dmdf_asinh}) for asinh magnitudes,
respectively.

The resultant expressions for the likelihood
are straightforward, but cumbersome;
they are more easily explored visually, as is done in \fig{likelihood}.
For the brighter source with a true flux of $\flux = 10 \sigma$ 
(where $\sigma$ is the background noise in flux units) 
the Gaussian likelihoods in terms of both magnitudes 
are consistent with the true Gaussian in flux units,
and of course almost any subsequent inferences would be similarly
accurate.
For the fainter sources with true fluxes of 
$\flux = 0.1 \sigma$
and 
$\flux = 2 \sigma$,
however, the differences
between the likelihood formaluations are readily apparent.
Most importantly,
negative measured fluxes
cannot be represented in 
the traditional logarithmic magnitude system,
rendering it useless in such low $\ston$ situations.
The situation for asinh magnitudes is more complicated:
with $m \propto \flux$ for small $|\flux|$, the likelihood
in the $\flux = 0.1 \sigma$ case is fairly accurate;
but, given the choice of softening parameter used here,
the are significant differences in the $\flux = 2 \sigma$ case.

Any conclusions drawn from 
magnitude-based Gaussian likelihoods will 
be discrepant, and in some cases catastrophically so.
In the case of the toy model adopted in \sect{upperlimits},
using a Gaussian in magnitude units can result in nonsensical 
inferences, as shown in \fig{modelcomp}.
The absurd rise of $\pq$ to near unity for unusually blue
sources 
(that are on the `far' side of the stellar locus from the quasars)
is because the weighted Gaussian in magnitude units for the 
brighter, well measured, stars is less than the down-weighted Gaussian
for the fainter, less accurately measured, quasars.
Despite the much lower prior for the quasars,
the stars' smaller error eventually dominates. 
The same calculation performed in flux units does not have this 
problem as, for these faint sources, 
the background error is the same for both in flux units.
For extremely blue or red sources which are a bad fit to both 
models the likelihood is almost irrelevant and the prior is the 
dominant factor, as it should be.

Operationally, the solution to these potential complications is to 
convert magnitude data into flux units for the purposes of 
any likelihood-based calculations,
the formulae for which are given below.

%%%%%%%%%%%%%%%%%%%%%%%%%%%%%%%%%%%%%%%%%%%%%%%%%%%%%%%%%%%%%%%%%%%%%%%%%%%%%%

\subsubsection{Magnitude to flux conversions}
\label{section:formulae}

Given the importance of performing statistical calculations in flux
units, formulae are needed to convert an estimated magnitude,
$\hat{m}$, and its associated error, $\Delta \hat{m}$, into 
an estimated flux, $\hat{\flux}$, and,
in the case of faint sources, the background noise, $\sigma$.
The correct conversions can be derived from the fact that 
$\hat{m}$ and $\Delta \hat{m}$
are inevitably calculated from 
$\hat{\flux}$ and $\sigma$ in the first place.
The problem would be more difficult if 
reported magnitudes and uncertainties were obtained using 
more complicated statistical arguments, 
but for the simple conversions commonly adopted all
that is required is to invert the relationships in 
\sects{logmag} and \ref{section:asinhmag}. 

Given $\hat{m}$ and $\Delta \hat{m}$ 
in conventional logarithmic magnitudes,
a straightforward subsitution into 
\eq{logmaginv} yields
\begin{equation}
\label{equation:ffromlogm}
\hat{\flux} = \flux_0 \exp\left[ - \frac{2 \ln(10)}{5} \hat{m} \right].
\end{equation}
Then using \eq{dmdf_log} 
gives 
\begin{eqnarray}
\label{equation:sigmafromlogm}
\sigma 
  & = & \flux_0 \frac{2 \ln(10)}{5} 
  \exp\left[- \frac{2 \ln(10)}{5} \hat{m} \right] \Delta \hat{m}
\nonumber \\
  & = & \hat{\flux} \frac{2 \ln(10)}{5} \, \Delta \hat{m}  
  \simeq \Delta \hat{m} \, \hat{\flux}.
\end{eqnarray}

Given $\hat{m}$ and $\Delta \hat{m}$
in terms of asinh magnitudes,
the corresponding flux estimate is given from
\eq{asinhmaginv} yields
\begin{equation}
\label{equation:ffromsinhm}
\hat{\flux} = 2 \times 10^{- 2 m_0 / 5} \flux_0
  \sinh\left[ \frac{2 \ln(10)}{5} (m_0 - \hat{m}) \right] 
\end{equation}
Then using \eq{dmdf_log} to change variables gives 
the background error in flux units as 
\begin{equation}
\label{equation:sigmafromsinhm}
\sigma = 
  \frac{4 \ln(10)}{5} 10^{-2 m_0 / 5} \flux_0
  \cosh\left[ \frac{2 \ln(10)}{5} (m_0 - \hat{m}) \right].
\end{equation}

For the calculation of $\pq$ in \sect{example},
\eqs{ffromlogm} and (\ref{equation:sigmafromlogm})
were used to convert the reported
UKIDSS $Y$- and $J$-band photometry to flux units
and 
\eqs{ffromsinhm} and (\ref{equation:sigmafromsinhm})
were used to convert the 
reported SDSS $\iprime$- and $\zprime$-band 
SDSS photometry to flux units.

%%%%%%%%%%%%%%%%%%%%%%%%%%%%%%%%%%%%%%%%%%%%%%%%%%%%%%%%%%%%%%%%%%%%%%%%%%%%%%

\section{Parameter fitting for the stellar population model}
\label{section:starfit}

In \sect{star} it was necessary to fit the parameters of an
empirical model of the intrinsic stellar colour and magnitude distribution 
to a sample of \sdssukidss\ point--sources extracted from the WSA.
Many methods exist for tackling this problem, 
although the fact that the observed distribution is
the result of convolving the underlying target distribution
with magnitude-dependent noise makes this task non-trivial in this case.
The overall theme of this paper would suggest taking 
a Bayesian approach, 
but the primary aim here is to find any function to describe
the intrinsic stellar distribution that is consistent with the observed data; 
the actual parameter values (and their uncertainties) are not of interest.
Hence a faster, if less principled, method was used.

After selecting $\sim 10^5$ bright, red \sdssukidss\ stars 
(with $\imyobs \geq 2.0$ and $15.0 \leq \hat{Y} \leq 19.5$),
the sample was
binned in $\hat{Y}$ and $\imyobs$, 
with a bin size of 0.1.
The data was hence reduced 
to the number of objects, $n_\bin$,
in each of $\nbin$ cells
(where the index $\bin$ covers the
two-dimensional parameter space).
For a given choice of the free parameters, $\poppars$,
describing the model under consideration
(as distinct from the parameters $\parameters_\str$ 
used to characterise a single star),
the expected number of stars in each cell,
$\bar{n}_\bin(\poppars)$, was calculated by
convolving the intrinstic population
with the appropriate photometric error distribution.
It also important to ensure 
that the value of 
$\bar{n}_\bin(\poppars)$ 
is not spuriously high due to the large number 
of faint undetected sources that could,
at least in theory, be scattered into the bin.
The potential problem with a simple treatment is that 
such numerous faint sources are typically beyond the 
confusion limit of the survey
and so to treat a single noise spike as being associated 
with each in turn results in an unrealistically high 
probability of a spurious source entering the survey.
The key is to adopt a self-consistent treatment in
which sources at or below the confusion limit are ignored
(\cf\ \citealt{Mortlock:2009}).

Irrespective of the method by which 
$\bar{n}_\bin(\poppars)$ is calculated,
$n_\bin$ is Poisson-distributed and uncorrelated between bins.
The log-likelihood of the full data-set is hence
\begin{eqnarray}
\label{equation:lnl}
\ln[\prob(\vect{n}| \poppars)]
  & = & \ln \left\{ \prod_{\bin = 1}^{\nbin}
      \frac{[\bar{n}_\bin(\poppars)]^{n_\bin} \exp[- \bar{n}_\bin(\poppars)] }
      {n_\bin!} \right\}
    \\
  & = & \sum_{\bin = 1}^{\nbin}
    n_\bin \ln[\bar{n}_\bin(\poppars)] - \bar{n}_\bin(\poppars)
    + {\rmn{constant}}.
    \nonumber  
\end{eqnarray}
The best fit parameters given in \eq{star} were found by
minimizing \eq{lnl} using 
the downhill simplex method \cite{Press_etal:2007}.

%%%%%%%%%%%%%%%%%%%%%%%%%%%%%%%%%%%%%%%%%%%%%%%%%%%%%%%%%%%%%%%%%%%%%%%%%%%%%%

\bsp
\label{lastpage}
\end{document}